# Wind-induced changes to surface gravity wave shape in deep to intermediate water

**Thomas Zdyrski[1]†, and Falk Feddersen[1]**

[1]Scripps Institution of Oceanography, UCSD, La Jolla, CA 92092-0209, USA

Wave shape (i.e. skewness or asymmetry) plays an important role in beach morphology evolution, remote sensing and ship safety. The wind's influence on ocean waves has been extensively studied theoretically in the context of growth, but most theories are phase averaged and cannot predict wave shape. Most laboratory and numerical studies similarly focus on wave growth. A few laboratory experiments have demonstrated that wind can change wave shape, and two-phase numerical simulations have also noted wind-induced wave-shape changes. However, the wind's effect on wave shape is poorly understood, and no theory for it exists. For weakly nonlinear waves, wave-shape parameters are the phase of the first harmonic relative to the primary frequency (or harmonic phase HP, zero for a Stokes wave) and relative amplitude of the first harmonic to the primary wave. Here, surface pressure profiles (denoted Jeffreys, Miles and generalized Miles) are prescribed based on wind–wave generation theories. Theoretical solutions are derived for quasi-periodic progressive waves and the wind-induced changes to the HP, relative harmonic amplitude, as well as the already known phase speed changes and growth rates. The wave-shape parameters depend upon the chosen surface pressure profile, pressure magnitude and phase relative to the wave profile and non-dimensional depth. Wave asymmetry is linked to the non-dimensional growth rate. Atmospheric large eddy simulations constrain pressure profile parameters. The HP predictions are qualitatively consistent with laboratory observations. This theory, together with the HP and relative harmonic amplitude observables, can provide insight into the actual wave surface pressure profile.

## 1. Introduction

The shape of surface gravity waves plays a role in many physical phenomena. Wave shape is described by the third-order statistical moments, skewness and asymmetry (e.g. Hasselmann 1962; Elgar *et al.* 1990). Wave skewness represents a wave's vertical asymmetry, while wave asymmetry corresponds to its horizontal asymmetry. These two parameters are integral in determining sediment transport direction (onshore *vs.* offshore) and magnitude (e.g. Drake & Calantoni 2001; Hsu & Hanes 2004; Gonzalez-Rodriguez & Madsen 2007), which play key roles in beach morphodynamics (e.g. Hoefel & Elgar 2003; Grasso *et al.* 2011). Wave shape is also pertinent in remote sensing, where wave skewness modulates the returned waveform in radar altimetry (e.g. Jackson 1979; Hayne 1980; Huang *et al.* 1983) and wave asymmetry affects the thermal emissions measured in polarimetric radiometry (e.g. Kunkee & Gasiewski 1997; Piepmeier & Gasiewski 2001; Johnson & Cai 2002). Additionally, these wave-shape parameters play a role in determining ship response to wave impacts (e.g. Soares *et al.* 2008; Oberhagemann *et al.* 2013). Waves propagating on a flat bottom are ordinarily symmetric, although a number of processes can create asymmetry. While some wave asymmetry-inducing phenomena, such as wave shoaling (e.g. Elgar & Guza 1985, 1986) and vertically sheared currents (e.g. Chen & Zou 2018, 2019), are well understood, the wind's effect on wave shape is still poorly understood.

† Email address for correspondence: tzdyrski@uscd.edu

The influence of wind on ocean waves has been extensively studied, although primarily in the context of wave growth. An initial investigation by Jeffreys (1925) was based on a sheltering hypothesis, where separated airflow resulted in reduced pressure on the wave's leeward side, causing wave growth. While conceptually simple, this mechanism has largely fallen out of favour because such separation only seems to occur near breaking (Banner & Melville 1976) or for steep waves under strong winds (e.g. Touboul & Kharif 2006; Tian & Choi 2013). Nevertheless, Jeffreys's theory has inspired some recent work; Belcher & Hunt (1993) developed a fully turbulent model wherein the sheltering effect causes a thickening of the boundary layer and wave growth, even without separation. Later treatments utilized different physical mechanisms such as resonant forcing by incoherent, turbulent eddies (Phillips 1957), vortex forcing from vertically sheared airflow (e.g. Miles 1957; Lighthill 1962) and non-separated sheltering (e.g. Belcher & Hunt 1993). Janssen (2004) provides an extensive overview of the relevant developments in wind–wave generation theory. When deriving energy and momentum fluxes from air to water, these seminal theories of wave growth (e.g. Phillips 1957; Lighthill 1962; Belcher & Hunt 1993) utilized a phase-averaging technique, which removes wave-shape information. Thus, although these wind–wave interaction theories focused on the wave growth rate, no theoretical work has investigated the effect of wind on wave shape in a physically consistent manner.

Measurements and numerical simulations have also been used to investigate the dependence of wave growth on wind speed. Field measurements (e.g. Longuet-Higgins 1962; Snyder 1966; Hasselmann *et al.* 1973) and laboratory experiments (e.g. Shemdin & Hsu 1967; Plant & Wright 1977; Mitsuyasu & Honda 1982) have been used to parameterize how quickly intermediate- and deep-water waves grow under various wind conditions, including short fetch (e.g. Lamont-Smith & Waseda 2008) and strong wind conditions (e.g. Troitskaya *et al.* 2012). Note that direct measurements of wave surface pressure (related to growth) are notoriously difficult (e.g. Donelan *et al.* 2005). Similarly, numerical simulations have also been used to predict wind-induced growth rates. Early numerical atmospheric models used the Reynolds-averaged Navier–Stokes equations (e.g. Gent & Taylor 1976; Al-Zanaidi & Hui 1984) to calculate the energy loss of the wind field. However, these early simulations could only approximate turbulence through a Reynolds-averaging process. Recent studies have analysed the turbulence behaviour in detail. Particle image velocimetry and laser-induced fluorescence have been used for turbulence measurements in laboratory experiments and have revealed turbulent structures above the waves (e.g. Veron *et al.* 2007; Buckley & Veron 2017, 2019). This turbulent behaviour has also been captured through direct numerical simulations of the governing equations (e.g. Yang & Shen 2009, 2010; Yang *et al.* 2018) and by parameterizing subgrid-scale processes in large eddy simulations (LES, e.g. Yang *et al.* 2013; Hara & Sullivan 2015; Hao *et al.* 2018). When solving for the atmospheric dynamics, many of these simulations prescribed a static sinusoidal wave shape while focusing on the evolution of the wind field, as well as energy and momentum transfers. Therefore, any wind-induced changes to wave shape were not captured.

While there has been much research regarding wind-induced wave growth, wave shape has seen relatively little work. Coupled air–water simulations (e.g. Liu *et al.* 2010; Hao & Shen 2019) and two-phase (air and water) simulations (e.g. Deike *et al.* 2017;

Zou & Chen 2017) have begun incorporating dynamically evolving waves into their analyses. These directly model the evolution of both the air and wave fields in a coupled manner in contrast to simulations prescribing a fixed wave shape. Furthermore, some also qualitatively consider how wave shape evolves under the influence of wind (e.g. Yan & Ma 2010; Xie 2014, 2017). However, theses analyses are focused on other parameters and do not quantify precisely how the wave shape changes. Additionally, there have been a small number of field measurements (e.g. Cox & Munk 1956) and laboratory experiments (Leykin *et al.* 1995; Feddersen & Veron 2005) that have directly investigated how wind affects wave shape. It was found that the skewness and asymmetry depended on wind speed for mechanically generated waves in relatively deep (Leykin *et al.* 1995) or intermediate and shallow (Feddersen & Veron 2005) water. In particular, the wave asymmetry (Leykin *et al.* 1995), skewness (Cox & Munk 1956) and energy ratio of the first harmonic (frequency $2f$) to the primary wave (frequency $f$) (Feddersen & Veron 2005) all increased with wind speed. It would be beneficial to develop a theory that explains these experimental findings.

In this paper, we develop a theory coupling wind to dynamically evolving intermediate- and deep-water waves ($kh \geqslant 1$ with $k$ the wavenumber and $h$ the water depth). We consider the fluid domain beneath a periodic, progressive wave that is forced by a prescribed, wave-dependent surface pressure profile. That is, the atmosphere is not treated dynamically. Determining the wind's effect on wave shape requires a nonlinear theory. As the surface boundary conditions for gravity waves are nonlinear, the equations are solved using a multiple-scale perturbation analysis where the wave steepness $\varepsilon := a_1 k$ (with $a_1$ the primary wave's amplitude) is small and new, slower time scales are introduced over which the nonlinearities act (see, for example, Ablowitz 2011). This formalism has been used to derive the canonical Stokes waves, which are periodic, progressive waves of permanent form in intermediate and deep water (Stokes 1880). By introducing a surface pressure-forcing term, we will derive solutions of the form

$$\eta = a_1 \exp(\mathrm{i}(kx - \omega t)) + a_2 \exp(\mathrm{i}[2(kx - \omega t) + \beta]) + \dots, \tag{1.1}$$

with the real part implied. Here, $\eta$ is the wave height, $\omega$ is the complex wave frequency, and $a_1 k = \varepsilon$ and $a_2 k = O(\varepsilon^2)$ are the non-dimensional amplitudes of the primary wave and first harmonic, respectively. We have defined a new parameter, the 'harmonic phase' (or HP) $\beta$, which is analogous to the biphase, a statistical tool (Elgar & Guza 1985). Both wave skewness and asymmetry depend on the HP $\beta$ and relative harmonic amplitude $a_2/(a_1^2 k)$. For example, both skewness and asymmetry are zero for linear waves with $a_2/(a_1^2 k) = 0$. For deep-water ($kh \gg 1$) Stokes waves without wind forcing, $a_2/(a_1^2 k) = 1/2$ gives non-zero skewness, but $\beta = 0$ yields no phase difference between the primary wave and first harmonic in (1.1). Indeed, unforced Stokes waves are exactly symmetric at all orders (Toland 2000).

Three surface pressure profiles, derived from the theories of Jeffreys (1925) and Miles (1957), are included in the perturbation expansion. Using the method of multiple scales, Stokes wave-like solutions are derived, giving the wave-shape (via $a_2/(a_1^2 k)$ and $\beta$) dependence on the wind-induced surface pressure profile. Additionally, wave growth will result from the fact that $\mathrm{Im}\{\omega\}$ is no longer zero (e.g. Miles 1957). These solutions reduce to unforced Stokes waves when the pressure forcing vanishes.

In § 2, we set up the equations and define the different pressure profiles used. Section 3 begins the general derivation covering a range of realistic pressure magnitudes, which is continued in appendix A. As a key aspect to the derivation, the non-dimensional pressure $p'$ is included in the leading-order equations ($p' = O(\varepsilon)$), which is the most general approach by allowing the substitution of $p' \to \varepsilon p'$ or $p' \to \varepsilon^2 p'$, generating weaker $p' = O(\varepsilon^2)$ and $p' = O(\varepsilon^3)$ solutions (cf. appendix A.6). Section 4 details the results of this analysis. In § 5, we clarify the solutions' time scale validity, relate the pressure parameters to LES simulations, compare our results to laboratory observations and compare the surface pressure form with existing data. Appendix A extends the general derivation to higher orders in $\varepsilon$ to demonstrate a weak amplitude dependence of the shape parameters.

## 2. Theoretical background

### 2.1. *Governing equations*

Here, we specify the equations governing the water dynamics. Homogeneous, incompressible fluids satisfy the incompressible continuity equation,

$$\nabla \cdot \boldsymbol{u} = 0, \tag{2.1}$$

within the fluid. We assume irrotational flow and write the water velocity $\boldsymbol{u}$ in terms of a velocity potential $\phi$ as $\boldsymbol{u} = \nabla\phi$. We define a coordinate system with $z = 0$ at the initial mean water level, positive $z$ upward and gravity pointing in the $-z$ direction. We assume planar wave propagation in the $+x$ direction and uniform in the $y$ direction. Then, the incompressibility condition becomes Laplace's equation,

$$\frac{\partial^2\phi}{\partial x^2} + \frac{\partial^2\phi}{\partial z^2} = 0. \tag{2.2}$$

Assuming uniform water depth with a flat bottom located at $z = -h$, we impose a no-flow bottom boundary condition

$$\frac{\partial\phi}{\partial z} = 0 \quad \text{at} \quad z = -h. \tag{2.3}$$

Finally, the standard surface boundary conditions (e.g. Whitham 2011) are the kinematic boundary condition

$$\frac{\partial\phi}{\partial z} = \frac{\partial\eta}{\partial t} + \frac{\partial\phi}{\partial x}\frac{\partial\eta}{\partial x} \quad \text{at} \quad z = \eta, \tag{2.4}$$

and the dynamic boundary condition

$$0 = \frac{p}{\rho_w} + g\eta + \frac{\partial\phi}{\partial t} + \frac{1}{2}\left(\frac{\partial\phi}{\partial x}^2 + \frac{\partial\phi}{\partial z}^2\right) \quad \text{at} \quad z = \eta. \tag{2.5}$$

Here, $g$ is the acceleration due to gravity, $\rho_w$ the water density, $\eta(x,t)$ the surface profile and $p(x,t)$ the surface pressure evaluated at $z = \eta$. Note that we have absorbed the Bernoulli constant from (2.5) into $\phi$ using its gauge freedom $\phi \to \phi + f(t)$ for arbitrary $f(t)$. In § 2.3 we specify the surface pressure profiles.

### 2.2. *Assumptions*

Our analysis is characterized by a number of non-dimensional parameters. The wave slope $\varepsilon := a_1 k$, assumed small, will order our perturbation expansion. Additionally, we will restrict our attention to intermediate and deep water by requiring that the non-dimensional depth $kh \gtrapprox 1$ so that the Ursell parameter is small, $\varepsilon/(kh)^3 \ll 1$. An additional parameter is the non-dimensional surface pressure magnitude induced by the wind discussed in §§ 2.3 and 2.4. We seek waves with wavelength $\lambda := 2\pi/k$ travelling in the $x$ direction that are periodic in $x$ and quasi-periodic in $t$

$$\eta(x,t) = \eta(x+\lambda,t) = \eta(\theta,t) \quad \text{and} \quad \phi(x,z,t) = \phi(x+\lambda,z,t) = \phi(\theta,z,t), \tag{2.6}$$

with $\theta$ defined for right-propagating waves ($\mathrm{Re}\{\omega(t)\} > 0$) as

$$\theta := kx - \int \mathrm{Re}\{\omega(t)\}\,\mathrm{d}t, \tag{2.7}$$

which is analogous to the standard $kx - \mathrm{Re}\{\omega\}t$, but allows for a complex, time dependent frequency $\omega(t)$. Additionally, we neglect surface tension $\sigma$ by restricting to wavelengths $\lambda \gg 2\,\mathrm{cm}$, implying a large Bond number ($\rho g/k^2\sigma \gg 1$). Furthermore, we assume no mean Eulerian current. Finally, we seek a solution of a single primary wave and its bound harmonics. Including additional primary waves permits us to study the wind's effect on sideband instabilities (e.g. Brunetti & Kasparian 2014) but is beyond the scope of this work.

In the dynamic boundary condition (2.5), we incorporated the normal stress (surface pressure) but neglected the shear stress as it is usually significantly smaller than the normal stress (e.g. Kendall 1970; Hara & Sullivan 2015; Husain *et al.* 2019). Additionally, we note that surface shear stresses cause a slight thickening of the boundary layer, which is equivalent to a pressure phase shift on the remainder of the water column (Longuet-Higgins 1969). Therefore, we can include the effect of shear stresses through a phase shift in the pressure relative to the wave profile. Hence, in this investigation we only consider pressures acting normal to the wave surface.

The irrotational assumption was motivated by our assumption that vorticity-generating wind shear stresses are small. Additionally, any such vorticity is constrained to a thin boundary layer just below the wave surface (Longuet-Higgins 1969). Finally, viscous forces vanish—necessary for Bernoulli's equation (2.5)—for any flow that is both irrotational and incompressible (with constant viscosity; e.g. Fang (2019)). Thus, we will assume irrotational, inviscid flow throughout the fluid interior.

### 2.3. *Surface pressure profiles*

Here, we define the surface pressure profiles used in the analysis. The Jeffreys (1925) theory yields a ('Jeffreys') surface pressure profile,

$$p_J(x,t) = s\rho_a U^2 \frac{\partial \eta(x,t)}{\partial x}, \tag{2.8}$$

with $\rho_a$ the air density, $U$ a characteristic wind speed and $s$ an empirical, unitless constant. Although the Miles mechanism is now favoured for gently sloping waves or weak winds (Tian & Choi 2013), the Jeffreys mechanism is still relevant for steep, strongly forced waves (e.g. Touboul & Kharif 2006). The simple, analytic form of the Jeffreys forcing also lends itself well to theoretical treatments. Indeed, many treatments (e.g. Banner & Song 2002; Kalmikov 2010; Brunetti *et al.* 2014) approximate the Miles forcing by a wave slope coherent pressure $p \propto \partial\eta/\partial x$ equivalent to our Jeffreys-type forcing (2.8).

The Miles (1957) theory of wind–wave growth gives a ('Miles') surface pressure profile of the form

$$p_M(x,t) = \left(\tilde{\alpha} + \mathrm{i}\tilde{\beta}\right)\rho_a U^2 k \eta_a(x,t), \tag{2.9}$$

with $\tilde{\alpha}$ and $\tilde{\beta}$ empirical, unitless constants. Additionally, $\eta_a$ the analytic representation of $\eta$, where the analytic representation of a real function $f(x)$ is $f(x) + \mathrm{i}\hat{f}(x)$ with $\hat{f}(x)$ the Hilbert transform of $f(x)$ (for our purposes, only two representations will be relevant: the analytic representation of $\cos(x)$ is $\exp(\mathrm{i}x)$ and that of $\sin(x)$ is $-\mathrm{i}\exp(\mathrm{i}x)$). This theory was developed for a linear, sinusoidal (i.e. primary) wave without harmonics. Note that (2.9) shifts each harmonic $\exp(\mathrm{i}mkx)$ by the same phase, $\tan^{-1}(\tilde{\beta}/\tilde{\alpha})$, but by a different distance, $m\tan^{-1}(\tilde{\beta}/\tilde{\alpha})/k$, distorting the pressure profile relative to $\eta$. This pressure profile gives no wave-shape change at leading order (appendix A.5) and, since wind-induced shape changes have been observed experimentally, they will not be discussed further here.

Another suitable generalization, capturing the motivation behind the Miles profile, is specifying the surface pressure as phase shifted relative to $\eta$. This prescription is more appropriate for nonlinear waves since all harmonics are shifted the same distance. Thus, we define another ('generalized Miles') surface pressure profile as

$$p_G(x,t) = r\rho_a U^2 k\eta(kx + \psi_P, t), \tag{2.10}$$

with $r$ a new, unitless constant and a new parameter, the 'wind phase' $\psi_P$, which corresponds to the phase shift between the wave and pressure profile, has been introduced. As the surface pressure is elevated on the wave's windward (relative to the leeward) side, $\psi_P > 0$ corresponds to wind blowing from the left, assuming $\psi_P \in (-\pi, \pi]$. Note that the wind phase $\psi_P$ is a free parameter for the pressure profile. Although $\psi_P$ likely depends on other factors such as wave age, determining such a relationship is outside the scope of this work. For a single primary wave, we treat $\psi_P$ as a fixed parameter (for a given wind speed) which is assumed known—possibly from experiments or simulations (cf. § 5.2).

To facilitate comparison, the various pressure profiles are written in a common form. Inspired by similarities in (2.8)–(2.10), we define a non-negative pressure magnitude constant, $P$, which implicitly encodes the wind speed. For instance, (2.8)–(2.10) suggest

$$P \propto \rho_a U^2, \tag{2.11}$$

although this form serves only as motivation, and the particular $U$ dependence will be immaterial to our analysis. Since the definition of $\varepsilon := a_1 k$ implies $k|\eta| = O(\varepsilon)$, we see from the various definitions (2.8)–(2.10) that

$$O(|p|) = O(\varepsilon P). \tag{2.12}$$

We will define $P_J$ for the Jeffreys profile such that

$$p_J(x,t) = \pm P_J \frac{\partial \eta(x,t)}{\partial x}, \tag{2.13}$$

with the plus sign for wind blowing from the left. Likewise, we will rewrite the generalized Miles profile as

$$p_G(x,t) = P_G k\eta(kx + \psi_P, t). \tag{2.14}$$

The constant $P$ is subscripted to denote Jeffreys ($P_J$) or generalized Miles ($P_G$) when the distinction is relevant. In § 3, these two surface pressure profiles, (2.13) and (2.14), are expanded in a Fourier series to yield simpler equations. Expanding an arbitrary function $f(x)$ in a Fourier series as the real part of $f(x) = \sum_{m=0} \hat{f}_m \exp(\mathrm{i}mkx)$ with $m \in \mathbb{N}$ yields

$$\hat{p}_{J,m}(t) = \pm \mathrm{i}kmP_J\hat{\eta}_m(t), \tag{2.15}$$

$$\hat{p}_{G,m}(t) = kP_G \exp(\mathrm{i}m\psi_P)\hat{\eta}_m(t). \tag{2.16}$$

Therefore, we will generically write

$$\hat{p}_m(t) = k\hat{P}_m\hat{\eta}_m(t), \tag{2.17}$$

with $\hat{P}_m = mP_J \exp(\mathrm{i}\psi_P)$ and $\psi_P = \pm\pi/2$ for Jeffreys and $\hat{P}_m = P_G \exp(\mathrm{i}m\psi_P)$ for generalized Miles profiles. Although we highlight these two forcing profiles, we stress the derivation's generality. The results apply to any pressure profile (2.17) that results from a convolution of $\eta(x,t)$ with a time-independent function $f(x)$, each yielding a specific $\hat{P}_m$. For example, $\hat{P}_m$ could be chosen to match numerical simulations (cf. § 5.4).

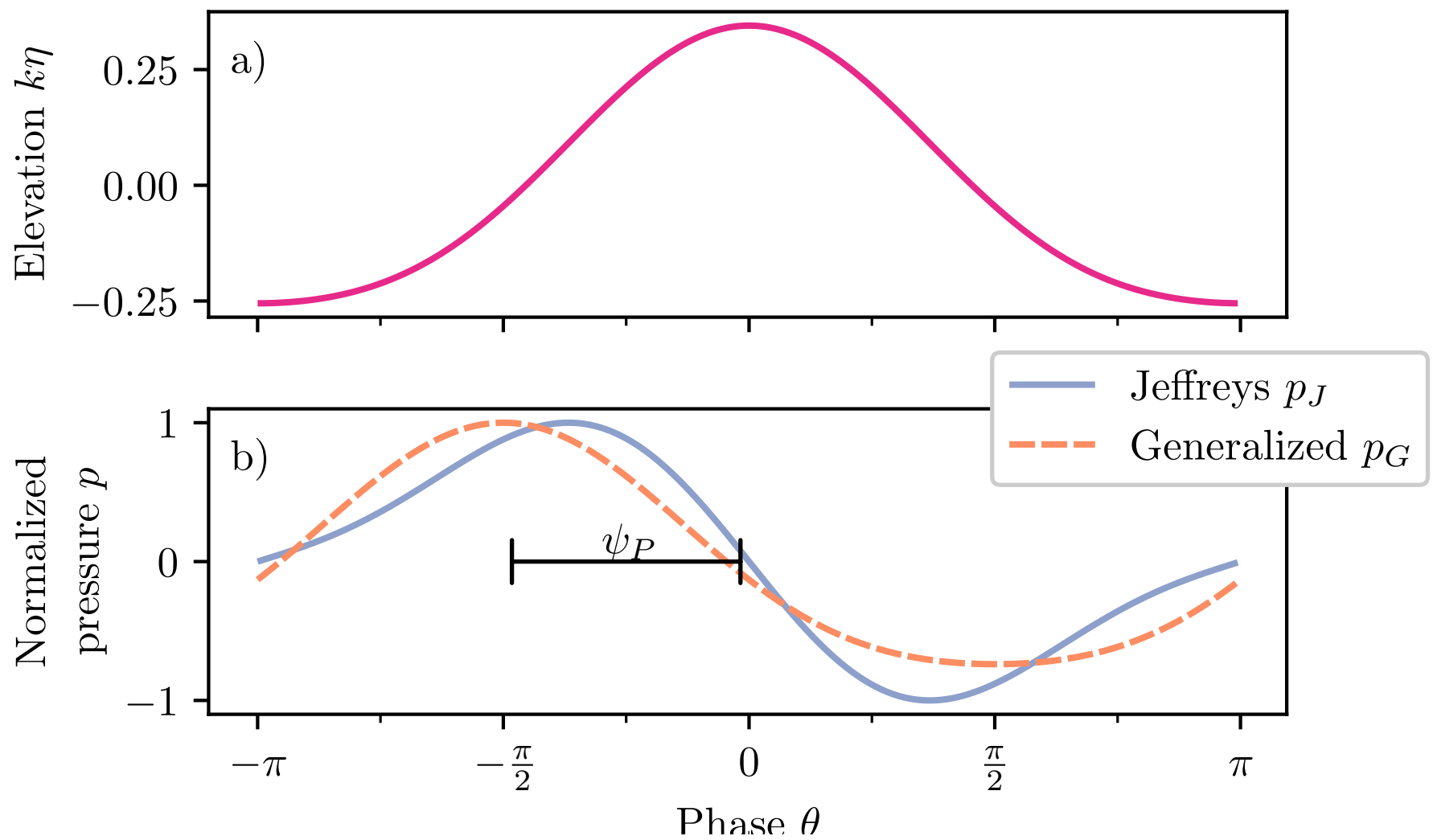

Figure 1. ($a$) Non-dimensional, right-propagating Stokes wave $k\eta$ (2.18) as a function of phase $\theta = kx - \omega t$ with $\varepsilon = 0.2$. ($b$) Normalized surface pressure profiles $p(\theta)$ as described in (2.13) and (2.14); see legend. The maximum pressure magnitude is normalized to unity (arbitrary units), and a value of $\psi_P = \pi/2$ was chosen to facilitate comparison with the Jeffreys profile with $\psi_P$ positive corresponding to wind blowing to the right.

To make these definitions concrete and contrast the different forcing types, a deep-water, second-order Stokes wave

$$k\eta(\theta) = \varepsilon\cos(\theta) + \frac{1}{2}\varepsilon^2\cos(2\theta) \tag{2.18}$$

is shown for $\varepsilon = 0.2$ in figure 1($a$) with phase $\theta = kx - \omega t$. The Stokes wave profile is used to compute both (unity normalized) surface pressure profiles (figure 1($b$)). These two pressure profiles, (2.13) and (2.14), are largely similar to each other, although differences arise due to the Stokes wave harmonics. The derivative in the Jeffreys profile (blue figure 1($b$)) multiplies each Fourier harmonic by its wavenumber, $mk$, enhancing higher frequencies. In contrast, the wind phase $\psi_P$, measured left from $\theta = 0$ to the pressure maximum, shifts the entire pressure waveform relative to the surface waveform $\eta$ for the generalized Miles profile (orange, figure 1($b$)). The LES numerical simulations of Hara & Sullivan (2015) and Husain *et al.* (2019) show $\psi_P \approx 3\pi/4$ for a variety of wind speeds (§ 5.2). However, in figure 1($b$), $\psi_P = \pi/2$ is chosen for the generalized Miles profiles to facilitate comparison with the Jeffreys case (for which $\psi_P = \pm\pi/2$).

### 2.4. *Determination of pressure magnitude P*

We will use existing experimental data to determine the magnitude of $P$ in various contexts. Assuming a logarithmic wind profile, Miles (1957) derived the wave-energy growth rate $\gamma$, normalized by the (unforced, linear, deep-water) wave frequency $f_0^\infty$, for the pressure profile $p_M$ (2.14)

$$\frac{\gamma}{f_0^\infty} = 2\pi\tilde{\beta}\frac{\rho_a}{\rho_w}\frac{U^2}{(c_0^\infty)^2} = 2\pi\frac{P_G}{\rho_w(c_0^\infty)^2}\sin(\psi_P), \tag{2.19}$$

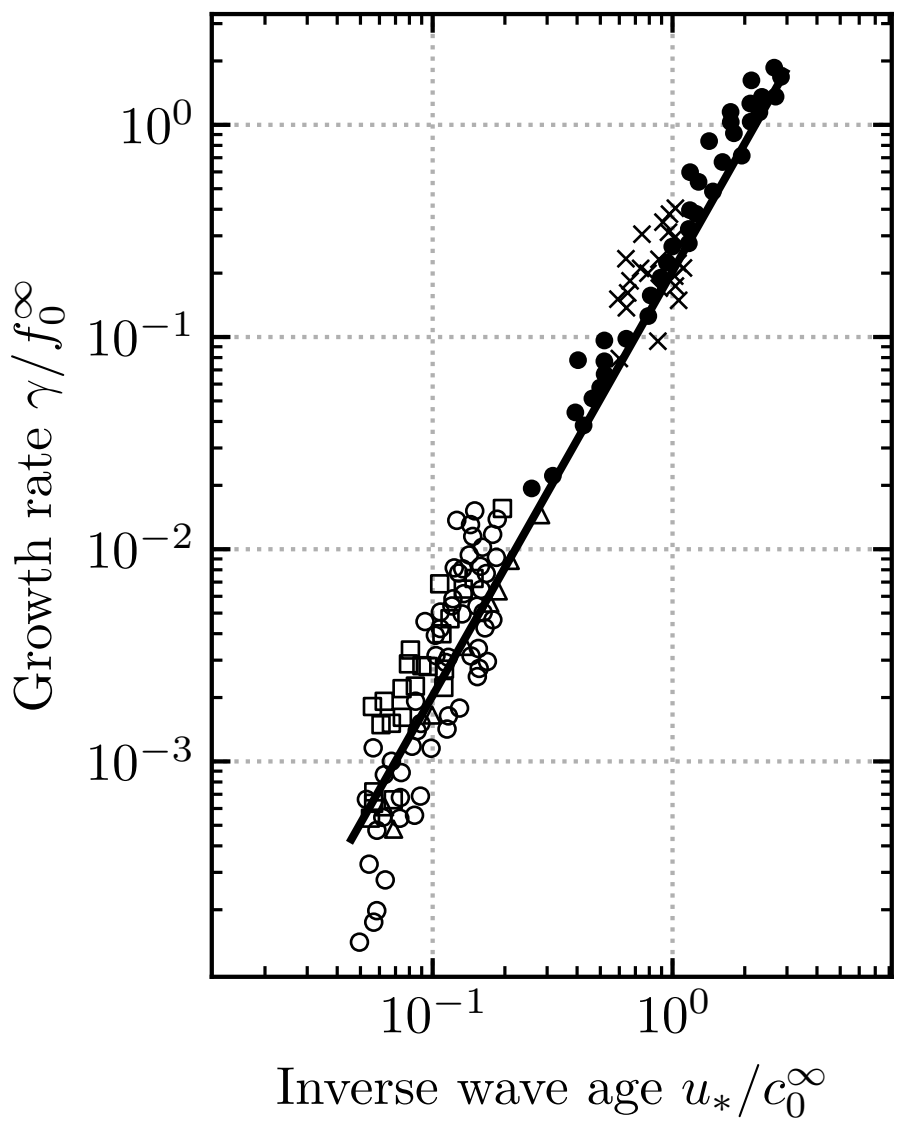

Figure 2. Non-dimensional, deep-water wave-energy growth rate $\gamma/f_0^\infty$ versus inverse wave age, $u_*/c_0^\infty$ with $u_*$ the wind's friction velocity and $c_0^\infty = \sqrt{g/k}$ the unforced, linear, deep-water phase speed. The filled symbols represent laboratory measurements while the hollow symbols represent field measurements (from Komen *et al.* 1994). The solid line represents the fit parameterized by Banner & Song (2002).

where, $c_0^\infty = \sqrt{g/k}$ is the unforced, linear, deep-water phase speed, $\rho_w$ is the water density and (2.11) is used to define $P_G$. Using the value $\psi_P = 3\pi/4$ from Hara & Sullivan (2015) and Husain *et al.* (2019) gives $P_G k/(\rho_w g) = 0.23(\gamma/f_0^\infty)$.

Furthermore, we use empirical data relating wind speed $U$ to growth rate to constrain the $P_G$ pressure magnitude constant in deep water. Figure 2 shows the energy growth rate $\gamma/f_0^\infty$ as a function of inverse wave age, $u_*/c_0^\infty$ with $u_*$ the friction velocity. The empirical observations of $\gamma/f_0^\infty$ versus $u_*/c_0^\infty$ in deep water collapse onto a curve permitting a conversion from $u_*/c_0^\infty$ to $\gamma/f_0^\infty$ and yielding $P_G k/(\rho_w g)$ (2.19).

Here, we consider $|p|k/(\rho_w g) = O(\varepsilon)$ to $O(\varepsilon^3)$, or $Pk/(\rho_w g) = O(1)$ to $O(\varepsilon^2)$—cf. (2.12). If we assume $\varepsilon \approx 0.1$, $\psi_P \approx 3\pi/4$, and $\rho_a/\rho_w = 1.225 \times 10^{-3}$, then (2.19) shows we are considering growth rates $\gamma/f_0^\infty \approx 4 \times 10^{-2}$ to 4. Referring to figure 2, we see these reside mostly in the laboratory measurement regime, corresponding to $u_*/c_0^\infty \approx 5 \times 10^{-1}$ to 5. We can approximate $U_{10}$ using logarithmic boundary layer theory (e.g. Monin & Obukhov 1954)

$$u_* = \frac{\kappa U_{10}}{\ln[(10\,\mathrm{m})/z_0]}, \tag{2.20}$$

with $\kappa \approx 0.4$ the von Kármán constant and $z_0 \approx 1.4 \times 10^{-5}$ the surface roughness parameter for 2 m long, 0.1 m high deep-water waves, as one might have in a wave tank (Taylor & Yelland 2001). Substituting these values, we find

$$U_{10} \approx 34 u_*, \tag{2.21}$$

yielding $U_{10}/c_0^\infty \approx 1 \times 10^1$ to $1 \times 10^2$, or $U_{10} \approx 3 \times 10^1\,\mathrm{m\,s^{-1}}$ to $3 \times 10^2\,\mathrm{m\,s^{-1}}$ assuming a deep-water dispersion relation.

It is interesting to examine the pressure-forcing magnitudes used previously. Phillips

(1957) modelled wave growth using a different mechanism, but the pressure forcing was included at the same order as $\eta$. That is, $|p|k/(\rho_w g) = O(\varepsilon)$, or $Pk/(\rho_w g) = O(1)$, implying $\gamma/f_0^\infty = O(1)$. Referring to figure 2, this corresponded to strongly forced waves and a fast wind ($u_*/c_0^\infty = O(1)$). Other theoretical works have used $|p|k/(\rho_w g) = O(\varepsilon^2)$ (e.g. Janssen 1982; Brunetti & Kasparian 2014; Brunetti *et al.* 2014) or $|p|k/(\rho_w g) = O(\varepsilon^3)$ (e.g. Leblanc 2007; Kharif *et al.* 2010; Onorato & Proment 2012), corresponding to $Pk/(\rho_w g) = O(\varepsilon)$ and $Pk/(\rho_w g) = O(\varepsilon^2)$, respectively. Thus, the choices of $Pk/(\rho_w g) = O(1)$ to $O(\varepsilon^2)$ are all relevant in the literature.

### 2.5. *Multiple-scale expansion*

As mentioned in § 2.2, we will utilize an asymptotic expansion in the small wave slope $\varepsilon := a_1 k$ to prevent secular terms in this singular perturbation expansion. While nonlinear wave theories often use an ordinary Stokes expansion (i.e. a Poincaré–Lindstedt, or strained coordinate, expansion), this does not permit the complex frequencies required for wave growth. Instead, we employ the method of multiple scales and replace $t$ by a series of slower time scales depending on $\varepsilon$ such that $t_0 = t$, $t_1 = t/\varepsilon$, etc, yielding

$$\frac{\partial}{\partial t} = \frac{\partial}{\partial t_0} + \varepsilon \frac{\partial}{\partial t_1} + \dots . \tag{2.22}$$

Additional time scales $t_2$ and $t_3$, inversely proportional to $\varepsilon^2$ and $\varepsilon^3$ respectively, are required in the $O(\varepsilon^4)$ derivation of appendix A. This is exclusively a temporal multiple-scale expansion. While a spatial multiple-scale analysis would also permit the study of surface pressure effects on modulational instabilities (Brunetti & Kasparian 2014), we solely focus on the wind-induced shape change of a single wave.

As discussed in § 2.4, the non-dimensional pressure forcing can have magnitudes ranging from $Pk/(\rho_w g) = O(1)$ to $O(\varepsilon^2)$. Writing $Pk/(\rho_w g) = O(\varepsilon^n)$ for $n = 0$ to 2, we will show that the derivation must be solved to $O(\varepsilon^{n+2})$ to demonstrate shape change. Many theoretical treatments using $Pk/(\rho_w g) = O(\varepsilon^2)$ only utilize $O(\varepsilon^3)$ equations like the nonlinear Schröedinger (NLS) equation (e.g. Kharif *et al.* 2010; Onorato & Proment 2012) or Davey–Stewartson equation (e.g. Leblanc 2007). Therefore, no shape change would be derived without going to higher order. In contrast, both Brunetti *et al.* (2014) and Brunetti & Kasparian (2014) coupled a moderately strong wind $Pk/(\rho_w g) = O(\varepsilon)$ to a slowly varying wave train and derived an $O(\varepsilon^3)$ forced NLS equation. Although solved at sufficiently high order to show wind-induced shape changes, neither reported results for the first harmonic. Instead, these focused on wind-forced wave packet evolution, with Brunetti *et al.* (2014) finding various envelope solitons for the primary wave and Brunetti & Kasparian (2014) deriving an enhancement of the primary wave's modulational instability.

In § 3 and appendix A, we include the pressure in the leading-order equations, i.e. $Pk/(\rho_w g) = O(1)$, which is the most general case. The leading-order contributions to the shape parameters $\beta$ and $a_2/(a_1^2 k)$ are found at $O(\varepsilon^2)$, while the higher-order corrections occur at $O(\varepsilon^4)$ (appendix A). From the full $O(\varepsilon^4)$ solution, shape changes for $Pk/(\rho_w g) = O(\varepsilon)$ or $Pk/(\rho_w g) = O(\varepsilon^2)$ can be found by substituting $P \to \varepsilon P$ or $P \to \varepsilon^2 P$, respectively (cf. appendix A.6).

### 2.6. *Non-dimensionalization*

Non-dimensional systems are useful in perturbation expansions. Here, a standard non-dimensionalization (e.g. Mei *et al.* 2005) is performed by defining new non-dimensional, order-unity primed variables

$$\left.\begin{aligned} x &= \frac{x'}{k}, & z &= \frac{z'}{k}, & \eta &= \varepsilon\frac{\eta'}{k}, \\ t &= \frac{t'}{\sqrt{gk}}, & h &= \frac{h'}{k}, & \varPhi &= \varepsilon\varPhi'\sqrt{\frac{g}{k^3}}, \end{aligned}\right\} \tag{2.23}$$

Notice the $\varepsilon$ factor in the equations for $\eta$ and $\phi$ since these are assumed small. Unlike in the standard Stokes wave problem, the surface pressure must also be non-dimensionalized. As shown in (2.12), $O(|p|) = \varepsilon O(P)$. Thus, we find $p$ and $P$ (as well as their Fourier transforms) are non-dimensionalized by

$$(p, \hat{p}) = O\left(\frac{\varepsilon Pk}{\rho_w g}\right)\frac{\rho_w g}{k}(p', \hat{p}'), \tag{2.24}$$

$$(P, \hat{P}_m) = O\left(\frac{Pk}{\rho_w g}\right)\frac{\rho_w g}{k}(P', \hat{P}'_m), \tag{2.25}$$

with $p'(x,t)$ and $P'$ (as well as their Fourier transforms) now order unity and dimensionless. For the remainder of the paper, primes will be dropped and all variables will be assumed non-dimensional and order unity, except where explicitly stated.

## 3. Derivation of wave-shape parameters

We now couple a prescribed surface pressure profile (2.17) to the nonlinear wave problem (2.2)–(2.5) to derive the wind's effect on wave shape. In this section, we will ultimately find an expression for the non-dimensional surface profile of the form

$$\eta = \varepsilon A_1(t_1, \ldots)\exp(\mathrm{i}(x - \omega_0 t_0)) + \varepsilon^2 A_2(t_1, \ldots)\exp(\mathrm{i}[2(x - \omega_0 t_0) + \beta]) + \ldots, \tag{3.1}$$

where the real part is implied and $\omega_0$ is the leading-order approximation to $\omega$ defined in (2.7). Note that we are not assuming this as a functional form for $\eta$, but are only giving a preview of our final result. A comparison of non-dimensional (3.1) with dimensional (1.1) shows we will have (ignoring the time dependence; cf. appendix A.4) $a_1 = \varepsilon A_1$, $a_2 = \varepsilon^2 A_2/k$, etc, so $A_2/A_1^2 = a_2/(a_1^2 k)$. Both the HP $\beta$ and $a_2/a_1^2$ encode information about the wave shape. We take the ratio $a_2/a_1^2$ because we will find that $a_2 \propto \exp(2\,\mathrm{Im}\{\omega_0\}t_0)$ while $a_1 \propto \exp(\mathrm{Im}\{\omega_0\}t_0)$. As we are mainly interested in the shape, the growth is removed by using the ratio $a_2/a_1^2$.

Now, expanding our non-dimensional variables in an asymptotic series of $\varepsilon$, we have

$$\eta = \sum_{n=1}^{\infty} \varepsilon^n \eta_n(x, t_0, t_1, \ldots), \tag{3.2}$$

$$\phi = \sum_{n=1}^{\infty} \varepsilon^n \phi_n(x, z, t_0, t_1, \ldots), \tag{3.3}$$

$$p = \sum_{n=1}^{\infty} \varepsilon^n p_n(x, t_0, t_1, \ldots). \tag{3.4}$$

Choosing $Pk/(\rho_w g) = O(1)$ gives $p_1 \neq 0$. Laplace's equation (2.2) and the bottom boundary condition (2.3) are linear and—unlike when spatial multiple scales are employed (e.g. Mei *et al.* 2005)—can be satisfied identically. Laplace's equation is solved via a Fourier transform and, with the bottom boundary condition, has solution (real part

implied)

$$\phi_n(x,z,t_0,t_1,\ldots) = \hat{\phi}_{n,0}(t_0,t_1,\ldots) + \frac{\cosh[m(z+h)]}{\sinh(mh)}\exp(\mathrm{i}mx)\hat{\phi}_{n,m}(t_0,t_1,\ldots), \quad (3.5)$$

with arbitrary $m \in \mathbb{N}_{>0}$ and arbitrary functions $\hat{\phi}_{n,0}(t_0,t_1,\ldots)$ and $\hat{\phi}_{n,m}(t_0,t_1,\ldots)$. Note that we imposed the no-mean-current condition by choosing $\langle u\rangle = \langle \partial_x \phi\rangle = 0$ at each order $n$, with $\langle\cdot\rangle$ the spatial average over one wavelength. Furthermore, to express the surface pressure profile $p_n$ in terms of the surface height $\eta_n$ (cf. (2.17)), all variables are written as Fourier series

$$\eta_n(x,t_0,t_1,\ldots) = \sum_{m=0}^{m=n} \exp(\mathrm{i}mx)\hat{\eta}_{n,m}(t_0,t_1,\ldots), \quad (3.6)$$

$$\phi_n(x,z,t_0,t_1,\ldots) = \sum_{m=1}^{m=n} \exp(\mathrm{i}mx)\hat{\phi}_{n,m}(t_0,t_1,\ldots)\frac{\cosh(m(z+h))}{\sinh(mh)} + \hat{\phi}_{n,0}(t_0,t_1,\ldots), \quad (3.7)$$

$$p_n(x,t_0,t_1,\ldots) = \sum_{m=0}^{m=n} \exp(\mathrm{i}mx)\hat{p}_{m,n}(t_0,t_1,\ldots). \quad (3.8)$$

Aside from the pressure expansion, this follows the standard Stokes expansion methodology (e.g. Ablowitz 2011). Other texts, such as Mei *et al.* (2005), treat the Stokes expansion using both slow time and spatial scales, but such spatial expansions are outside the scope of this paper (cf. § 2.2). Recall that we previously related (cf. (2.17)) the Fourier transform of the surface pressure to the surface profile,

$$\hat{p}_{m,n}(t_0,t_1,\ldots) = \hat{P}_m\hat{\eta}_{m,n}(t_0,t_1,\ldots). \quad (3.9)$$

Thus, $p$ has higher-order corrections because $\eta$ has higher-order Stokes-like corrections.

We now expand the kinematic (2.4) and dynamic (2.5) boundary conditions in $\varepsilon$ and collect terms order by order.

$O(\varepsilon)$ :

$$\frac{\partial \eta_1}{\partial t_0} - \frac{\partial \phi_1}{\partial z} = 0 \quad (3.10)$$

$$\eta_1 + \frac{\partial \phi_1}{\partial t_0} + p_1 = 0, \quad (3.11)$$

$O(\varepsilon^2)$ :

$$\frac{\partial \phi_2}{\partial z} - \frac{\partial \eta_2}{\partial t_0} = \frac{\partial \eta_1}{\partial t_1} + \frac{\partial \eta_1}{\partial x}\frac{\partial \phi_1}{\partial x} - \eta_1\frac{\partial^2 \phi_1}{\partial z^2}, \quad (3.12)$$

$$\eta_2 + \frac{\partial \phi_2}{\partial t_0} + p_2 = -\frac{\partial \phi_1}{\partial t_1} - \eta_1\frac{\partial \phi_1}{\partial z t_0} - \frac{1}{2}\left(\frac{\partial \phi_1}{\partial x}\right)^2 - \frac{1}{2}\left(\frac{\partial \phi_1}{\partial z}\right)^2, \quad (3.13)$$

$O(\varepsilon^3)$ :

$$\begin{aligned}\frac{\partial \phi_3}{\partial z}-\frac{\partial \eta_3}{\partial t_0}=\frac{\partial \eta_2}{\partial t_1}+\frac{\partial \eta_1}{\partial t_2}+\frac{\partial \eta_2}{\partial x}\frac{\partial \phi_1}{\partial x}+\frac{\partial \eta_1}{\partial x}\frac{\partial \phi_2}{\partial x}+\eta_1\frac{\partial \eta_1}{\partial x}\frac{\partial^2 \phi_1}{\partial z\partial x}-\eta_1\frac{\partial^2 \phi_2}{\partial z^2}\\ -\frac{1}{2}\eta_1^2\frac{\partial^3\phi_1}{\partial z^3}-\eta_2\frac{\partial^2\phi_1}{\partial z^2},\end{aligned} \tag{3.14}$$

$$\begin{aligned}\eta_3+\frac{\partial \phi_3}{\partial t_0}+p_3=-\frac{\partial \phi_1}{\partial t_2}-\frac{\partial \phi_2}{\partial t_1}-\frac{1}{2}\eta_1^2\frac{\partial^3\phi_1}{\partial z^2\partial t_0}-\eta_1\frac{\partial^2\phi_2}{\partial z\partial t_0}-\eta_2\frac{\partial^2\phi_1}{\partial z\partial t_0}\\ -\eta_1\frac{\partial^2\phi_1}{\partial z\partial t_1}-\frac{\partial \phi_1}{\partial x}\frac{\partial \phi_2}{\partial x}-\eta_1\frac{\partial \phi_1}{\partial x}\frac{\partial^2\phi_1}{\partial x\partial z}-\frac{\partial \phi_1}{\partial z}\frac{\partial \phi_2}{\partial z}-\eta_1\frac{\partial \phi_1}{\partial z}\frac{\partial^2\phi_1}{\partial z^2},\end{aligned} \tag{3.15}$$

$O(\varepsilon^4)$ :

$$\begin{aligned}\frac{\partial \phi_4}{\partial z}-\frac{\partial \eta_4}{\partial t_0}=-\frac{\partial \eta_1}{\partial t_3}-\frac{\partial \eta_2}{\partial t_2}-\frac{\partial \eta_3}{\partial t_1}-\frac{\partial \eta_1}{\partial x}\frac{\partial \phi_3}{\partial x}-\frac{\partial \eta_2}{\partial x}\frac{\partial \phi_2}{\partial x}-\frac{\partial \eta_3}{\partial x}\frac{\partial \phi_1}{\partial x}+\eta_3\frac{\partial^2\phi_1}{\partial z^2}\\ -\left(\frac{\partial \eta_1}{\partial x}\frac{\partial^2\phi_1}{\partial x\partial z}-\frac{\partial^2\phi_2}{\partial z^2}\right)\eta_2-\left(\frac{\partial \eta_1}{\partial x}\frac{\partial^2\phi_2}{\partial x\partial z}+\frac{\partial \eta_2}{\partial x}\frac{\partial^2\phi_1}{\partial x\partial z}-\frac{\partial^2\phi_3}{\partial z^2}\right)\eta_1+\frac{\partial^3\phi_1}{\partial z^3}\eta_1\eta_2\\ -\left(\frac{1}{2}\frac{\partial \eta_1}{\partial x}\frac{\partial^2\phi_1}{\partial x\partial z}-\frac{1}{2}\frac{\partial^3\phi_2}{\partial z^3}\eta_1^2-\frac{1}{6}\frac{\partial^4\phi_1}{\partial z^4}\eta_1^3\right),\end{aligned} \tag{3.16}$$

$$\begin{aligned}\eta_4+\frac{\partial \phi_4}{\partial t_0}+p_4=-\frac{\partial \phi_1}{\partial t_3}-\frac{\partial \phi_2}{\partial t_2}-\frac{\partial \phi_3}{\partial t_1}-\frac{\partial \phi_1}{\partial x}\frac{\partial \phi_3}{\partial x}-\frac{1}{2}\left(\frac{\partial \phi_2}{\partial x}\right)^2-\frac{\partial \phi_1}{\partial z}\frac{\partial \phi_3}{\partial z}\\ -\frac{1}{2}\left(\frac{\partial \phi_2}{\partial z}\right)^2-\frac{\partial^2\phi_1}{\partial t_0\partial z}\eta_3-\left(\frac{\partial^2\phi_1}{\partial t_1\partial z}+\frac{\partial^2\phi_2}{\partial t_0\partial z}+\frac{\partial \phi_1}{\partial x}\frac{\partial^2\phi_1}{\partial x\partial z}+\frac{\partial \phi_1}{\partial z}\frac{\partial^2\phi_1}{\partial z^2}\right)\eta_2\\ -\left(\frac{\partial^2\phi_1}{\partial t_2\partial z}+\frac{\partial^2\phi_2}{\partial t_1\partial z}+\frac{\partial^2\phi_3}{\partial t_0\partial z}+\frac{\partial \phi_1}{\partial x}\frac{\partial^2\phi_2}{\partial x\partial z}+\frac{\partial \phi_2}{\partial x}\frac{\partial^2\phi_1}{\partial x\partial z}+\frac{\partial \phi_1}{\partial z}\frac{\partial^2\phi_2}{\partial z^2}+\frac{\partial \phi_2}{\partial z}\frac{\partial^2\phi_1}{\partial z^2}\right)\eta_1\\ -\frac{\partial^2\phi_1}{\partial t_0\partial z}\eta_1\eta_2-\left(\frac{1}{2}\frac{\partial^2\phi_1}{\partial t_1\partial z}+\frac{1}{2}\frac{\partial^2\phi_2}{\partial t_0\partial z}+\frac{1}{2}\frac{\partial \phi_1}{\partial x}\frac{\partial^2\phi_1}{\partial x\partial z}+\frac{1}{2}\left(\frac{\partial^2\phi_1}{\partial x\partial z}\right)^2+\frac{1}{2}\frac{\partial \phi_1}{\partial z}\frac{\partial^3\phi_1}{\partial z^3}\right.\\ \left.+\frac{1}{2}\left(\frac{\partial^2\phi_1}{\partial z^2}\right)^2\right)\eta_1^2-\frac{1}{6}\frac{\partial^2\phi_1}{\partial t_0\partial z}\eta_1^3.\end{aligned} \tag{3.17}$$

We solve these equations to $O(\varepsilon^2)$ here and $O(\varepsilon^4)$ in appendix A.

### 3.1. *The $O(\varepsilon)$ equations*

Proceeding to first order in $\varepsilon$, the linearized boundary conditions are

$$\frac{\partial \phi_1}{\partial z}-\frac{\partial \eta_1}{\partial t_0}=0, \tag{3.18}$$

$$\frac{\partial \phi_1}{\partial t_0}+\eta_1+p_1=0. \tag{3.19}$$

Inserting the Fourier transforms (3.6)–(3.8) and the pressure profile (2.17) gives

$m = 1$ Fourier component:

$$\hat{\phi}_{1,1} - \frac{\partial \hat{\eta}_{1,1}}{\partial t_0} = 0, \tag{3.20}$$

$$\frac{\partial \hat{\phi}_{1,1}}{\partial t_0} \coth(h) + \hat{\eta}_{1,1} + \hat{P}_1 \hat{\eta}_{1,1} = 0, \tag{3.21}$$

$m = 0$ Fourier component:

$$-\frac{\partial \hat{\eta}_{1,0}}{\partial t_0} = 0, \tag{3.22}$$

$$\frac{\partial \hat{\phi}_{1,0}}{\partial t_0} + \hat{\eta}_{1,0} + \hat{P}_0 \hat{\eta}_{1,0} = 0. \tag{3.23}$$

The $m = 0$ Fourier equations are solved by $\hat{\eta}_{1,0} = \hat{\phi}_{1,0} = 0$ when placing the initial mean water level $\langle \eta \rangle$ at $z = 0$. Combining the $m = 1$ equations (3.20) and (3.21) to eliminate $\hat{\eta}_{1,1}$ gives

$$\frac{\partial^2 \hat{\phi}_{1,1}}{\partial t_0^2} \coth(h) + \left(1 + \hat{P}_1\right) \hat{\phi}_{1,1} = 0. \tag{3.24}$$

This is the usual, finite-depth, linear operator on $\hat{\phi}_{1,1}$ modified by the presence of $\hat{P}_1$, showing that $\hat{\phi}_{1,1}(t_0, t_1, \ldots)$ is harmonic. Using a bit of foresight to define the constants, we write

$$\hat{\phi}_{1,1} = -\mathrm{i}\omega_0 A_1(t_1) \exp(-\mathrm{i}\omega_0 t_0), \tag{3.25}$$

giving

$$\phi_1 = -\mathrm{i}\omega_0 A_1(t_1) \exp(\mathrm{i}(x - \omega_0 t_0)) \frac{\cosh(z + h)}{\sinh(h)}, \tag{3.26}$$

where

$$\omega_0 = \pm \sqrt{\tanh(h) \left(1 + \hat{P}_1\right)}. \tag{3.27}$$

We choose the (+) sign, corresponding to waves propagating to the right. While $A_1(t_1)$ and $\exp(-\mathrm{i}\omega_0 t_0)$ always appear together and could be simply left as a single, $t_0$-dependent variable $A(t_0, t_1) \in \mathbb{C}$, we find it instructive to explicitly write the $t_0$-dependence. Inserting this into the surface boundary conditions gives equations for $\eta_1$,

$$\frac{\partial \hat{\eta}_1}{\partial t_0} = -\mathrm{i}\omega_0 A_1(t_1) \exp(-\mathrm{i}\omega_0 t_0), \tag{3.28}$$

$$\hat{\eta}_1 + \hat{P}_1 \hat{\eta}_1 = \coth(h) \omega_0^2 A_1(t_1) \exp(-\mathrm{i}\omega_0 t_0). \tag{3.29}$$

This gives

$$\eta_1 = A_1(t_1) \exp(\mathrm{i}(x - \omega_0 t_0)). \tag{3.30}$$

It is instructive to consider the real and imaginary parts of $\omega_0$

$$\mathrm{Re}\{\omega_0\} = \sqrt{\frac{\tanh(h)}{2}} \sqrt{1 + \mathrm{Re}\left\{\hat{P}_1\right\} + \sqrt{1 + \left|\hat{P}_1\right|^2 + 2\,\mathrm{Re}\left\{\hat{P}_1\right\}}}, \tag{3.31}$$

$$\mathrm{Im}\{\omega_0\} = \mathrm{sgn}\left(\mathrm{Im}\left\{\hat{P}_1\right\}\right) \sqrt{\frac{\tanh(h)}{2}} \sqrt{-1 - \mathrm{Re}\left\{\hat{P}_1\right\} + \sqrt{1 + \left|\hat{P}_1\right|^2 + 2\,\mathrm{Re}\left\{\hat{P}_1\right\}}}. \tag{3.32}$$

Notice that the pressure causes growth ($\mathrm{Im}\{\omega_0\} > 0$) for wind in the direction of the waves ($\mathrm{Im}\{\hat{P}_1\} > 0$) and decay ($\mathrm{Im}\{\omega_0\} < 0$) for opposing wind ($\mathrm{Im}\{\hat{P}_1\} < 0$). Likewise, observe that an applied pressure, $\hat{P}_1 \neq 0$, modifies the dispersion relation (3.31). This phenomenon was also derived by Jeffreys (1925) and Miles (1957) for $Pk/(\rho_w g) = O(\varepsilon)$, which we can reproduce by substituting $\hat{P}_1 \to \varepsilon \hat{P}_1$ in (3.31) and (3.32).

### 3.2. *The $O(\varepsilon^2)$ equations*

Proceeding to second order, the kinematic and dynamic boundary conditions are

$$\frac{\partial \phi_2}{\partial z} - \frac{\partial \eta_2}{\partial t_0} = \frac{\partial \eta_1}{\partial t_1} + \frac{\partial \eta_1}{\partial x}\frac{\partial \phi_1}{\partial x} - \eta_1 \frac{\partial^2 \phi_1}{\partial z^2}, \tag{3.33}$$

$$\frac{\partial \phi_2}{\partial t_0} + \eta_2 + p_2 = -\frac{\partial \phi_1}{\partial t_1} - \eta_1 \frac{\partial^2 \phi_1}{\partial z \partial t_0} - \frac{1}{2}\left(\frac{\partial \phi_1}{\partial x}\right)^2 - \frac{1}{2}\left(\frac{\partial \phi_1}{\partial z}\right)^2. \tag{3.34}$$

By inserting the Fourier transforms (3.6)–(3.8), we can express $p_2$ using (3.9). Inserting the first-order solutions (3.26) and (3.30) and collecting harmonics yields

$m = 1$ Fourier component:

$$\hat{\phi}_{2,1} - \frac{\partial \hat{\eta}_{2,1}}{\partial t_0} = \frac{\partial A_1}{\partial t_1} \exp(-\mathrm{i}\omega_0 t_0), \tag{3.35}$$

$$\frac{\partial \hat{\phi}_{2,1}}{\partial t_0} \coth(h) + (1 + \hat{P}_1)\hat{\eta}_{2,1} = \mathrm{i}\omega_0 \frac{\partial A_1}{\partial t_1} \exp(-\mathrm{i}\omega_0 t_0)\coth(h), \tag{3.36}$$

$m = 2$ Fourier component:

$$2\hat{\phi}_{2,2} - \frac{\partial \hat{\eta}_{2,2}}{\partial t_0} = \mathrm{i}\omega_0 A_1^2 \exp(-2\mathrm{i}\omega_0 t_0)\coth(h), \tag{3.37}$$

$$\frac{\partial \hat{\phi}_{2,2}}{\partial t_0} \coth(2h) + (1 + \hat{P}_2)\hat{\eta}_{2,2} = \frac{1}{4}\omega_0^2 A_1^2 \exp(-2\mathrm{i}\omega_0 t_0)\big(2 - \operatorname{csch}^2(h)\big), \tag{3.38}$$

$m = 0$ Fourier component:

$$-\frac{\partial \hat{\eta}_{2,0}}{\partial t_0} = 0, \tag{3.39}$$

$$\frac{\partial \hat{\phi}_{2,0}}{\partial t_0} + \hat{\eta}_{2,0} = \frac{1}{4}\Big(2\,\mathrm{Re}\{\omega_0^2\} - |\omega_0|^2\big(2 + \operatorname{csch}^2(h)\big)\Big)|A_1|^2|\exp(-\mathrm{i}\omega_0 t_0)|^2. \tag{3.40}$$

Eliminating the various $\hat{\eta}_{2,m}$ to get equations solely in terms of $\hat{\phi}_{2,m}$ gives

$m = 1$ Fourier component:

$$\frac{\partial^2 \hat{\phi}_{2,1}}{\partial t_0^2} \coth(h) + (1 + \hat{P}_1)\hat{\phi}_{2,1} = 2\Big(1 + \hat{P}_1\Big)\frac{\partial A_1}{\partial t_1} \exp(-\mathrm{i}\omega_0 t_0), \tag{3.41}$$

$m = 2$ Fourier component:

$$\begin{aligned}\frac{\partial^2 \hat{\phi}_{2,2}}{\partial t_0^2} \coth(2h) + 2(1 + \hat{P}_2)\hat{\phi}_{2,2} = -\mathrm{i}\frac{1}{2}\omega_0 A_1^2 \Big\{ &[2 - \operatorname{csch}^2(h)]\omega_0^2 \\ &- 2\Big(1 + \hat{P}_2\Big)\coth(h)\Big\} \exp(2\mathrm{i}\omega_0 t_0),\end{aligned} \tag{3.42}$$

$m = 0$ Fourier component:

$$\frac{\partial^2 \hat{\phi}_{2,0}}{\partial t_0^2} = \frac{1}{2}\Big(2\,\mathrm{Re}\{\omega_0^2\} - |\omega_0|^2\big(2 + \operatorname{csch}^2(h)\big)\Big)|A_1|^2 \exp(2\,\mathrm{Im}\{\omega_0\}t_0)\,\mathrm{Im}\{\omega_0\}. \tag{3.43}$$

Preventing secular terms in $\hat{\phi}_{2,1}$ requires that $\partial_{t_1} A_1 = 0$. This is consistent with standard, unforced Stokes waves: Stokes corrections to the unforced wave frequency first occur at $O(\varepsilon^2)$, meaning we would only expect $A_1$ to have a $t_2$-dependence (which we also observe, cf. appendix A.1). Solving (3.41)–(3.43) for $\hat{\phi}_{2,m}$ and transforming back to $\phi_2$ via (3.5) gives

$$\phi_2 = \mathrm{i}\frac{\omega_0}{4} A_1^2 \coth(h) \frac{\left(2 - \operatorname{csch}^2(h)\right)\omega_0^2 - 2\left[1 + \hat{P}_2\right]\coth(h)}{\left(2 + \operatorname{csch}^2(h)\right)\omega_0^2 - \left[1 + \hat{P}_2\right]\coth(h)} \exp(2\mathrm{i}(x - \omega_0 t_0)) \frac{\cosh[2(z+h)]}{\sinh(2h)}$$
$$+ \frac{1}{8\,\mathrm{Im}\{\omega_0\}}\left(2\,\mathrm{Re}\{\omega_0^2\} - |\omega_0|^2\left(2 + \operatorname{csch}^2(h)\right)\right)|A_1|^2 (\exp(2\,\mathrm{Im}\{\omega_0\} t_0) - 1). \tag{3.44}$$

We have included a constant term $-1$ in $\exp(2\,\mathrm{Im}\{\omega_0\}t_0) - 1$ so that $\phi_2$ remains finite if $P \to 0$ (i.e. $\mathrm{Im}\{\omega_0\} \to 0$). We have also dropped the homogeneous solution, which would only amount to redefining the linear solution, $A_1$.

The surface boundary conditions are now solely equations for $\hat{\eta}_{2,m}$

$$\frac{\partial \hat{\eta}_{2,2}}{\partial t_0} = -\mathrm{i}\frac{1}{2}\omega_0^3 A_1^2 \exp(-2\mathrm{i}\omega_0 t_0) \frac{\left(2 + 3\operatorname{csch}^2(h)\right)\coth(h)}{\left(2 + \operatorname{csch}^2(h)\right)\omega_0^2 - \left[1 + \hat{P}_2\right]\coth(h)} \tag{3.45}$$

$$\left[1 + \hat{P}_2\right]\hat{\eta}_{2,2} = \frac{1}{4}\left[1 + \hat{P}_2\right] A^2 \exp(-2\mathrm{i}\omega_0 t_0) \frac{(2 + 3\operatorname{csch}^2(h))\coth(h)\omega_0^2}{\left(2 + \operatorname{csch}^2(h)\right)\omega_0^2 - \left[1 + \hat{P}_2\right]\coth(h)}, \tag{3.46}$$

and $\hat{\eta}_{2,0} = \hat{\eta}_{2,1} = 0$. These have the solution

$$\eta_2 = \frac{1}{4} A_1^2 \exp(2\mathrm{i}(x - \omega_0 t_0)\left(2 + 3\operatorname{csch}^2(h)\right)\coth(h)\left(1 - \coth^2(h)\left[\frac{\hat{P}_2 - \hat{P}_1}{1 + \hat{P}_1}\right]\right)^{-1} \tag{3.47}$$

Note that we chose $\hat{\eta}_{2,0} = 0$ since we imposed $\overline{\eta} = 0$ at $t = 0$ with our choice of the mean water level as our initial datum in § 2.1. It is interesting to note that, when $\overline{\eta} = 0$ initially, it remains zero for all times. This implies that the mean water level does not change over time. Another choice of datum occasionally used (e.g. Laitone 1962) is the mean energy level (MEL), defined such that $\overline{\partial_t \phi} = 0$ (e.g. Song *et al.* 2013). However, even if we chose $\overline{\partial_t \phi} = 0$ initially by adding a constant $A$ to $\eta_2$ and a term $-At_0$ to $\phi_2$, (3.44) shows that the MEL would still vary with time.

Redimensionalizing, we find

$$\eta = \varepsilon\frac{A_1}{k}\exp(\mathrm{i}(x - \omega_0 t_0)) + \varepsilon^2\frac{A_1^2}{k}\exp(2\mathrm{i}(x - \omega_0 t_0))C_{2,2} + O(\varepsilon^3), \tag{3.48}$$

where the complex $C_{2,2}$ is the pressure-induced (or wind-induced) correction to the first harmonic

$$C_{2,2} := \frac{1}{4}\left(2 + 3\operatorname{csch}^2(h)\right)\coth(h)\left(1 - \coth^2(h)\left[\frac{\hat{P}_2 - \hat{P}_1}{1 + \hat{P}_1}\right]\right)^{-1}. \tag{3.49}$$

Note that $A_1^2|C_{2,2}|/k$ is the quantity denoted $A_2$ in (3.1).

We have now found the primary wave $\hat{\eta}_{m=1} = \varepsilon\hat{\eta}_{1,m} + O(\varepsilon^3)$ and first harmonic $\hat{\eta}_{m=2} = \varepsilon^2\hat{\eta}_{2,2} + O(\varepsilon^3)$. Therefore, the amplitudes of the primary wave and first harmonic

are respectively

$$a_1 := |\hat{\eta}_{m=1}| = \varepsilon \frac{|A_1(t_2)|}{k} \exp(\mathrm{Im}\{\omega_0\} t_0) + O\big(\varepsilon^3\big), \tag{3.50}$$

$$a_2 := |\hat{\eta}_{m=2}| = \varepsilon^2 \frac{|A_1^2(t_2)|}{k} \exp(2\,\mathrm{Im}\{\omega_0\} t_0) |C_{2,2}| + O\big(\varepsilon^3\big). \tag{3.51}$$

Hence, in order to cancel the $t_0$-dependence, we define the relative harmonic amplitude shape parameter as

$$\frac{a_2}{a_1^2 k} := \left| \frac{\hat{\eta}_{m=2}}{\hat{\eta}_{m=1}^2 k} \right|. \tag{3.52}$$

With this definition, (3.48) becomes

$$\eta = a_1 \exp(\mathrm{i}(x - \mathrm{Re}\{\omega_0\} t_0)) + a_2 \exp(\mathrm{i}[2(x - \mathrm{Re}\{\omega_0\} t_0) + \beta]) + O\big(\varepsilon^3\big), \tag{3.53}$$

where we have absorbed the complex phase of $A_1$ into $\exp(\mathrm{i}x)$ (redefining the $x = 0$ location) and defined the harmonic phase $\beta$ as the complex angle of $\hat{\eta}_{m=2}/\hat{\eta}_{m=1}^2$

$$\beta := \tan^{-1}\left( \frac{\mathrm{Im}\{\hat{\eta}_{m=2}/\hat{\eta}_{m=1}^2\}}{\mathrm{Re}\{\hat{\eta}_{m=2}/\hat{\eta}_{m=1}^2\}} \right). \tag{3.54}$$

In general, both $\beta$ and $a_2/(a_1^2 k)$ will have an expansion in $\varepsilon$ since $\hat{\eta}_{m=2}$ will have higher-order corrections. For instance, the HP $\beta$ has expansion $\beta = \beta_0 + \varepsilon\beta_1 + \dots$. Inserting our solution (3.48) into (3.54) gives $\beta_0$, which is just the complex angle of $C_{2,2}$ at this order

$$\begin{aligned} \beta_0 &= \tan^{-1}\left( \frac{\mathrm{Im}\left\{ \left[ \frac{\hat{P}_2 - \hat{P}_1}{1+\hat{P}_1} \right] \right\}}{\tanh^2(h) - \mathrm{Re}\left\{ \left[ \frac{\hat{P}_2 - \hat{P}_1}{1+\hat{P}_1} \right] \right\}} \right) \\ &= \tan^{-1}\left( \frac{\mathrm{Im}\left\{ \left[ \hat{P}_2 - \hat{P}_1 \right] \left( 1 + \hat{P}_1^* \right) \right\}}{\left| 1 + \hat{P}_1 \right|^2 \tanh^2(h) - \mathrm{Re}\left\{ \left[ \hat{P}_2 - \hat{P}_1 \right] \left( 1 + \hat{P}_1^* \right) \right\}} \right), \end{aligned} \tag{3.55}$$

with an asterisk representing the complex conjugate. Similarly, using (3.52) shows that the leading-order term of $a_2/(a_1^2 k)$ is just $|C_{2,2}|$

$$\frac{a_2}{a_1^2 k} = |C_{2,2}| = \frac{2 + 3\,\mathrm{csch}^2(h)}{4} \coth(h) \left| 1 - \coth^2(h) \left[ \frac{\hat{P}_2 - \hat{P}_1}{1 + \hat{P}_1} \right] \right|^{-1}. \tag{3.56}$$

Without wind ($\hat{P}_1 = \hat{P}_2 = 0$), $C_{2,2}$ is real and equals $(2 + 3\,\mathrm{csch}^2(h)) \coth(h)/4$, or $1/2$ in deep water. Thus, $\hat{P}_1 = \hat{P}_2 = 0$ reproduces the usual Stokes waves values of $a_2/(a_1^2 k) = 1/2$ in deep water and $\beta = 0$.

Asymmetry and skewness are common shape parameters that depend on $\beta$ and $a_2/(a_1^2 k)$. The skewness S and asymmetry A are defined as

$$\mathrm{S} := \frac{\langle \eta^3 \rangle}{\langle \eta^2 \rangle^{3/2}}, \tag{3.57}$$

$$\mathrm{A} := \frac{\langle \mathcal{H}\{\eta\}^3 \rangle}{\langle \eta^2 \rangle^{3/2}}, \tag{3.58}$$

with $\langle \cdot \rangle$ the spatial average over one wavelength and $\mathcal{H}\{\cdot\}$ the Hilbert transform (in $x$). The average of any Fourier component $\exp(\mathrm{i}mx)$ over a wavelength is zero for all $m \neq 0 \in \mathbb{N}$. Therefore, only combinations wherein the $x$-dependence cancels will contribute.

Inserting our solution for $\eta$ (3.53) into the skewness and asymmetry definitions (3.57) and (3.58) yields

$$\mathrm{S} = \frac{3}{\sqrt{2}}\varepsilon|A_1|\exp(\mathrm{Im}\{\omega_0\}t_0)\frac{a_2}{a_1^2k}\cos(\beta_0) + O(\varepsilon^2), \tag{3.59}$$

$$\mathrm{A} = -\frac{3}{\sqrt{2}}\varepsilon|A_1|\exp(\mathrm{Im}\{\omega_0\}t_0)\frac{a_2}{a_1^2k}\sin(\beta_0) + O(\varepsilon^2). \tag{3.60}$$

By solving the kinematic and dynamic boundary conditions to $O(\varepsilon^2)$, we have generated the leading-order terms for $\beta$ and $a_2/(a_1^2k)$. We continue this analysis by solving to $O(\varepsilon^4)$ in appendix A, deriving the first non-trivial correction to $\beta$ (A 58), $a_2/(a_1^2k)$ (A 57) and the complex frequency $\omega$ (A 64). Additionally, going to $O(\varepsilon^4)$ also extends our solutions to weaker wind conditions. As outlined in § 2.5, we can substitute $P \to \varepsilon P$ or $P \to \varepsilon^2 P$ to generate shape parameters for weaker winds $Pk/(\rho_w g) = O(\varepsilon)$ or $Pk/(\rho_w g) = O(\varepsilon^2)$, respectively (appendix A.6). In this way, we find the shape parameters' dependence on our non-dimensional parameters ($kh$, $\varepsilon$, $P$ and $\psi_P$) and demonstrate weak time and amplitude dependence over a range of wind conditions from strong $Pk/(\rho_w g) = O(1)$ to relatively weak $Pk/(\rho_w g) = O(\varepsilon^2)$.

## 4. Results

Now, we present the main results of this theory. The harmonic phase $\beta$, harmonic magnitudes $a_1$ and $a_2$, and complex frequency $\omega$ depend on the four non-dimensional parameters: the wave steepness $\varepsilon := a_1 k$, water depth $kh$, pressure magnitude constant $Pk/(\rho_w g)$ and wind phase $\psi_P$. To reduce the non-dimensional parameter range, we keep a fixed $\varepsilon = 0.2$. Recall (§ 2.2) the requirement of $\varepsilon/(kh)^3 \leqslant 1$, such that the expansion remains properly ordered, implies $kh \geqslant 0.5$, although we keep $kh \geqslant 1$. Note that taking $kh$ to $\infty$ yields solutions on infinite depth. The pressure magnitude constant $P$ is $P_J$ or $P_G$, corresponding to the choice of pressure profile. For both solutions, taking $P \to 0$ recovers the unforced Stokes wave.

For the remainder of the paper, we will revert to dimensional variables. In particular, the pressure constant $P$ is dimensional again and not necessarily order unity. Replacing the multiple time scales with the true time $t$ in our solution (3.53), we obtain a surface height profile $\eta$ of the form

$$k\eta = (a_1k)\exp(\mathrm{i}\theta) + (a_1k)^2\frac{a_2}{a_1^2k}\exp(\mathrm{i}(2\theta+\beta)) + \dots, \tag{4.1}$$

with the real part implied and $\theta$ defined in (2.7). Note that the growth of the harmonics means that these solutions are only valid for finite time (cf. § 5.1).

### 4.1. *Harmonic phase, relative harmonic amplitude and wave shape*

The wave shape is a function of the harmonic phase $\beta$, quantifying the relative phase shift between the primary wave and first harmonic, and the relative harmonic ratio $a_2/(a_1^2k)$. The solutions for these parameters are extended to $O(\varepsilon^2)$ in appendix A.3, applying to all pressure profiles satisfying (2.17) with magnitude $Pk/(\rho_w g) = O(1)$ to $O(\varepsilon^2)$. We now specialize these results to the two pressure profiles of interest.

The full, $O(\varepsilon^2)$-accurate Jeffreys harmonic phase $\beta_J$ (A 58) is depicted in figures 3(*a*), 4(*a*) and 5(*a*). To develop a better understanding of its functional dependence, we can consider simpler, limiting cases. For very small wave steepnesses, $\varepsilon \lll 1$, the leading-

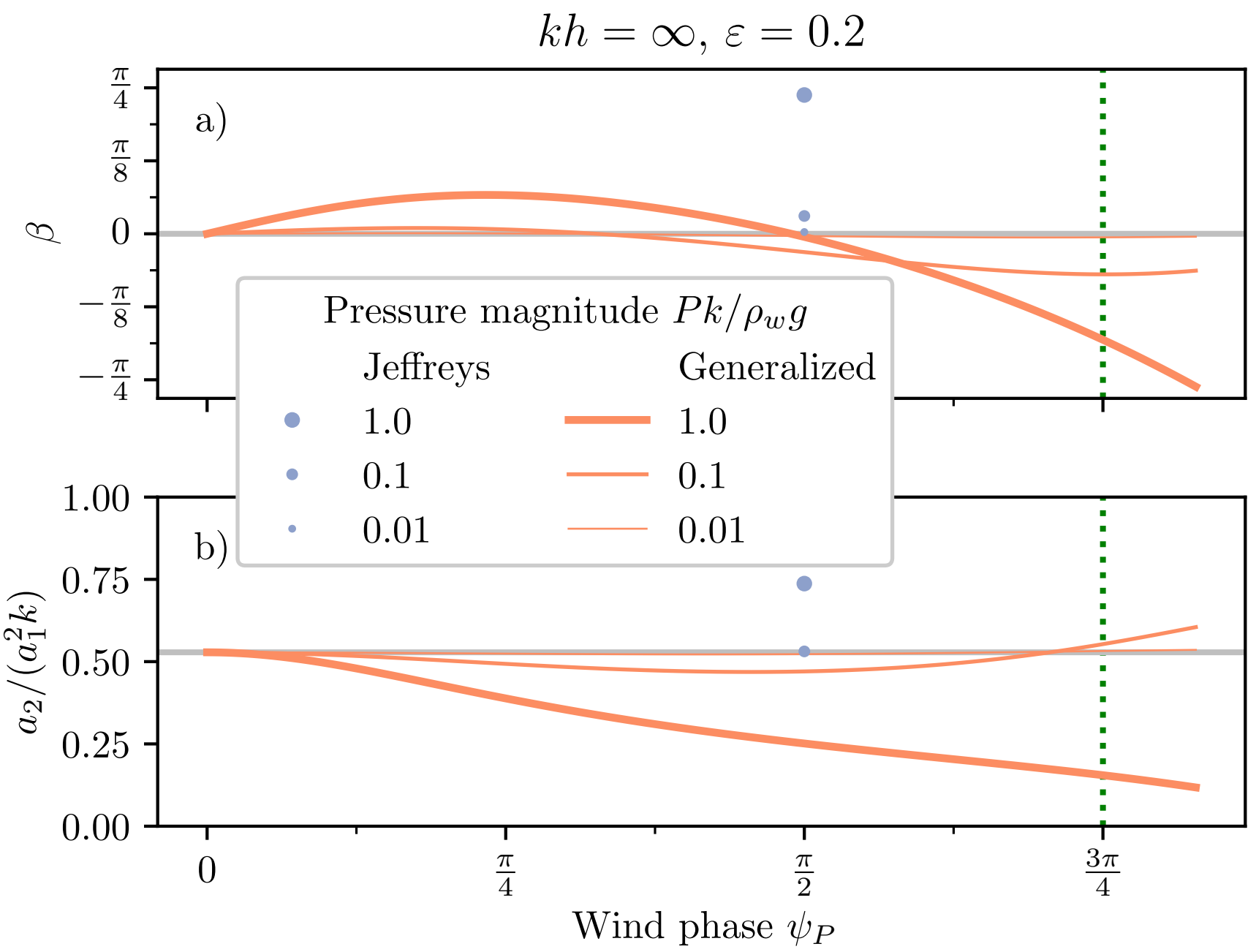

Figure 3. ($a$) Harmonic phase $\beta$ (3.54) and ($b$) relative harmonic amplitude $a_2/(a_1^2 k)$ (3.52) versus wind phase $\psi_P$. Results are shown for Jeffreys and generalized Miles profiles with $kh = \infty$, $\varepsilon = 0.2$ and pressure magnitude constants $Pk/(\rho_w g) = 0.01$, 0.1 and 1, as indicated in the legend. The Jeffreys $\beta_J$ is only shown at $\psi_P = \pi/2$ as that is its implied $\psi_P$. All results are plotted using the full, $O(\varepsilon^2)$-accurate expressions (A 57) and (A 58). The grey lines are the results for a fourth-order unforced Stokes wave, and the green dotted line represents $\psi_P = 3\pi/4$ used in many of the other plots and supported by numerical simulations from Hara & Sullivan (2015) and Husain *et al.* (2019).

order correction (3.55) is

$$\beta_J = \pm \tan^{-1}\left(\frac{Pk/(\rho_w g)}{\tanh^2(kh) - \operatorname{sech}^2(kh) P^2 k^2/(\rho_w^2 g^2)}\right) + O(\varepsilon), \tag{4.2}$$

with the $\pm$ corresponding to the sign of $\psi_P = \pm\pi/2$ in the pressure profile. If, instead of assuming $\varepsilon \lll 1$, we expand (A 58) considering a weak pressure forcing $Pk/(\rho_w g) \ll 1$, we find

$$\beta_J = \pm \frac{Pk}{\rho_w g} \coth^2(kh) + O(\varepsilon^3). \tag{4.3}$$

The full, $O(\varepsilon^2)$-accurate generalized Miles $\beta_G$ (A 58) is also depicted in figures 3($a$), 4($a$)

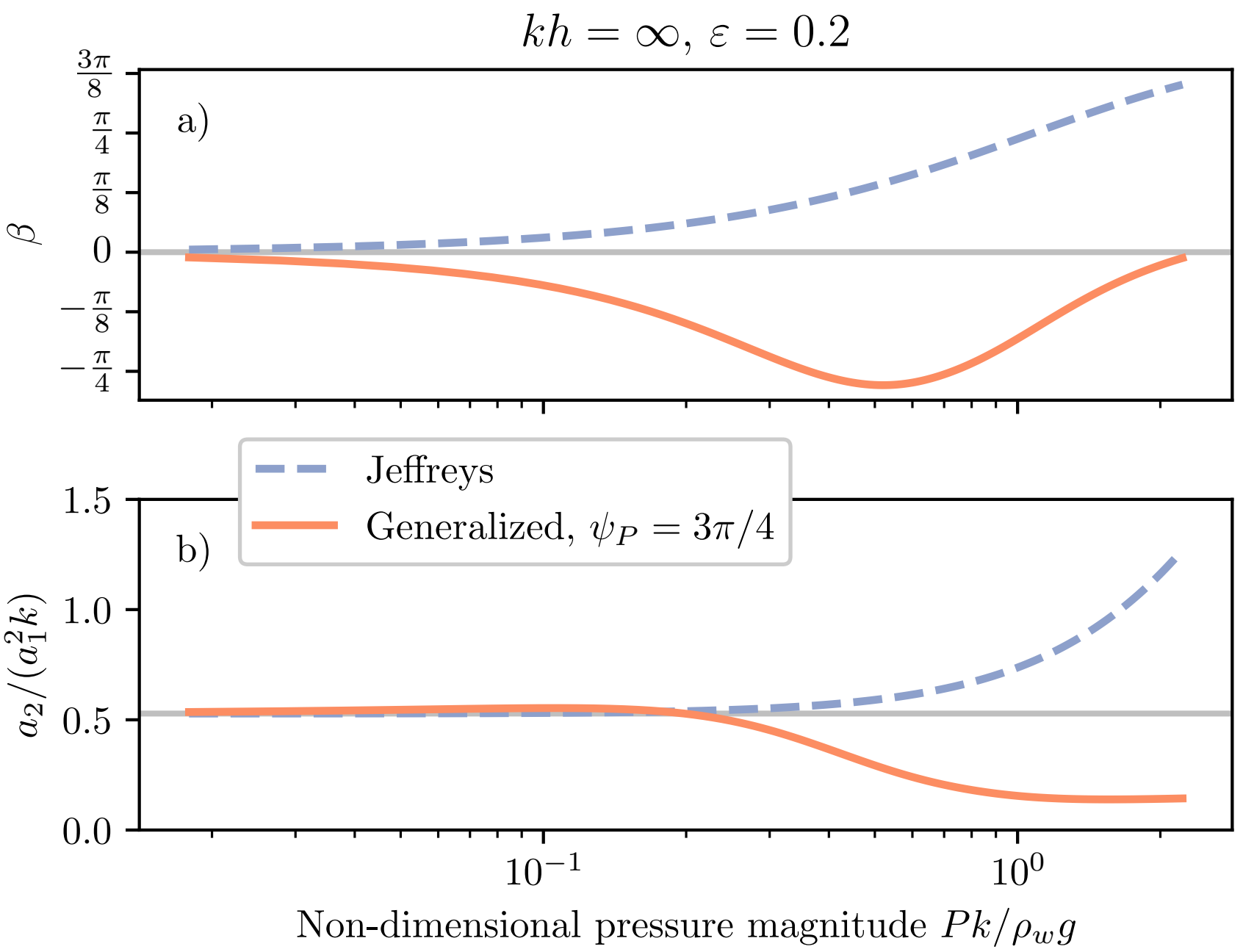

Figure 4. (*a*) Harmonic phase $\beta$ (3.54) and (*b*) relative harmonic amplitude $a_2/(a_1^2 k)$ (3.52) versus non-dimensional pressure magnitude constant $Pk/(\rho_w g)$. Results are shown for Jeffreys and generalized Miles profiles, as indicated in the legend, with $kh = \infty$, $\varepsilon = 0.2$ and $\psi_P = 3\pi/4$ (for generalized Miles). All results are plotted using the full, $O(\varepsilon^2)$-accurate expressions (A 57) and (A 58), which include the $Pk/(\rho_w g) \ll 1$ limits for $\beta$ (4.3) and (4.5), as well as for $a_2/(a_1^2 k)$ (4.7) and (4.9). The grey lines are the results for a fourth-order unforced Stokes wave.

and 5(*a*). For very small $\varepsilon \lll 1$, we have the approximation

$$\beta_G = \tan^{-1}\left(\left[2\cos(\psi_P) - 1 + \frac{Pk}{\rho_w g}\right]\frac{Pk}{\rho_w g}\sin(\psi_P)\left[-\left(\frac{Pk}{\rho_w g}\right)^2\cos(\psi_P) - 1\right.\right.$$
$$\left.\left. - \frac{Pk}{\rho_w g}(\cos(2\psi_P) + \cos(\psi_P)) + \left(1 + 2\frac{Pk}{\rho_w g}\cos(\psi_P) + \left(\frac{Pk}{\rho_w g}\right)^2\right)(2 - \operatorname{sech}^2(kh))\right]^{-1}\right)$$
$$+ O(\varepsilon). \tag{4.4}$$

Instead of requiring $\varepsilon \lll 1$, we can expand (A 58) while considering a weak pressure forcing $Pk/(\rho_w g) \ll 1$ to find

$$\beta_G = \frac{Pk}{\rho_w g}(\sin(2\psi_P) - \sin(\psi_P))\coth^2(kh) + \frac{1}{2}\left(\frac{Pk}{\rho_w g}\right)^2\coth^4(kh)\Big(\sin(4\psi_P)$$
$$- 4\sin(3\psi_P) + 3\sin(2\psi_P) + 2\operatorname{sech}^2(kh)[\sin(3\psi_P) - \sin(2\psi_P)]\Big) + O(\varepsilon^3). \tag{4.5}$$

Next, we consider the relative harmonic amplitude, $a_2/(a_1^2 k)$. The full, $O(\varepsilon^2)$-accurate Jeffreys relative harmonic amplitude (A 57) is shown in figures 3(*b*), 4(*b*) and 5(*b*), but

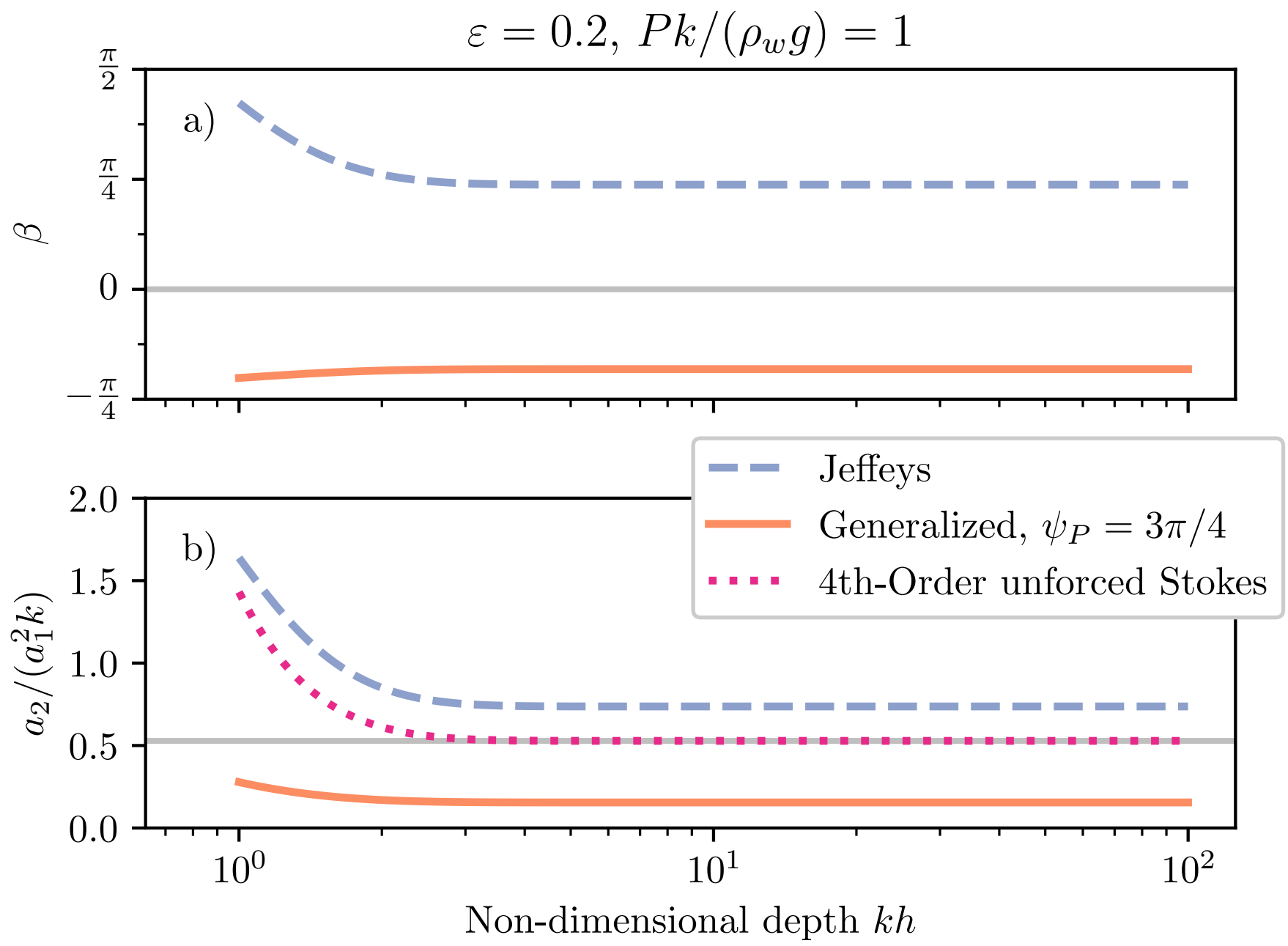

Figure 5. (*a*) Harmonic phase $\beta$ (3.54) and (*b*) relative harmonic amplitude $a_2/(a_1^2k)$ (3.52) versus non-dimensional depth $kh$. Results are shown for Jeffreys and generalized Miles profiles, as well as unforced (i.e. no wind) Stokes waves, with $\varepsilon = 0.2$, pressure magnitude constant $Pk/(\rho_w g) = 1$ and $\psi_P = 3\pi/4$ (for generalized Miles). All results are plotted using the full, $O(\varepsilon^2)$-accurate expressions (A 57) and (A 58). The grey lines are the results for a fourth-order unforced Stokes wave with $kh = \infty$.

we can approximate it for very small $\varepsilon \lll 1$ as (3.56)

$$\left(\frac{a_2}{a_1^2k}\right)_J = \frac{2 + 3\operatorname{csch}^2(kh)}{4}\coth(kh)\sqrt{\frac{1 + P^2k^2/(\rho_w^2 g^2)}{1 + P^2k^2/(\rho_w^2 g^2)\operatorname{csch}^4(kh)}}. \tag{4.6}$$

Inserting a weak wind $Pk/(\rho_w g) \ll 1$ in (A 57) instead of requiring $\varepsilon \lll 1$ yields

$$\left(\frac{a_2}{a_1^2k}\right)_J = \frac{2 + 3\operatorname{csch}^2(kh)}{4}\coth(kh)\left(1 + \frac{1 - \operatorname{csch}^4(kh)}{2}\left(\frac{Pk}{\rho_w g}\right)^2\right) + (a_1k)^2\mathcal{A} + O(\varepsilon^3), \tag{4.7}$$

with $\mathcal{A}$ only a function of $kh$ and defined in (A 55). We now have the direct appearance of the amplitude $a_1k$ with an implicit time dependence due to growth. The full, $O(\varepsilon^2)$-accurate generalized Miles $a_2/(a_1^2k)_G$ (A 57) is also plotted in figures 3(*b*), 4(*b*) and 5(*b*). We can simplify $a_2/(a_1^2k)_G$ by assuming a very small wave steepness $\varepsilon \lll 1$

$$\left(\frac{a_2}{a_1^2k}\right)_G = \frac{2 + 3\operatorname{csch}^2(kh)}{4}\coth(kh)\left|1 - \coth^2(kh)\frac{[\exp(\mathrm{i}\psi_P) - 1]Pk/(\rho_w g)}{\exp(-\mathrm{i}\psi_P) + Pk/(\rho_w g)}\right|^{-1} + O(\varepsilon). \tag{4.8}$$

Instead of assuming very small $\varepsilon$, we can approximate (A 57) by assuming $Pk/(\rho_w g) \ll 1$

to give

$$\left(\frac{a_2}{a_1^2 k}\right)_G = \frac{2 + 3\,\mathrm{csch}^2(kh)}{4}\coth(kh)\Bigg(1 + \frac{Pk}{\rho_w g}[\cos(2\psi_P) - \cos(\psi_P)]\coth^2(kh) + \frac{1}{2}\left(\frac{Pk}{\rho_w g}\right)^2 (\cos(\psi_P) - 1)\Big\{3\coth^2(kh)\cos(3\psi_P) - 4\cos(2\psi_P) - 4\cos(\psi_P) - 3 - \mathrm{csch}^2(kh)\Big\}\coth^2(kh)\Bigg) + (a_1 k)^2 \mathcal{A} + O\big(\varepsilon^3\big). \tag{4.9}$$

Note that we see a weak amplitude dependence (i.e. $a_1 k$ terms) appearing in some of these results, such as (4.7) and (4.9). This amplitude dependence is implicitly present in figures 3–5 since they show the full, $O\big(\varepsilon^2\big)$ results (A 57) and (A 58) which encode this dependence. However, we do not show the $\beta$ and $a_2/(a_1^2 k)$ dependence on $\varepsilon$ as these effects are $O\big(\varepsilon^2\big)$, or approximately 4 % of the leading-order effects in figures 3–5.

Figure 3 shows the influence of wind phase $\psi_P$ on $\beta$ and $a_2/(a_1^2 k)$ for both the Jeffreys and generalized Miles profiles with $kh = \infty$ and $\varepsilon = 0.2$ for a range of pressure magnitudes $Pk/(\rho_w g) = 0.01$, 0.1 and 1. For the strongest pressure forcing $Pk/(\rho_w g) = 1$, both the Jeffreys and generalized Miles profiles induce a harmonic phase magnitude $|\beta|$ up to $\pi/4$ (figure 3(*a*)). The Jeffreys value of $\beta_J = \pi/4$ is placed at $\psi_P = \pi/2$ to correspond with its restriction that $\psi_P = \pm\pi/2$. The generalized Miles HP $\beta$ increases from zero at $\psi_P = 0$ (figure 3(*a*)) to roughly $\pi/16$ for the largest pressure, before decreasing to approximately $-\pi/4$ and passing through zero near $\psi_P = \pi/2$. The weaker pressure forcings show a much reduced $\beta$ range, cross $\beta = 0$ at somewhat smaller values of $\psi_P$ and yield much smaller $\beta$ for large wind phase angles. The angle $\psi_P = 3\pi/4$ is denoted by a dashed line in figure 3, and this $\psi_P$ is utilized hereafter, as suggested by Hara & Sullivan (2015) and Husain *et al.* (2019).

The relative harmonic amplitude shows opposing behaviour for the two forcing types in figure 3(*b*). The Jeffreys $a_2/(a_1^2 k)_J = 0.7$ for the strongest wind is enhanced relative to the deep-water Stokes value $a_2/(a_1^2 k) = 1/2$, while the generalized Miles value is suppressed $a_2(a_1^2 k)_G \leqslant 1/2$ for most values of $\psi_P$. As in figure 3(*a*), the weaker pressure magnitudes give correspondingly smaller changes to $a_2/(a_1^2 k)$, although the small $Pk/(\rho_w g)$ do slightly enhance $a_2/(a_1^2 k)$ for large $\psi_P$. It is worth noting that the strongest pressure $Pk/(\rho_w g) = 1$ suppresses the first harmonic $a_2$ as $\psi_P \to \pi$, making the wave more linear. However, as discussed in § 5.2, $\psi_P \approx \pi$ is usually observed for very weak winds. As the $Pk/(\rho_w g) = 0.1$ and 0.01 lines in figure 3(*b*) show, weaker winds show no such linearization. Note that figure 3 only depicts $\psi_P \geqslant 0$ since $\beta$ (3.55) is antisymmetric and $a_2/(a_1^2 k)$ (3.56) is symmetric about $\psi_P = 0$. This is seen by noticing $\psi_P \to -\psi_P \implies \hat{P}_m \to \hat{P}_m^*$.

The wave-shape parameters show a particularly rich dependence on the pressure magnitude $Pk/(\rho_w g)$ (figure 4). While both Jeffreys and generalized Miles yield non-zero harmonic phase $\beta$ for small pressures (figure 4(*a*)), they have opposite responses for large $Pk/(\rho_w g)$. The Jeffreys profile increases steadily, reaching $3\pi/8$ for $Pk/(\rho_w g) = 3$. Instead, the generalized Miles profile first decreases, reaching a minimum of approximately $-\pi/4$ at $Pk/(\rho_w g) = 0.6$ and then increasing to small, positive values. The relative harmonic amplitude shows (figure 4(*b*)) virtually no change from the deep-water Stokes value of 1/2 until $Pk/(\rho_w g) = 0.3$. Then, the Jeffreys profile increases rapidly, attaining $a_2/(a_1^2 k)_J = 1.7$ for $Pk/(\rho_w g) = 3$. Contrarily, the generalized Miles profile decreases and asymptotes to $a_2/(a_1^2 k)_G \approx 0.2$.

Finally, the non-dimensional depth $kh$ also modulates the wind's effect on wave shape.

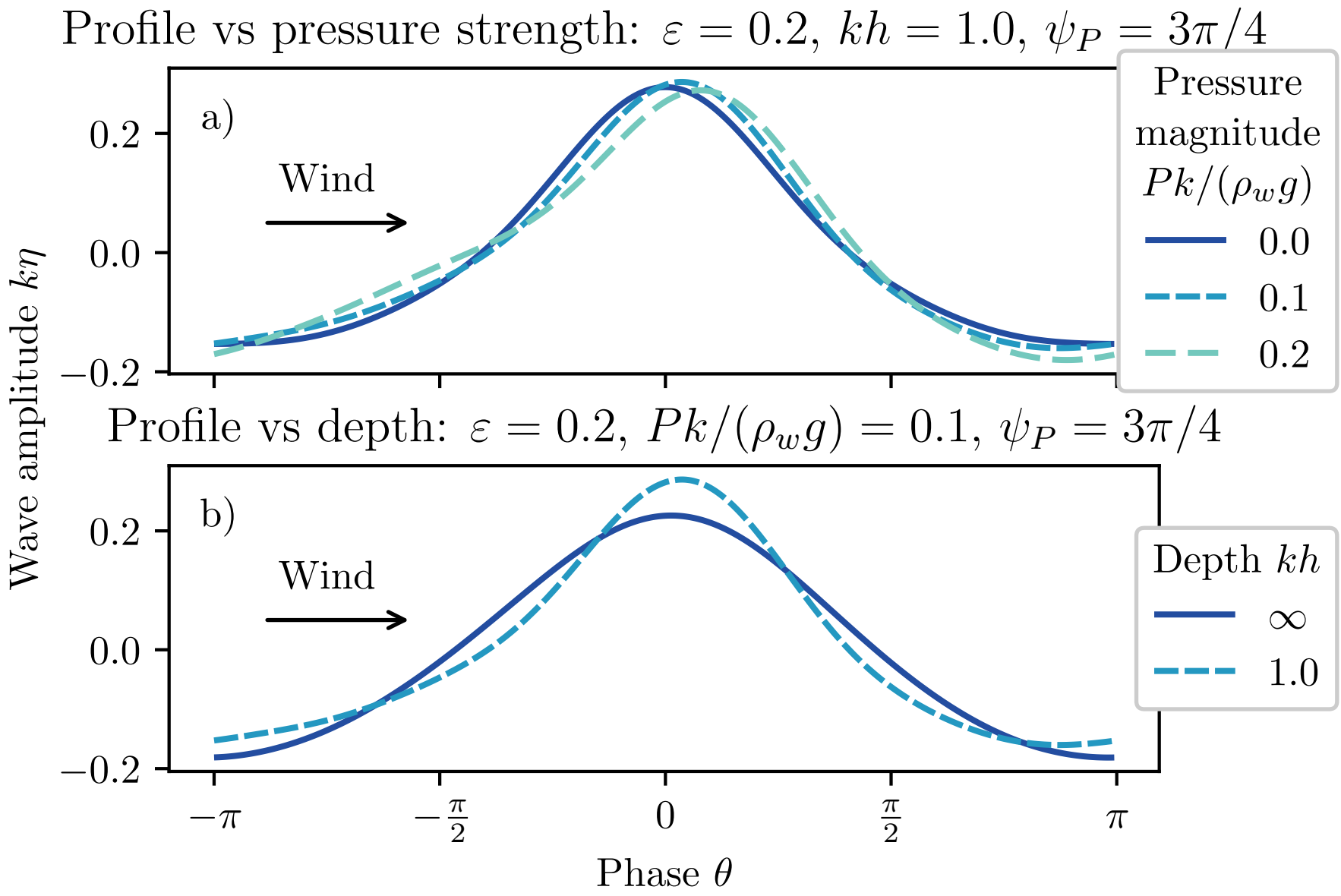

Figure 6. Wave profile $k\eta$ versus phase $\theta$ for $\varepsilon = 0.2$, $\psi_P = 3\pi/4$ and the generalized Miles pressure profile for (*a*) $kh = 1.0$ and variable $Pk/(\rho_w g)$ (see legend) and (*b*) $Pk/(\rho_w g) = 0.1$ and variable $kh$ (see legend).

For the chosen values of $Pk/(\rho_w g) = 1$ and $\psi_P = 3\pi/4$, the generalized Miles $\beta_G \approx -\pi/4$ while Jeffreys $\beta_J \approx +\pi/4$ for large $kh$ (figure 5(*a*)). However, as $kh$ decreases, both values grow in magnitude with $\beta_J$ increasing faster, nearly reaching $\beta_J = \pi/2$ at $kh = 1$. Thus, the shallower depth $kh$ strongly enhances the effect of wind on $\beta$. The wind's influence on $a_2/(a_1^2 k)$ is less pronounced. Notice that the unforced Stokes wave also has a depth dependence for $a_2/(a_1^2 k)$ (dashed line in figure 5(*b*)). Although the relative harmonic amplitude is enhanced for small $kh$ in all three cases (Jeffreys, generalized and unforced Stokes), both pressure profiles grow slower than the unforced Stokes wave. That is, the pressure forcing appears to counteract shoaling-induced $a_2/(a_1^2 k)$ enhancement to some extent. Figure 5(*b*) also highlights the importance of restricting to $kh \geqslant 1$. As $kh$ decreases, $a_2$ becomes large compared to $a_1$ and the perturbation expansion could become disordered. This figure highlights a trend where Jeffreys and generalized Miles profiles exhibit opposite responses to the wind: namely, Jeffreys yields positive $\beta$ and an enhanced $a_2/(a_1^2 k)$, while generalized Miles gives a negative $\beta$ and a suppressed $a_2/(a_1^2 k)$. This difference is also apparent in figure 4, wherein $\beta$ and $a_2/(a_1^2 k)$ increase with increasing pressure magnitude for Jeffreys, while they decrease (at least initially) for the generalized Miles profile. This can be attributed to different choices of $\psi_P$ ($\psi_P = \pi/2$ for Jeffreys, but $\psi_P = 3\pi/4$ for generalized Miles), as well as different effects on higher harmonics, including the derivative in the Jeffreys profile enhancing higher harmonics.

Both the harmonic phase and the relative harmonic amplitude determine the wave shape. We consider their combined influence by plotting the surface profile under the action of the generalized Miles pressure profile, with $\varepsilon = 0.2$ to emphasize the pressure-induced shape changes. Figure 6(*a*) shows how the surface profile $\eta$ versus phase $\theta$ varies with $Pk/(\rho_w g) = 0$, 0.1 and 0.2 for wind blowing to the right. The $Pk/(\rho_w g) = 0$ profile has skewness (3.59) $S = 0.6$ and asymmetry (3.60) $A = 0$, as expected for a $kh = 1$ Stokes wave. The $Pk/(\rho_w g) = 0.1$ profile deviates only slightly from the unforced profile. However, the $Pk/(\rho_w g) = 0.2$ profile shows a noticeable horizontal asymmetry, with both skewness $S = 0.4$ and asymmetry $A = 0.3$ that are fundamentally different from a Stokes wave. This follows from figure 4(*a*) with $kh = \infty$ since $Pk/(\rho_w g) = 0.1$ generates a somewhat small $\beta_G \approx -12°$, while $\beta_G \approx -27°$ is significantly larger for $Pk/(\rho_w g) = 0.2$. Instead, (4.5) can be used when $kh = 1$ to calculate $\beta_G \approx -19°$ for $Pk/(\rho_w g) = 0.1$ and $\beta_G \approx -45°$ for $Pk/(\rho_w g) = 0.2$. Note that the larger pressure magnitudes cause the crest to shrink. This is to be expected, since the magnitude of the first harmonic $a_2/a_1^2 k$ decreases as $Pk/(\rho_w g)$ increases for the generalized Miles profile (figure 3(*b*)). We can also see that increasing the depth $kh$ decreases the influence of wind on asymmetry (figure 6(*b*)). The $kh = \infty$ profile ($S = 0.2$, $A = 0.04$) is less asymmetric than the $kh = 1$ profile, in agreement with figure 5.

### 4.2. *Phase speed and growth rate*

In addition to influencing wave shape, the pressure-forcing terms also affect the phase speed, as predicted by Jeffreys (1925) and Miles (1957). We normalize the phase speed $c = \mathrm{Re}\{\omega\}/k$ by the unforced, linear phase speed $c_0 = \sqrt{g\tanh(kh)/k}$. The complete fractional phase speed change $\varDelta c/c_0$ is given in (A 65). If we consider very small waves $\varepsilon \lll 1$, then (A 65) simplifies considerably

$$\begin{aligned}\frac{\varDelta c}{c_0} &= \frac{|c| - |c|\big|_{P=0}}{c_0}\\ &= \frac{1}{\sqrt{2}}\sqrt{1 + \frac{Pk}{\rho_w g}\cos(\psi_P) + \sqrt{1 + \left(\frac{Pk}{\rho_w g}\right)^2 + 2\frac{Pk}{\rho_w g}\cos(\psi_P)}} - 1 + O\big(\varepsilon^2\big),\end{aligned} \tag{4.10}$$

with $\psi_P = \pm\pi/2$ for the Jeffreys profile. If, instead of very small waves, we assume the forcing is weak, $Pk/(\rho_w g) \ll 1$, we find

$$\begin{aligned}\frac{\varDelta c}{c_0} &= \frac{1}{2}\frac{Pk}{\rho_w g}\cos(\psi_P) - \frac{1}{8}\left(\frac{Pk}{\rho_w g}\right)^2\cos(2\psi_P)\\ &\quad + \frac{8\cosh^4(kh) - 8\cosh^2(kh) + 9}{16\sinh^4(kh)}\left((a_1 k)^2 - (a_1 k)^2\bigg|_{P=0}\right) + O\big(\varepsilon^3\big),\end{aligned} \tag{4.11}$$

For these limiting cases, we find that both surface pressure profiles generate the same change to the phase speed. This is unsurprising since, at leading order, both pressure profiles are equivalent (if $\psi_P = \pm\pi/2$). The $a_1^2$ term is the amplitude dispersion due to nonlinearity described by Stokes (1880).

As shown in § 3, the different harmonics grow at different rates. Here, we will discuss the growth rate of the primary wave. It is conventional to describe the energy growth rate, $\gamma := \partial_t E/E$, rather than the amplitude growth rate, $\partial_t \eta/\eta = \mathrm{Im}\{\omega\}$. However, since $E \propto \eta^2$, they are related as $\gamma = 2\,\mathrm{Im}\{\omega\}$. The complete non-dimensional growth rate $\gamma/f_0$ is given in (A 66). For very small waves, $\varepsilon \lll 1$, (A 66) simplifies to

$$\begin{aligned}\frac{\gamma}{f_0} &= \frac{4\pi\,\mathrm{Im}\{\omega\}}{c_0 k} = 2\sqrt{2}\pi\,\mathrm{sgn}\left(\frac{Pk}{\rho_w g}\sin(\psi_P)\right)\\ &\quad\times\sqrt{-1 - P\cos(\psi_P)k/(\rho_w g) + \sqrt{1 + P^2k^2/(\rho_w^2 g^2) + 2P\cos(\psi_P)k/(\rho_w g)}} + O\big(\varepsilon^2\big),\end{aligned} \tag{4.12}$$

with $f_0 = \mathrm{Re}\{\omega_0\}/(2\pi) = c_0 k/(2\pi)$ the unforced, linear wave frequency. Instead of assuming very small waves, if we consider weak wind forcing $Pk/(\rho_w g) \ll 1$, we find

$$\frac{\gamma}{f_0} = 2\pi \frac{Pk}{\rho_w g}\sin(\psi_P) - \frac{\pi}{2}\left(\frac{Pk}{\rho_w g}\right)^2 \sin(2\psi_P) + O\big(\varepsilon^3\big). \tag{4.13}$$

Both Jeffreys (1925)—with $\psi_P = \pi/2$—and Miles (1957) calculated the growth rate to leading order for weak pressure forcing $Pk/(\rho_w g) = O(\varepsilon)$; (4.13) matches their results. Naturally, if $P \to 0$, we find $\gamma \to 0$, as there is no growth.

Notice that, for both the Jeffreys and generalized Miles profiles, the HP $\beta$ and growth rate are related for very small waves ($\varepsilon \lll 1$) with weak wind ($Pk/(\rho_w g) \ll 1$) as

$$\begin{aligned}\begin{pmatrix}\beta_{0,J}\\ \beta_{0,G}\end{pmatrix} &= \frac{Pk}{\rho_w g}\begin{pmatrix}\pm 1\\ \sin(2\psi_P) - \sin(\psi_P)\end{pmatrix}\coth^2(kh) + O\left(\varepsilon\frac{Pk}{\rho_w g}\right)\\ &= \frac{1}{2\pi}\frac{\gamma}{f_0}\begin{pmatrix}1\\ (2\cos(\psi_P) - 1)\end{pmatrix}\coth^2(kh) + O\left(\varepsilon\frac{Pk}{\rho_w g}\right).\end{aligned} \tag{4.14}$$

The connection with wave asymmetry (related to $\beta$) suggests a deeper link between wave growth and wave shape. This is potentially analogous to shoaling, weakly nonlinear waves that both grow and becomes asymmetric.

## 5. Discussion

### 5.1. *Time scale validity*

Here, we discuss the time scale validity of our results. As asymptotic expansions must stay ordered to remain consistent, the solution's $O(\varepsilon)$ term must be larger than the $O\big(\varepsilon^2\big)$ term as $\varepsilon \to 0$. However, as shown in (3.50) and (3.51), the $O\big(\varepsilon^2\big)$ first harmonic grows faster than the $O(\varepsilon)$ primary wave, resulting in a disordered expansion in finite time. As the first harmonic grows faster than the primary wave by a factor of $\exp(\mathrm{Im}\{\omega_0\}t)$, consistency requires that this exponent remain $O(1)$ as $\varepsilon \to 0$, i.e.

$$t\,\mathrm{Im}\{\omega_0\} = O(1). \tag{5.1}$$

Since $O(\mathrm{Im}\{\omega_0\}) = O(Pk/(\rho_w g))$, redimensionalizing shows our results are restricted to

$$\frac{t}{T_0^\infty} \leqslant O\left(\frac{Pk}{\rho_w g}\right)^{-1}, \tag{5.2}$$

with the characteristic, unforced, linear, deep-water wave period $T_0^\infty = 2\pi/\sqrt{gk}$. For the case considered here with $Pk/(\rho_w g) = O(1)$, this implies the solution may only be valid for a few characteristic waves periods $T_0^\infty$. However, for weaker winds, the temporal range of validity is extended. For $Pk/(\rho_w g) = O\big(\varepsilon^2\big)$, the solution is well ordered for time intervals $O\big(T_0^\infty/\varepsilon^2\big)$, assuming the solution is calculated to $O\big(\varepsilon^3\big)$ accuracy with a frequency $\omega$ accurate to order $\varepsilon^2$.

The shape parameters $\beta$ and $a_2/(a_1^2 k)$ change very little over time. To leading order, the primary wave (3.50) grows like $\hat{\eta}_{m=1} \propto \exp(\mathrm{Im}\{\omega_0\}t_0)$ while the first harmonic (3.51) goes as $\hat{\eta}_{m=2} \propto \exp(2\,\mathrm{Im}\{\omega_0\}t_0)$. By dividing $\hat{\eta}_{m=2}/(\hat{\eta}_1^2 k)$ in (3.52) and (3.54), our shape parameters $\beta$ (3.55) and $a_2/(a_1^2 k)$ (3.56) are constant for time intervals of the length $O(T_0^\infty)$. Even the higher-order corrections (appendix A) for $a_2/(a_1^2 k)$ (A 57) and $\beta$ (A 58) show very little temporal variation during the valid time scales where $t\,\mathrm{Im}\{\omega_0\} = O(1)$. In contrast, the skewness (3.59) and asymmetry (3.60) show a stronger time scale dependence, with $\exp(t\,\mathrm{Im}\{\omega_0\})$ appearing at leading order. Nevertheless, our restriction that $t\,\mathrm{Im}\{\omega_0\} = O(1)$ ensures that, over the solution's range of temporal validity, the skewness and asymmetry do not vary substantially.

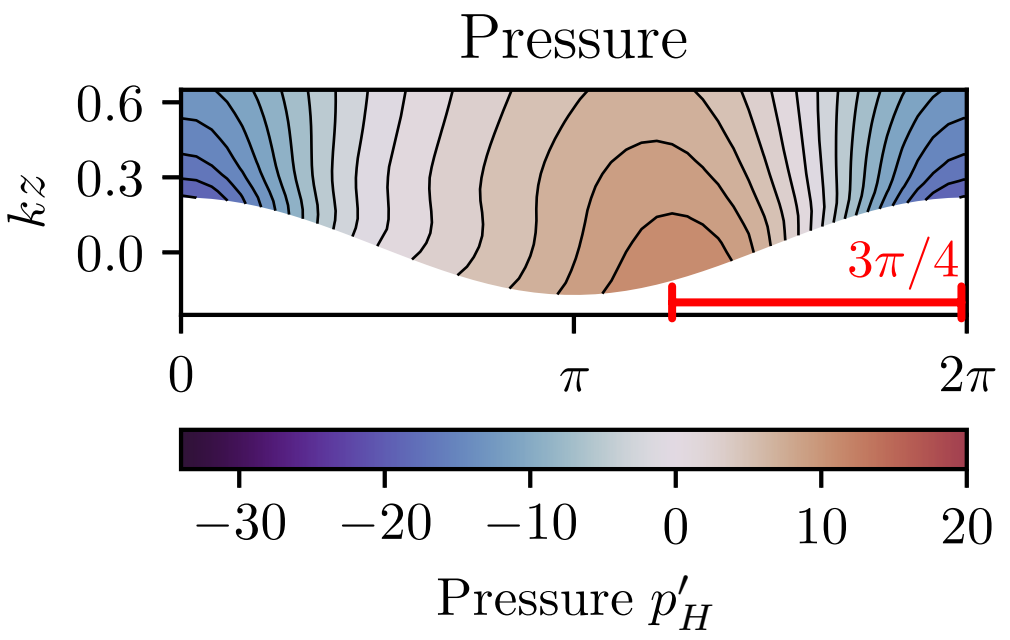

Figure 7. LES modelled non-dimensional, perturbation air pressure over a right-propagating linear surface gravity wave as a function of non-dimensional phase $kx$ and height $kz$. This simulation has non-dimensional surface roughness $kz_0 = 1.35 \times 10^{-3}$, wave steepness $\varepsilon = 0.2$ and inverse wave age $u_*/c_0^\infty = 0.71$. The red line denotes the wind phase $\psi_P$, as measured from the wave crest to the high pressure location. Reproduced from figure 2b of Husain *et al.* (2019).

### 5.2. *Using LES to constrain the surface pressure*

LES simulations of the airflow over a single, static, sinusoidal (i.e. no harmonics), deep-water wave by Husain *et al.* (2019) (see also Hara & Sullivan 2015) allow estimation of the two unknown parameters: pressure magnitude $Pk/(\rho_w g)$ and wind phase $\psi_P$. The Husain *et al.* (2019) simulations were based on the laboratory experiments of Buckley & Veron (2016) and explored a variety of surface roughnesses $kz_0$, wave steepnesses $\varepsilon$ and wind speeds $u_*/c_0^\infty$. We consider the simulation (Husain *et al.* 2019) with intermediate surface roughness $kz_0 = 1.35 \times 10^{-3}$, appreciable wave slope $\varepsilon = 0.2$ and young waves $u_*/c_0^\infty = 0.71$ (figure 7). The non-dimensional surface perturbation pressure $p'_{\mathrm{H}}$ varies over a range of $\pm 20$ with the maximum shifted $\approx 3\pi/4$ windward of the crest (red bar in figure 7), yielding our choice of $\psi_P \approx 3\pi/4$. The Husain *et al.* (2019) value of $\psi_P \approx 3\pi/4$ is also qualitatively consistent with the surface pressure and wind phase reported by Donelan *et al.* (2006).

Note that $\psi_P$ appears to be a function of wind speed (or pressure magnitude). Donelan *et al.* (2006), Hara & Sullivan (2015) and Husain *et al.* (2019) suggest $\psi_P \approx 3\pi/4$ for inverse wave ages $u_c/c_0 \approx 0.19$ to 0.71. In contrast, numerical simulations find $\psi_P \approx \pi$ for very small inverse wave ages $u_*/c_0 \leqslant 0.09$ (e.g. Sullivan *et al.* 2000; Yang & Shen 2010). According to figure 2, this corresponds to a growth rate $\gamma/f_0^\infty \leqslant 10^{-3} \lessapprox \varepsilon^4$ for $\varepsilon = 0.2$. Given that our analysis is limited to $Pk/(\rho_w g) = O(\varepsilon^2)$ or stronger (cf. appendix A.6), these weak winds are outside the scope of our analysis.

In regards to the pressure magnitude, Husain *et al.* (2019) non-dimensionalized pressure with the air density and friction velocity,

$$p'_{\mathrm{H}} = \frac{p}{\rho_a u_*^2}, \tag{5.3}$$

whereas we non-dimensionalized $p'$ by $\rho_w$, $g$ and $k$. Thus, converting $p'_{\mathrm{H}}$ to $p'$ we find

$$p' = \frac{pk}{\rho_w g} = \frac{p}{\rho_a u_*^2}\frac{u_*^2}{(c_0^\infty)^2}\frac{\rho_a}{\rho_w} = \frac{u_*^2}{(c_0^\infty)^2}\frac{\rho_a}{\rho_w}p'_{\mathrm{H}} \approx 5.0 \times 10^{-4} p'_{\mathrm{H}}. \tag{5.4}$$

With $u_*/c_0^\infty = 0.71$ (Husain *et al.* 2019) and $\rho_a/\rho_w \approx 10^{-3}$, $p' \approx 10^{-2}$ and $|p'| \approx 7\times10^{-3}$. Using their value of $\varepsilon = 0.2$ then gives $|p|k/(\rho_w g) \approx \varepsilon^3$, or $Pk/(\rho_w g) \approx \varepsilon^2$. Interestingly, the non-dimensional pressure magnitude for this simulation is consistent with that inferred from the $u_*/c_0^\infty$ versus $\gamma/f_0$ relationship (figure 2), where we see that $u_*/c_0^\infty = 0.7 \implies \gamma/f_0 = 0.1$. Using (2.19) and $\psi_P = 3\pi/4$ gives $Pk/(\rho_w g) = \gamma/[2\pi f_0 \sin(\psi_P)] = 2\times10^{-2}$. That is, $Pk/(\rho_w g) \approx \varepsilon^2$. This can be compared to our results for weak wind $Pk/(\rho_w g) \ll 1$, such as (4.5) and (4.9) truncated to $\varepsilon O(Pk/(\rho_w g)) = O(\varepsilon^3)$. Thus, the results of Husain *et al.* (2019) provide an estimate for $\psi_P$ and a $Pk/(\rho_w g)$ consistent with our theoretical development. However, the appropriate, specific pressure profile (Jeffreys or generalized Miles) remains to be determined; cf. § 5.4.

### 5.3. *Comparison of theory to laboratory wave-shape observations*

Here, we compare our predicted harmonic phase to the laboratory experiments in Leykin *et al.* (1995). We cannot compare to Feddersen & Veron (2005) as their $kh \leqslant 1.2$, and the $u_*/c_0$ to $\gamma/f_0$ relationship (figure 2) needed for determining $Pk/(\rho_w g)$ is for deep water. In Leykin *et al.* (1995), laboratory wind-generated surface gravity waves with $\varepsilon \approx 0.15$ and $kh = 2.5$ had a quasi-linear relationship between the biphase $\beta$ at the peak frequency (the statistical analogue of our harmonic phase $\beta$) and the inverse wave age $u_*/c_0$ (figure 8). For comparison, our pressure magnitude $Pk/(\rho_w g)$ must be converted to an inverse wave age $u_*/c_0$ (§ 2.4). We assume the deep-water relationship between $u_*/c_0$ and $\gamma/f_0$ (figure 2) holds for $kh = 2.5$, which is parameterized (Banner & Song 2002) as (figure 2, solid line)

$$\frac{\gamma}{f_0} = 32.5(2\pi)\frac{\rho_a}{\rho_w}\left(\frac{u_*}{c_0}\right)^2. \tag{5.5}$$

Using (2.19), we can relate $\gamma/f_0$ to $Pk/(\rho_w g)$ for deep water to give

$$\frac{Pk}{\rho_w g} = \frac{32.5}{\sin(\psi_P)}\frac{\rho_a}{\rho_w}\left(\frac{u_*}{c_0}\right)^2 \tag{5.6}$$

allowing comparison between theory and laboratory observations.

Using (5.6), the measured inverse wave ages $u_*/c_0 = 0.5$ to $1.5$ correspond to pressure magnitudes $Pk/(\rho_w g) = 0.01$ to $0.1$, or $Pk/(\rho_w g) = O(\varepsilon^2)$ to $O(\varepsilon)$. Therefore, our results for weak forcing $Pk/(\rho_w g) \ll 1$ are applicable here (cf. appendix A.6). Assuming a generalized Miles pressure profile with $\psi_P = 3\pi/4$, the predicted and measured $\beta$ are in qualitative agreement (compare red curve to symbols in figure 8). We emphasize that (5.6), relying on the conversion between $u_*/c_0$ and $\gamma/f_0$ from figure 2, is only approximate and is of questionable applicability for water depth $kh = 2.5$. If the conversion coefficient were a factor of 3 larger, the results would match reasonably well. We also note that the relatively high wind speeds ($u_*$ up to $1.7\,\mathrm{m\,s^{-1}}$) likely caused additional physical processes, such as whitecapping or microbreaking, to occur. Such dissipative processes are not considered in our theoretical treatment.

### 5.4. *The surface pressure profile*

Most theoretical treatments of wind-induced wave growth utilize a linear theory with monochromatic waves (e.g. Miles 1957; Belcher & Hunt 1993; Young & Wolfe 2014). In this scenario, for the same $\psi_P$, the pressure profiles considered are identical at leading order and one need not distinguish between, for instance, the Jeffreys or generalized Miles profiles. However, when considering higher-order corrections to the higher harmonics, differences arise and care must be taken when choosing the pressure profile.

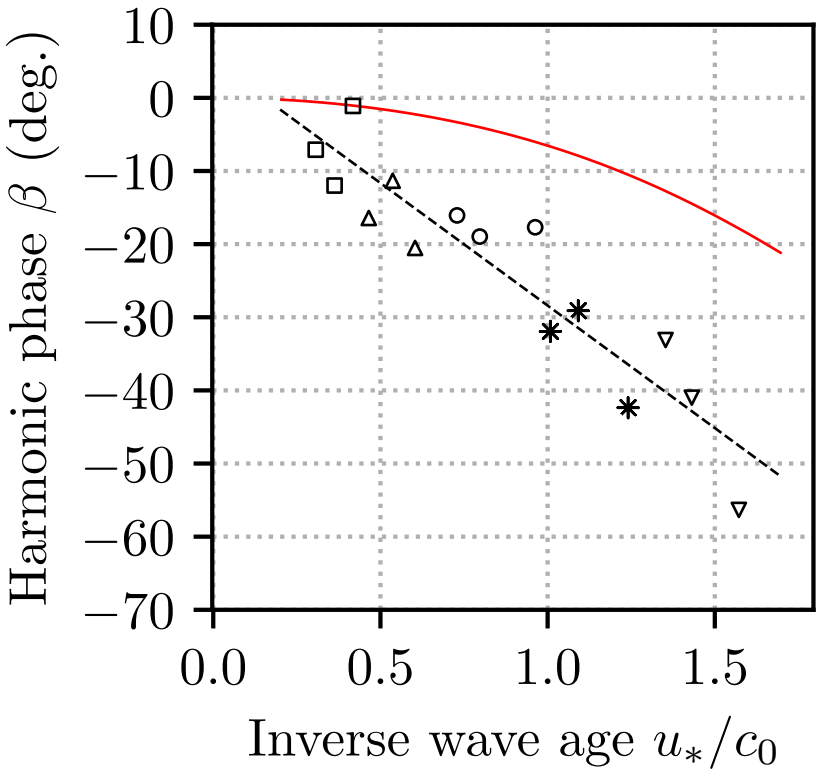

Figure 8. Harmonic phase $\beta$ versus inverse wave age $u_*/c_0$ (symbols) for the Leykin *et al.* (1995) laboratory experiments. The black, dashed line is the Leykin *et al.* (1995) linear fit. Theoretical HP $\beta$ (solid red) are given for the generalized Miles pressure profile with $\psi_P = 3\pi/4$, $kh = 2.5$ and $\varepsilon = 0.15$, and conversion of $u_*/c_0$ to $Pk/(\rho_w g)$ is given by (5.6) (cf. § 5.3).

Direct measurements of the surface pressure profile are challenging and rare (Donelan *et al.* 2006). However, our theory can offer insight by comparing the profiles' differing effects on wave-shape parameters to simulations and measurements of wind-forced waves, which have found a non-zero $\beta$ (Leykin *et al.* 1995; Feddersen & Veron 2005). Both Feddersen & Veron (2005) and Leykin *et al.* (1995) measure a harmonic phase $\beta < 0$ for co-aligned wind and waves. However, the Jeffreys profile gives a positive $\beta$ while the generalized Miles profile with $\psi_P \approx 3\pi/4$ gives a negative $\beta$ (figures 3(*a*) and 4(*a*)). Additionally, the Jeffreys requirement of $\psi_P = \pm\pi/2$ appears inconsistent with numerical simulations showing $\psi_P \approx 3\pi/4$ (Hara & Sullivan 2015; Husain *et al.* 2019). Among the profiles considered here, the generalized Miles case best reproduces the results of wave-shape experiments.

Throughout the derivation, we have maintained a rather general surface pressure profile $p(x,t)$, namely any time-independent convolution with $\eta$ (i.e. $\hat{p}_m \propto \hat{\eta}_m$, cf. § 2.3). Coupled air–water simulations (e.g. Liu *et al.* 2010; Hao & Shen 2019) offer the possibility of extracting realistic wave shapes and surface pressures, which could then be compared to our theory. However, LES atmospheric simulations over purely sinusoidal waves yield surface pressure profiles that are not purely sinusoidal (e.g. Hara & Sullivan 2015, figure 7 or Husain *et al.* 2019, figure 6). Although this is counter to our assumption that $\hat{p}_m \propto \hat{\eta}_m$, it could be remedied by extending our small $\varepsilon$ theory to allow pressures with Fourier representations $\hat{p}_m = k\hat{P}_m\hat{\eta}_m + k^2\sum_n \hat{P}_{m,n}\hat{\eta}_n\hat{\eta}_m + \ldots$. Additional surface pressure complexity is likely generated if LES atmospheric simulations used a Stokes wave profile instead of a single sinusoid. Finally, allowing the wind, via surface pressure profiles, to affect wave shape, as we have done, likely induces further changes back to the airflow and surface pressure profile. That is, the air and water phases are coupled. Although this study relied on prescribed surface pressures, it lays the groundwork for a weakly nonlinear coupled theory. Future work will attempt to couple the wind and waves directly, providing insight into the surface pressure profile and the related wave shape and growth.

## 6. Summary

Here, we derive a theory for the wind's effect on the shape of surface gravity waves. The influence of the wind on ocean waves has been studied in great detail theoretically, numerically and observationally in the context of wave growth. A few laboratory and numerical experiments have shown that wind can also influence wave shape, although no theory for this effect exists. Two key, weakly nonlinear wave-shape parameters are the harmonic phase $\beta$, encoding the relative phase between the primary wave and first harmonic (zero for unforced Stokes waves), and the relative harmonic amplitude $a_2/(a_1^2 k)$. These two parameters can also be converted to the more conventional skewness and asymmetry. Motivated by prior wind–wave generation theories, two surface pressure profiles (Jeffreys and generalized Miles) based on convolutions with the wave profile $\eta$ are prescribed. A multiple-scale perturbation analysis is performed for the small wave steepness $\varepsilon := a_1 k$. The deep- to intermediate-water theoretical solutions are derived for quasi-periodic progressive waves yielding the wind-induced changes to $\beta$ and $a_2/(a_1^2 k)$ as well as higher-order corrections to the previously known growth and phase speed changes. These parameters are functions of the four non-dimensional parameters: the wave steepness $a_1 k$, depth $kh$, pressure magnitude $Pk/(\rho_w g)$ and wind phase $\psi_P$. By substituting the pressure magnitude $P$ with $P \to \varepsilon P$ or $P \to \varepsilon^2 P$, our derivation permits a variety of pressure magnitudes (i.e. wind speeds).

The relative harmonic ratio $a_2/(a_1^2 k)$ displays a strong dependence on the forcing type, enhanced for Jeffreys but suppressed for generalized Miles. The harmonic phase $\beta$ has more complicated behaviour, including a local minimum for the generalized Miles case as a function of the pressure magnitude. Despite restricting our analysis to intermediate and deep water, we find decreasing $kh$ enhances the wind's effect on wave shape. This suggests pressure forcing could play a larger role in wave shape for shallow-water waves. We also find direct relationships between growth rates and $\beta$ for the pressure profiles considered. Atmospheric large eddy simulations constrain both the pressure magnitude $P$ and wind phase $\psi_P$. Using the constrained $\psi_P$, our HP predictions are qualitatively consistent with laboratory observations. Only the generalized Miles profile could reproduce the observed sign for $\beta$, suggesting that generalized Miles surface pressure profiles best represent the actual wave surface pressure profile. Future studies will investigate the shallow-water limit. Other avenues for future work include dynamically coupling the air and wave field. Such an approach would obviate the need to impose a specified pressure profile, increasing the applicability of the theory.

We are grateful to W. R. Young, N. Pizzo and A. B. Villas Bôas for discussions on this work. The computations in this paper were performed by using MAPLE™ (a division of Waterloo Maple Inc. 2018). We thank the National Science Foundation (OCE-1558695) and the Mark Walk Wolfinger Surfzone Processes Research Fund for their support of this work. Declaration of interests. The authors report no conflict of interest.

## Appendix A. Strong forcing: $Pk/(\rho_w g) = O(1)$ continued

### A.1. *The* $O(\varepsilon^3)$ *equations*

In § 3, we derived the leading-order contributions to the HP $\beta$ and relative amplitude $a_2/(a_1^2 k)$. Now, we will extend this derivation to the next non-zero correction. This will reveal a weak amplitude and time dependence to these shape parameters. Furthermore, by finding $\beta$ and $a_2/(a_1^2 k)$ accurate to $O(\varepsilon^2)$, we can substitute $P \to \varepsilon P$ yielding solutions with $Pk/(\rho_w g) = O(\varepsilon)$, or $P \to \varepsilon^2$ generating $Pk/(\rho_w g) = O(\varepsilon^2)$ results (appendix A.6). However, the expressions begin to become unwieldy. Therefore, we will only sketch the derivation. The third-order equations give

$$\frac{\partial \phi_3}{\partial z} - \frac{\partial \eta_3}{\partial t_0} = \frac{\partial A_1}{\partial t_2} \exp(\mathrm{i}(x - \omega_0 t_0)) + A_1 |A_1|^2 \, \mathrm{KIN}_{3,1} \exp(\mathrm{i}(x - \omega_0 t_0)) |\exp(-\mathrm{i}\omega_0 t_0)|^2 + A_1^3 \, \mathrm{KIN}_{3,3} \exp(3\mathrm{i}(x - \omega_0 t_0)), \tag{A 1}$$

$$\frac{\partial \phi_3}{\partial t_0} + \eta_3 + p_3 = \mathrm{i}\omega_0 \frac{\partial A_1}{\partial t_2} \exp(\mathrm{i}(x - \omega_0 t_0)) \coth(h) + A_1 |A_1|^2 \, \mathrm{DYN}_{3,1} \exp(\mathrm{i}(x - \omega_0 t_0)) |\exp(-\mathrm{i}\omega_0 t_0)|^2 + A_1^3 \, \mathrm{DYN}_{3,3} \exp(3\mathrm{i}(x - \omega_0 t_0)), \tag{A 2}$$

with the real part implied. Here, $\mathrm{KIN}_{3,1}, \mathrm{KIN}_{3,3}, \mathrm{DYN}_{3,1}, \mathrm{DYN}_{3,3} \in \mathbb{C}$ are constants that do not depend on $A_1$, $x$, $t_n$ or $z$ (these dependencies have been explicitly factored out) and are composed entirely of known quantities from previous orders. In general, $\mathrm{KIN}_{n,m}$ and $\mathrm{DYN}_{n,m}$ are the constants (depending on $h$, $\psi_P$ and $\hat{P}_m$ only) for the $n$th order, $m$th Fourier component (i.e., $\exp(\mathrm{i}mkx)$) term from the kinematic or dynamic boundary condition, respectively. See the other supplementary material for their expressions.

Once again, inserting our Fourier transforms (3.6)–(3.8), we find

$m = 1$ Fourier component:

$$\hat{\phi}_{3,1} - \frac{\partial \hat{\eta}_{3,1}}{\partial t_0} = \frac{\partial A_1}{\partial t_2} \exp(-\mathrm{i}\omega_0 t_0) + A_1 |A_1|^2 \, \mathrm{KIN}_{3,1} \exp(-\mathrm{i}\omega_0 t_0) |\exp(-\mathrm{i}\omega_0 t_0)|^2, \tag{A 3}$$

$$\coth(h) \frac{\partial \hat{\phi}_{3,1}}{\partial t_0} + (1 + \hat{P}_1) \hat{\eta}_{3,1} = \mathrm{i}\omega_0 \frac{\partial A_1}{\partial t_2} \exp(-\mathrm{i}\omega_0 t_0) \coth(h) + A_1 |A_1|^2 \, \mathrm{DYN}_{3,1} \exp(-\mathrm{i}\omega_0 t_0) |\exp(-\mathrm{i}\omega_0 t_0)|^2, \tag{A 4}$$

$m = 3$ Fourier component:

$$3\hat{\phi}_{3,3} - \frac{\partial \hat{\eta}_{3,3}}{\partial t_0} = A_1^3 \, \mathrm{KIN}_{3,3} \exp(-3\mathrm{i}\omega_0 t_0), \tag{A 5}$$

$$\coth(3h) \frac{\partial \hat{\phi}_{3,3}}{\partial t_0} + (1 + \hat{P}_3) \hat{\eta}_{3,3} = A_1^3 \, \mathrm{DYN}_{3,3} \exp(-3\mathrm{i}\omega_0 t_0). \tag{A 6}$$

Eliminating $\hat{\phi}_{3,m}$ gives

$m = 1$ Fourier component:

$$\coth(h) \frac{\partial^2 \hat{\eta}_{3,1}}{\partial t_0^2} + (1 + \hat{P}_1) \hat{\eta}_{3,1} = -\left(-\mathrm{i}\omega_0 + \frac{\partial}{\partial t_0}\right) \frac{\partial A_1}{\partial t_2} \exp(-\mathrm{i}\omega_0 t_0) \coth(h) + A_1 |A_1|^2 [(\mathrm{i}\omega_0 - 2\,\mathrm{Im}\{\omega_0\}) \coth(h) \, \mathrm{KIN}_{3,1} + \mathrm{DYN}_{3,1}] \exp(-\mathrm{i}\omega_0 t_0) |\exp(-\mathrm{i}\omega_0 t_0)|^2, \tag{A 7}$$

$m = 3$ Fourier component:

$$\coth(3h) \frac{\partial^2 \hat{\eta}_{3,3}}{\partial t_0^2} + 3(1 + \hat{P}_3) \hat{\eta}_{3,3} = 3A_1^3 [\mathrm{i}\omega_0 \coth(3h) \, \mathrm{KIN}_{3,3} + \mathrm{DYN}_{3,3}] \exp(-3\mathrm{i}\omega_0 t_0). \tag{A 8}$$

Notice that we did not evaluate the $\partial/\partial t_0$ derivative in the $(\partial/\partial t_0 - \mathrm{i}\omega_0)$ of (A 7). We will discuss this momentarily.

Preventing secular terms requires that coefficients of $\exp(-\mathrm{i}\omega_0 t_0)$ for $m = 1$ vanish. Thus, we require

$$\coth(h)\left(-\mathrm{i}\omega_0 + \frac{\partial}{\partial t_0}\right)\frac{\partial A_1}{\partial t_2}\exp(-\mathrm{i}\omega_0 t_0)$$
$$= A_1|A_1|^2\exp(-\mathrm{i}\omega_0 t_0)\exp(2\,\mathrm{Im}\{\omega_0\}t_0)[(\mathrm{i}\omega_0 - 2\,\mathrm{Im}\{\omega_0\})\coth(h)\,\mathrm{KIN}_{3,1} + \mathrm{DYN}_{3,1}]. \quad \text{(A 9)}$$

Here, we encounter an issue: given that $A_1(t_2, t_3, \ldots)$ is explicitly not a function of $t_0$, there is no (non-trivial) way to satisfy the $t_0$ dependence of this compatibility condition.

We encounter this issue because the growth on the fast time scale affects the period of the slower time scales. This could be dealt with formally if we had allowed the fast time scale $t_0$ to modulate the slower time scales by defining our multiple-scale expansion with additional, fast time scale dependencies

$$\frac{\mathrm{d}t_0'}{\mathrm{d}t} = 1, .\frac{\mathrm{d}t_1'}{\mathrm{d}t} = \varepsilon\mu_1(t_0'), .\frac{\mathrm{d}t_2'}{\mathrm{d}t} = \varepsilon^2\mu_2(t_0'), \ldots, .\frac{\mathrm{d}t_n'}{\mathrm{d}t} = \varepsilon^n\mu_n(t_0'), \quad \text{(A 10)}$$

with the primes to make our new time scales distinct from the originally defined ones. Then, we can choose the form of $\mu_n$ to remove secular terms. This modified multiple-scale approach is similar to the one specified in Pedersen (2006).

Using this freedom to remove these problematic secularities, we would find that

$$\mu_n(t_0') = \exp(n\,\mathrm{Im}\{\omega_0\}t_0'). \quad \text{(A 11)}$$

This method would eliminate the need to be careful about the $(\partial/\partial t_0' - \mathrm{i}\omega_0)\ \partial A_1/\partial t_2'$ term previously mentioned, and would eliminate the $\exp(2\,\mathrm{Im}\{\omega_0\}t_0')$ term we are attempting to deal with currently. Later, to re-express the solution in terms of $t$, a simple integration yields

$$t_n' = \frac{\varepsilon^n}{n\,\mathrm{Im}\{\omega_0\}}(\exp(n\,\mathrm{Im}\{\omega_0\}t) - 1). \quad \text{(A 12)}$$

where we required that $t_n' = 0$ at $t = 0$. Note that $t_0'$ is not a special case; treating $n$ as a continuous variable and taking the limit $n \to 0$ recovers $t_0' = t$.

Note that, since our previous solutions had no $t_1$ dependence, making this change to $t_2$ does not alter any of our previous conclusions. Furthermore, we will see that only the even time scales ($t_2$, $t_4$, etc) need this treatment. Since we are only considering time scales up to $t_3$, we will only make this replacement for $t_2$.

Making this redefinition, our compatibility conditions becomes

$$\coth(h)\left(-\mathrm{i}\omega_0 + \frac{\partial}{\partial t_0}\right)\frac{\partial A_1}{\partial t_2'}\exp(-\mathrm{i}\omega_0 t_0)\exp(2\,\mathrm{Im}\{\omega_0\}t_0)$$
$$= A_1|A_1|^2\exp(-\mathrm{i}\omega_0 t_0)\exp(2\,\mathrm{Im}\{\omega_0\}t_0)[(\mathrm{i}\omega_0 - 2\,\mathrm{Im}\{\omega_0\})\coth(h)\,\mathrm{KIN}_{3,1} + \mathrm{DYN}_{3,1}], \quad \text{(A 13)}$$

which simplifies to

$$\begin{aligned}\frac{\partial A_1}{\partial t_2'} &= A_1|A_1|^2\frac{(\mathrm{i}\omega_0 - 2\,\mathrm{Im}\{\omega_0\})\,\mathrm{KIN}_{3,1} + \tanh(h)\,\mathrm{DYN}_{3,1}}{-2\mathrm{i}\omega_0 + 2\,\mathrm{Im}\{\omega_0\}} \\ &:= -\mathrm{i}A_1|A_1|^2\,\mathrm{COMB}_{3,1},\end{aligned} \quad \text{(A 14)}$$

where we defined

$$\mathrm{COMB}_{3,1} := \mathrm{i}\frac{(\mathrm{i}\omega_0 - 2\,\mathrm{Im}\{\omega_0\})\,\mathrm{KIN}_{3,1} + \tanh(h)\,\mathrm{DYN}_{3,1}}{-2\mathrm{i}\omega_0 + 2\,\mathrm{Im}\{\omega_0\}}. \quad \text{(A 15)}$$

Now, if we assume a solution of the form

$$A_1(t_2') = \rho(t_2')\exp(\mathrm{i}\psi(t_2')), \tag{A 16}$$

with $\rho(t_2'), \psi(t_2') \in \mathbb{R}$, (A 15) yields

$$\frac{\partial\rho}{\partial t_2'} + \mathrm{i}\rho\frac{\partial\psi}{\partial t_2'} = -\mathrm{i}\rho^3\,\mathrm{COMB}_{3,1}\,. \tag{A 17}$$

Collecting real and imaginary parts and solving yields

$$\begin{aligned} A_1(t_2') = A_1' \exp\left[\mathrm{i}\frac{1}{2}\ln\left(1 - 2|A_1'|^2 t_2' \,\mathrm{Im}\{\mathrm{COMB}_{3,1}\}\right)\frac{\mathrm{Re}\{\mathrm{COMB}_{3,1}\}}{\mathrm{Im}\{\mathrm{COMB}_{3,1}\}}\right] \\ \div\sqrt{1 - 2|A_1'|^2 t_2'\,\mathrm{Im}\{\mathrm{COMB}_{3,1}\}}, \end{aligned} \tag{A 18}$$

with $A_1'(t_3) \in \mathbb{C}$. Later, converting back to $t$ will give

$$\begin{aligned} A_1(t) = A_1' \exp\left\{\frac{\mathrm{i}}{2}\ln\left[1 - \varepsilon^2|A_1'|^2(\exp(2\,\mathrm{Im}\{\omega_0\}t) - 1)\frac{\mathrm{Im}\{\mathrm{COMB}_{3,1}\}}{\mathrm{Im}\{\omega_0\}}\right]\frac{\mathrm{Re}\{\mathrm{COMB}_{3,1}\}}{\mathrm{Im}\{\mathrm{COMB}_{3,1}\}}\right\} \\ \div\sqrt{1 - \varepsilon^2|A_1'|^2\left(\exp(2\,\mathrm{Im}\{\omega_0\}t) - 1\right)\frac{\mathrm{Im}\{\mathrm{COMB}_{3,1}\}}{\mathrm{Im}\{\omega_0\}}}. \end{aligned} \tag{A 19}$$

Note that if $p \to 0$, then $\mathrm{COMB}_{3,1}$ reduces to the real quantity

$$\mathrm{COMB}_{3,1}\Bigg|_{p=0} = \omega_0\frac{8\cosh^4(h) - 8\cosh^2(8) + 9}{16\sinh^4(h)}. \tag{A 20}$$

With the compatibility condition solved, the $m = 1$ equation reduces to the homogeneous equation. For simplicity, we will choose $\hat{\eta}_{3,1} = 0$.

Substituting (A 14) and our solution for $\hat{\eta}_{3,1}$ into the surface boundary conditions allows us to solve for $\hat{\phi}_{3,1}$. Assuming a solution of the form

$$\hat{\phi}_{3,1} = C_{3,1}A_1|A_1|^2\exp(-\mathrm{i}\omega_0 t_0)\exp(2\,\mathrm{Im}\{\omega_0\}t_0), \tag{A 21}$$

yields

$$C_{3,1} = \frac{-\mathrm{i}\omega_0\,\mathrm{KIN}_{3,1} + \tanh(h)\,\mathrm{DYN}_{3,1}}{-2\mathrm{i}\omega_0 + 2\,\mathrm{Im}\{\omega_0\}}. \tag{A 22}$$

The second harmonic ($m = 3$) equation is solved for $\hat{\eta}_{3,3}$ as usual. Then, substituting this solution into the surface boundary conditions permits solving for $\hat{\phi}_{3,3}$.

Thus, we have the solutions

$$\begin{aligned} \phi_3 = C_{3,1}A_1|A_1|^2\exp(2\,\mathrm{Im}\{\omega_0\}t_0)\exp(\mathrm{i}(x - \omega_0 t_0))\frac{\cosh(z+h)}{\sinh(h)} \\ + C_{3,3}'A_1^3\exp(3\mathrm{i}(x - \omega_0 t_0))\frac{\cosh[3(z+h)]}{\sinh(3h)}, \end{aligned} \tag{A 23}$$

$$\eta_3 = C_{3,3}A_1^3\exp(3\mathrm{i}(x - \omega_0 t_0)), \tag{A 24}$$

with

$$C_{3,1} = \frac{-\mathrm{i}\omega_0 \,\mathrm{KIN}_{3,1} + \tanh(h)\,\mathrm{DYN}_{3,1}}{-2\mathrm{i}\omega_0 + 2\,\mathrm{Im}\{\omega_0\}}, \tag{A 25}$$

$$C'_{3,3} = \frac{\left(1+\hat{P}_3\right)\mathrm{KIN}_{3,3} - 3\mathrm{i}\omega_0\,\mathrm{DYN}_{3,3}}{-9\omega_0^2\coth(3h) + 3\left(1+\hat{P}_3\right)}, \tag{A 26}$$

$$C_{3,3} = 3\frac{\mathrm{i}\omega_0\coth(3h)\,\mathrm{KIN}_{3,3} + \mathrm{DYN}_{3,3}}{-9\omega_0^2\coth(3h) + 3\left(1+\hat{P}_3\right)}. \tag{A 27}$$

With no correction to the first harmonic $\hat{\eta}_{m=2}$, we continue to the next order.

## A.2. *The* $O(\varepsilon^4)$ *equations*

Finally, going to fourth order, we have

$$\begin{aligned}\frac{\partial\phi_4}{\partial z} - \frac{\partial\eta_4}{\partial t_0} = {} & \frac{\partial A_1}{\partial t_3}\exp(\mathrm{i}(x-\omega_0 t_0)) \\ & + \mathrm{KIN}_{4,0}\,|A_1\exp(-\mathrm{i}\omega_0 t_0)|^4 + \mathrm{KIN}_{4,2}\,A_1^2\exp(2\mathrm{i}(x-\omega_0 t_0))|A_1\exp(-\mathrm{i}\omega_0 t_0)|^2 \\ & + \mathrm{KIN}_{4,4}\,A_1^4\exp(4\mathrm{i}(x-\omega_0 t_0)),\end{aligned} \tag{A 28}$$

$$\begin{aligned}\frac{\partial\phi_4}{\partial t_0} + \eta_4 + p_4 = {} & \mathrm{i}\omega_0\frac{\partial A_1}{\partial t_3}\exp(\mathrm{i}(x-\omega_0 t_0))\coth(h) \\ & + \mathrm{DYN}_{4,0}\,|A_1\exp(-\mathrm{i}\omega_0 t_0)|^4 + \mathrm{DYN}_{4,2}\,A_1^2\exp(2\mathrm{i}(x-\omega_0 t_0))|A_1\exp(-\mathrm{i}\omega_0 t_0)|^2 \\ & + \mathrm{DYN}_{4,4}\,A_1^4\exp(4\mathrm{i}(x-\omega_0 t_0))\end{aligned} \tag{A 29}$$

Here, $\mathrm{KIN}_{4,0}, \mathrm{KIN}_{4,2}, \mathrm{KIN}_{4,4}, \mathrm{DYN}_{4,0}, \mathrm{DYN}_{4,2}, \mathrm{DYN}_{4,4} \in \mathbb{C}$ are constants that do not depend on $A_1$, $x$, $t_n$ or $z$ (these dependencies have been explicitly factored out) and are composed entirely of known quantities from previous orders. See the other supplementary material for their expressions.

Inserting the Fourier transforms (3.6)–(3.8) gives

$m = 2$ Fourier component:

$$2\hat{\phi}_{4,2} - \frac{\partial\hat{\eta}_{4,2}}{\partial t_0} = A_1^2|A_1|^2\,\mathrm{KIN}_{4,2}\exp(2\mathrm{i}(x-\omega_0 t_0))|\exp(-\mathrm{i}\omega_0 t_0)|^2, \tag{A 30}$$

$$\frac{\partial\hat{\phi}_{4,2}}{\partial t_0}\coth(2h) + (1+\hat{P}_2)\hat{\eta}_{4,2} = A_1^2|A_1|^2\,\mathrm{DYN}_{4,2}\exp(2\mathrm{i}(x-\omega_0 t_0))|\exp(-\mathrm{i}\omega_0 t_0)|^2, \tag{A 31}$$

$m = 4$ Fourier component:

$$4\hat{\phi}_{4,4} - \frac{\partial\hat{\eta}_{4,4}}{\partial t_0} = A_1^4\,\mathrm{KIN}_{4,4}\exp(4\mathrm{i}(x-\omega_0 t_0)), \tag{A 32}$$

$$\frac{\partial\hat{\phi}_{4,4}}{\partial t_0}\coth(4h) + (1+\hat{P}_4)\hat{\eta}_{4,4} = A_1^4\,\mathrm{DYN}_{4,4}\exp(4\mathrm{i}(x-\omega_0 t_0)), \tag{A 33}$$

$m = 0$ Fourier component:

$$-\frac{\partial \hat{\eta}_{4,0}}{\partial t_0} = |A_1|^4 \, \mathrm{KIN}_{4,0} \, |\exp(-\mathrm{i}\omega_0 t_0)|^4, \tag{A 34}$$

$$\frac{\partial \hat{\phi}_{4,0}}{\partial t_0} + \hat{\eta}_{4,0} = |A_1|^4 \, \mathrm{DYN}_{4,0} \, |\exp(-\mathrm{i}\omega_0 t_0)|^4, \tag{A 35}$$

$m = 1$ Fourier component:

$$\hat{\phi}_{4,1} - \frac{\partial \hat{\eta}_{4,1}}{\partial t_0} = \frac{\partial A_1}{\partial t_3} \exp(-\mathrm{i}\omega_0 t_0), \tag{A 36}$$

$$\frac{\partial \hat{\phi}_{4,1}}{\partial t_0} \coth(h) + (1 + \hat{P}_1)\hat{\eta}_{4,1} = \mathrm{i}\omega_0 \frac{\partial A_1}{\partial t_3} \exp(-\mathrm{i}\omega_0 t_0) \coth(h). \tag{A 37}$$

Again, eliminating $\hat{\eta}_4$ gives

$m = 2$ Fourier component:

$$\frac{\partial^2 \hat{\phi}_{4,2}}{\partial t_0^2} \coth(2h) + 2(1 + \hat{P}_2)\hat{\phi}_{4,2} = A_1^2 |A_1|^2 \exp(-2\mathrm{i}\omega_0 t_0) \exp(2 \, \mathrm{Im}\{\omega_0\} t_0) \times \Big[(1 + \hat{P}_2) \, \mathrm{KIN}_{4,2} + 2(-\mathrm{i}\omega_0 + \mathrm{Im}\{\omega_0\}) \, \mathrm{DYN}_{4,2}\Big]. \tag{A 38}$$

$m = 4$ Fourier component:

$$\frac{\partial^2 \hat{\phi}_{4,4}}{\partial t_0^2} \coth(4h) + 4(1 + \hat{P}_4)\hat{\phi}_{4,4} = A_1^4 \exp(-4\mathrm{i}\omega_0 t_0) \Big[(1 + \hat{P}_4) \, \mathrm{KIN}_{4,4} - 4\mathrm{i}\omega_0 \, \mathrm{DYN}_{4,4}\Big]. \tag{A 39}$$

$m = 0$ Fourier component:

$$\frac{\partial^2 \hat{\phi}_{4,0}}{\partial t_0^2} = |A_1|^4 \exp(4 \, \mathrm{Im}\{\omega_0\} t_0) [\mathrm{KIN}_{4,0} + 4 \, \mathrm{Im}\{\omega_0\} \, \mathrm{DYN}_{4,0}]. \tag{A 40}$$

$m = 1$ Fourier component:

$$\frac{\partial^2 \hat{\phi}_{4,1}}{\partial t_0^2} \coth(h) + (1 + \hat{P}_1)\hat{\phi}_{4,1} = 2\Big(1 + \hat{P}_1\Big) \frac{\partial A_1}{\partial t_3} \exp(-\mathrm{i}\omega_0 t_0). \tag{A 41}$$

Preventing secular terms requires that $\partial_{t_3} A_1 = 0$. These can be solved as usual for $\hat{\phi}_{4,m}$. Using the surface boundary conditions, the solutions for $\hat{\eta}_{4,m}$ can then be determined as well.

The only terms worth discussing are the zero modes, $\hat{\phi}_{4,0}$ and $\hat{\eta}_{4,0}$. While $\hat{\eta}_{4,0}$ has physical meaning (this is a component of the setup or setdown), $\hat{\phi}_{4,0}$ has a gauge freedom. We may add a constant term (in $x$, $z$ and $t_0$), as well as a term proportional to $t_0$, without affecting any observables. Using this freedom, we will choose these two free constants such that the $\hat{\eta}_{4,0} \to 0$ and $\hat{\phi}_{4,0} \to 0$ as $P \to 0$.

The solutions at this order are

$$\phi_4 = C'_{4,2}A_1^2|A_1|^2 \exp(2\mathrm{i}(x-\omega_0 t_0))\exp(2\,\mathrm{Im}\{\omega_0\}t_0)\frac{\cosh[2(z+h)]}{\sinh(2h)} + C'_{4,4}A_1^4\exp(4\mathrm{i}(x-\omega_0 t_0))\frac{\cosh[4(z+h)]}{\sinh(4h)} + C'_{4,0}\left(|A_1|^4\exp(4\,\mathrm{Im}\{\omega_0\}t_0) - \left|\tilde{A}_1\right|^4\right) + t_0 C_{4,0}\left|\tilde{A}_1\right|^4, \tag{A 42}$$

$$\eta_4 = C_{4,2}A_1^2|A_1|^2\exp(2\mathrm{i}(x-\omega_0 t_0))\exp(2\,\mathrm{Im}\{\omega_0\}t_0) + C_{4,4}A_1^4\exp(4\mathrm{i}(x-\omega_0 t_0)) + C_{4,0}\left(|A_1|^4\exp(4\,\mathrm{Im}\{\omega_0\}t_0) - \left|\tilde{A}_1\right|^4\right), \tag{A 43}$$

with

$$C_{4,0} = -\frac{\mathrm{KIN}_{4,0}}{4\,\mathrm{Im}\{\omega_0\}} = 0, \tag{A 44}$$

$$C_{4,2} = \frac{(\mathrm{i}\omega_0 - \mathrm{Im}\{\omega_0\})\coth(2h)\,\mathrm{KIN}_{4,2} + \mathrm{DYN}_{4,2}}{2(-\mathrm{i}\omega_0 + \mathrm{Im}\{\omega_0\})^2\coth(2h) + \left(1+\hat{P}_2\right)}, \tag{A 45}$$

$$C_{4,4} = \frac{\mathrm{i}\omega_0\coth(4h)\,\mathrm{KIN}_{4,4} + \mathrm{DYN}_{4,4}}{-4\omega_0^2\coth(4h) + \left(1+\hat{P}_4\right)}, \tag{A 46}$$

$$C'_{4,0} = \frac{\mathrm{KIN}_{4,0} + 4\,\mathrm{Im}\{\omega_0\}\,\mathrm{DYN}_{4,0}}{16\,\mathrm{Im}\{\omega_0\}^2}, \tag{A 47}$$

$$C'_{4,2} = \frac{\left(1+\hat{P}_2\right)\mathrm{KIN}_{4,2} + 2(-\mathrm{i}\omega_0 + \mathrm{Im}\{\omega_0\})\,\mathrm{DYN}_{4,2}}{4(-\mathrm{i}\omega_0 + \mathrm{Im}\{\omega_0\})^2\coth(2h) + 2\left(1+\hat{P}_2\right)}, \tag{A 48}$$

$$C'_{4,4} = \frac{\left(1+\hat{P}_4\right)\mathrm{KIN}_{4,4} - 4\mathrm{i}\omega_0\,\mathrm{DYN}_{4,4}}{-16\omega_0^2\coth(4h) + 4\left(1+\hat{P}_4\right)}. \tag{A 49}$$

Here, $\tilde{A}_1 := A_1\big|_{P=0}$ is the additive 'constant' we were permitted from the $m=0$ equation; note that $\tilde{A}_1$ could still be a function of slower time scales $t_1$, $t'_2$, etc. As mentioned previously, a term, linear in $t_0$, was included in $\hat{\phi}_{4,0}$. This was necessary in order to include the $\tilde{A}_1$ term in $\hat{\eta}_{4,0}$, ensuring that the $C_{4,0}$ setdown term vanishes as $t \to 0$, as required by our choice of $z=0$ datum at the initial mean water level. In addition, note that $\mathrm{KIN}_{4,0} = 0$ (cf. the supplementary material) implies the setdown term $C_{4,0}$ is identically zero for all times. For reference, the full solution for $\eta$ is

$$\eta = \mathrm{Re}\Big\{\varepsilon A_1\exp(\mathrm{i}(x-\omega_0 t_0)) + \varepsilon^2 A_1^2 C_{2,2}\exp(2\mathrm{i}(x-\omega_0 t_0)) + \varepsilon^3 A_1^3 C_{3,3}\exp(3\mathrm{i}(x-\omega_0 t_0)) + \varepsilon^4\Big(A_1^4 C_{4,4}\exp(4\mathrm{i}(x-\omega_0 t_0)) + A_1^2|A_1|^2\exp(2\,\mathrm{Im}\{\omega_0\}t_0)\exp(2\mathrm{i}(x-\omega_0 t_0))\Big)\Big\} + O(\varepsilon^5), \tag{A 50}$$

with $A_1(t_2)$ given by (A 18). At this order, we have a correction to the first harmonic $\hat{\eta}_{m=2}$, which will modify our shape parameters.

### A.3. *Shape parameters*

Now, we can calculate the shape parameters when pressure enters at leading order. Recall that we are seeking two parameters—the HP $\beta$, and the relative harmonic amplitude, $a_2/(a_1^2 k)$ (with $a_2$ the amplitude of the complete first harmonic, and $a_1$ the amplitude of the complete primary wave).

The primary wave is simply

$$\eta_{m=1} = \varepsilon A_1 \exp(\mathrm{i}(x - \omega_0 t_0)) + O\big(\varepsilon^5\big), \tag{A 51}$$

with $A_1(t_2')$ given by (A 18).

The first harmonic has two components. We calculated the $O\big(\varepsilon^2\big)$ contribution in (3.47), and the $O\big(\varepsilon^4\big)$ contribution in (A 43). Combining these, we have the first harmonic

$$\begin{aligned}\eta_{m=2} = \varepsilon^2 A_1^2 \exp(2\mathrm{i}(x - \omega_0 t_0)) C_{2,2} + \varepsilon^4 A_1^2 |A_1|^2 \exp(2\mathrm{i}(x - \omega_0 t_0)) \exp(2\,\mathrm{Im}\{\omega_0\} t_0) C_{4,2} \\ + O\big(\varepsilon^5\big)\end{aligned} \tag{A 52}$$

with $C_{2,2}$ defined in (3.49) as

$$C_{2,2} := \frac{1}{4}\big(2 + 3\,\mathrm{csch}^2(h)\big) \coth(h) \frac{1 + \hat{P}_1}{1 + \hat{P}_1 - \coth^2(h)\Big[\hat{P}_2 - \hat{P}_1\Big]}, \tag{A 53}$$

and $C_{4,2}$ defined in (A 45) as

$$C_{4,2} = \frac{(\mathrm{i}\omega_0 - \mathrm{Im}\{\omega_0\}) \coth(2h)\,\mathrm{KIN}_{4,2} + \mathrm{DYN}_{4,2}}{2(-\mathrm{i}\omega_0 + \mathrm{Im}\{\omega_0\})^2 \coth(2h) + \Big(1 + \hat{P}_2\Big)}. \tag{A 54}$$

See the other supplementary material for the full expression. Note that if $p \to 0$, then $C_{4,2}$ reduces to the real quantity

$$\begin{aligned}C_{4,2}\bigg|_{p=0} := \mathcal{A} = \frac{\tanh(h)}{384}\big(272 + 856\,\mathrm{csch}^2(h) + 512\,\mathrm{csch}^4(h) \\ - 558\,\mathrm{csch}^6(h) - 567\,\mathrm{csch}^8(h) - 81\,\mathrm{csch}^{10}(h)\big).\end{aligned} \tag{A 55}$$

To find the relative harmonic amplitude and HP, we will need to calculate the ratio of the first harmonic, $\hat{\eta}_{m=2}$, to the primary wave, $\hat{\eta}_{m=1}$, squared (cf. (3.52) and (3.54)):

$$\frac{\hat{\eta}_{m=2}}{\hat{\eta}_{m=1}^2} = C_{2,2} + \varepsilon^2 |A_1|^2 \exp(2\,\mathrm{Im}\{\omega_0\} t_0) C_{4,2} + O\big(\varepsilon^3\big). \tag{A 56}$$

Now, the relative harmonic amplitude (3.52), $a_2/(a_1^2 k)$, is the magnitude of this quantity,

$$\begin{aligned}\frac{a_2}{a_1^2 k} &= \Big|C_{2,2} + \varepsilon^2 |A_1|^2 \exp(2\,\mathrm{Im}\{\omega_0\} t_0) C_{4,2}\Big| + O\big(\varepsilon^3\big) \\ &= |C_{2,2}| \left(1 + \varepsilon^2 |A_1|^2 \exp(2\,\mathrm{Im}\{\omega_0\} t_0) \frac{\mathrm{Re}\big\{C_{4,2} C_{2,2}^*\big\}}{|C_{2,2}|^2}\right) + O\big(\varepsilon^3\big),\end{aligned} \tag{A 57}$$

with an asterisk representing the complex conjugate. We can see that the $O\big(\varepsilon^2\big)$ correction grows as a function of the fast time scale, $t_0$, as well as the slow time scale, $t_2'$ (through its $A_1(t_2)$ dependence).

Likewise, the HP $\beta$ is the complex angle (3.54) of (A 56)

$$\beta := \tan^{-1}\left(\frac{\operatorname{Im}\left\{C_{2,2} + \varepsilon^2|A_1|^2\exp(2\operatorname{Im}\{\omega_0\}t_0)C_{4,2}\right\}}{\operatorname{Re}\left\{C_{2,2} + \varepsilon^2|A_1|^2\exp(2\operatorname{Im}\{\omega_0\}t_0)C_{4,2}\right\}}\right) + O\big(\varepsilon^3\big)$$
$$\approx \beta_0 + \varepsilon^2|A_1|^2\exp(2\operatorname{Im}\{\omega_0\}t_0)\frac{\operatorname{Re}\{C_{2,2}\}\operatorname{Im}\{C_{4,2}\} - \operatorname{Im}\{C_{2,2}\}\operatorname{Re}\{C_{4,2}\}}{\operatorname{Re}\{C_{2,2}\}^2 + \operatorname{Im}\{C_{2,2}\}^2} + O\big(\varepsilon^3\big), \tag{A 58}$$

with $\beta_0$ given in (3.55) by

$$\beta_0 = \tan^{-1}\left(\frac{\operatorname{Im}\left\{\left[\hat{P}_2 - \hat{P}_1\right]\left(1 + \hat{P}_1^*\right)\right\}}{\left|1 + \hat{P}_1\right|^2\tanh^2(h) - \operatorname{Re}\left\{\left[\hat{P}_2 - \hat{P}_1\right]\left(1 + \hat{P}_1^*\right)\right\}}\right). \tag{A 59}$$

Notice that $\beta$ also has a weak time dependence appearing at $O\big(\varepsilon^2\big)$. Additionally, both $\beta$ and $a_2/(a_1^2k)$ display a weak amplitude, $\varepsilon|A_1|$, dependence. Finally, as given in (3.57) and (3.58), the skewness S and asymmetry A of a wave are defined as

$$\mathrm{S} := \frac{\langle\eta^3\rangle}{\langle\eta^2\rangle^{3/2}}, \tag{A 60}$$

$$\mathrm{A} := \frac{\langle\mathcal{H}\{\eta\}^3\rangle}{\langle\eta^2\rangle^{3/2}}, \tag{A 61}$$

with $\langle\cdot\rangle$ the spatial average over one wavelength and $\mathcal{H}\{\cdot\}$ the Hilbert transform (in $x$). In § 3, we only calculated the $O(\varepsilon)$ contribution for brevity. Using the full solution (A 50) for $\eta$ would yield a solution accurate up to and including $O\big(\varepsilon^3\big)$ terms.

### A.4. *Complex frequency*

After deriving our solutions (A 51) and (A 52), it is useful to repackage them in a more conventional notation. Therefore, we will gather the entire time dependence into a complex phase $\varTheta \in \mathbb{C}$, from which we can extract a complex, time dependent frequency $\omega(t) \in \mathbb{C}$ giving both propagation and growth. From (A 19), we can write the entire $t$-dependence of $A_1(t)$ as a complex phase

$$A_1(t) = A_1'\exp\left\{\mathrm{i}\frac{1}{2}\frac{\mathrm{COMB}_{3,1}}{\operatorname{Im}\{\mathrm{COMB}_{3,1}\}}\ln\left[1 - \varepsilon^2|A_1'|^2(\exp(2\operatorname{Im}\{\omega_0\}t) - 1) \times \frac{\operatorname{Im}\{\mathrm{COMB}_{3,1}\}}{\operatorname{Im}\{\omega_0\}}\right]\right\} + O\big(\varepsilon^4\big). \tag{A 62}$$

Therefore, the entire complex phase $\varTheta$ of the first harmonic $\eta_{m=1} = A_1'\exp(\mathrm{i}\varTheta)$ is

$$\varTheta := kx - \omega_0 t + \mathrm{i}\frac{1}{2}\frac{\mathrm{COMB}_{3,1}}{\operatorname{Im}\{\mathrm{COMB}_{3,1}\}}\ln\left[1 - \varepsilon^2|A_1'|^2(\exp(2\operatorname{Im}\{\omega_0\}t) - 1) \times \frac{\operatorname{Im}\{\mathrm{COMB}_{3,1}\}}{\operatorname{Im}\{\omega_0\}}\right] + O\big(\varepsilon^4\big). \tag{A 63}$$

Now, we define the full, complex frequency as

$$\omega := -\frac{\partial\varTheta}{\partial t} = \omega_0 + \varepsilon^2|A_1'|^2\exp(2\operatorname{Im}\{\omega_0\}t)\,\mathrm{COMB}_{3,1} + O\big(\varepsilon^4\big). \tag{A 64}$$

Notice that the time dependence of $\omega$ is a manifestation of the (time-dependent) amplitude dispersion of unforced Stokes waves. Then, the phase speed is the real part of $\omega$,

$$c := \mathrm{Re}\{\omega\} = \mathrm{Re}\{\omega_0\} + \varepsilon^2 |A_1'|^2 \exp(2\,\mathrm{Im}\{\omega_0\}t)\,\mathrm{Re}\{\mathrm{COMB}_{3,1}\} + O(\varepsilon^4), \tag{A 65}$$

while the growth rate is the imaginary

$$\gamma := \mathrm{Im}\{\omega\} = \mathrm{Im}\{\omega_0\} + \varepsilon^2 |A_1'|^2 \exp(2\,\mathrm{Im}\{\omega_0\}t)\,\mathrm{Im}\{\mathrm{COMB}_{3,1}\} + O(\varepsilon^4). \tag{A 66}$$

It is natural to define the (dimensional) harmonic amplitudes $a_n$ of (A 50) as containing the growth time dependence

$$a_1(t) := |\hat{\eta}_{m=1}| = \varepsilon \frac{|A_1'|}{k} \exp(\mathrm{Im}\{\Theta\}) + O(\varepsilon^5), \tag{A 67}$$

$$\begin{aligned} a_2(t) := |\hat{\eta}_{m=2}| &= \varepsilon^2 \frac{|A_1'|^2}{k} \exp(2\,\mathrm{Im}\{\Theta\}) \left[1 + \varepsilon^2 |A_1|^2 \exp(2\,\mathrm{Im}\{\omega_0\}t)\right] + O(\varepsilon^5) \\ &= \varepsilon^2 \frac{|A_1'|^2}{k} \exp(2\,\mathrm{Im}\{\Theta\}) \left[1 + (a_1 k)^2\right] + O(\varepsilon^5), \end{aligned} \tag{A 68}$$

where we made the approximation $\mathrm{Im}\{\omega_0\}t_0 \approx \mathrm{Im}\{\Theta\}$ in the final line. This leaves the propagation time dependence given by the (real) phase

$$\theta := \mathrm{Re}\{\Theta\} = kx - \int \mathrm{Re}\{\omega\}\,\mathrm{d}t, \tag{A 69}$$

such that the dimensional solution is

$$k\eta = (a_1 k)\exp(\mathrm{i}\theta) + (a_1 k)^2 \frac{a_2}{a_1^2 k} \exp(\mathrm{i}(2\theta + \beta)) + \dots . \tag{A 70}$$

### A.5. *Miles profile*

The Miles surface pressure defined in (2.9) has a Fourier representation, similar to (2.15) and (2.16), given by

$$\hat{p}_{M,m}(t) = kP_M \exp(\mathrm{i}\,\mathrm{sgn}(m)\psi_P)\hat{\eta}_m(t), \tag{A 71}$$

or $\hat{P}_m = P_M \exp(\mathrm{i}\,\mathrm{sgn}(m)\psi_P)$, with $P_M$ the constant $P$ for the Miles profile. For this profile, the leading-order correction to the first harmonic $C_{2,2}$ (3.48) reduces to the unforced Stokes result ($\hat{P}_1 = \hat{P}_2 = 0$). Indeed, the leading-order HP $\beta_0$ (3.55) vanishes for any pressure profile of the form $\hat{P}_2 = \alpha + \hat{P}_1(1+\alpha)$, with $\alpha \in \mathbb{R}$ (Miles is $\alpha = 0$). Thus, the Miles pressure profile has no impact on leading-order wave shape. Note that, for the Miles profile, the higher-order correction $C_{4,2}$ differs from the unforced case, giving a small $O(\varepsilon^2)$ change to the shape parameters. Given that leading-order wind-induced shape changes have been measured (e.g. Leykin *et al.* 1995; Feddersen & Veron 2005), the Miles profile appears to be an inappropriate pressure profile.

### A.6. *Weaker wind forcing*

In appendices A.1 and A.2 we performed the derivation up to $O(\varepsilon^4)$ with a strong pressure forcing $Pk/(\rho_w g) = O(1)$. This yielded expressions (A 57), (A 58) and (A 64) for $a_2/(a_1^2 k)$ $\beta$, and $\omega \in \mathbb{C}$ accurate to $O(\varepsilon^3)$. However, it is occasionally useful to consider weaker winds, such as $Pk/(\rho_w g) = O(\varepsilon)$ or $O(\varepsilon^2)$, as discussed in § 2.4. These results can be generated by substituting $P \to \varepsilon P$ or $P \to \varepsilon^2 P$, respectively, into (A 57), (A 58) and (A 64) and dropping terms $O(\varepsilon^3)$ or higher. We have also performed the derivation assuming *a priori* that $Pk/(\rho_w g) = O(\varepsilon^2)$ (not included here), which gives identical results to $O(\varepsilon^2)$ to the more general solution (§ 3 and appendix A) after converting back to the true time $t$. This further confirms the wide parameter range of the $Pk/(\rho_w g) = O(1)$ derivation (§ 3 and appendix A).

# Other supplemental material

## 1. The $\mathcal{O}\left(\varepsilon^3\right)$ coefficients

Here, we give the full expressions for third-order coefficients, $\mathrm{KIN}_{3,m}$ and $\mathrm{DYN}_{3,m}$, defined in **????** as the coefficients of the lower-order terms' $m$-th harmonics in the kinematic and dynamic boundary conditions, respectively. Recall that $\hat{P}_m$ was defined in **??** as the pressure's Fourier coefficient multiplying the $m$-th harmonic of the wave profile: $\hat{p}_m(t) = k\hat{P}_m\hat{\eta}_m(t)$.

$$\begin{aligned}
&\mathrm{KIN}_{3,1} = \\
&\frac{1}{4}\left(\mathrm{i}\left(\left(\left(\left(\frac{3}{4}+\hat{P}_1-\frac{\hat{P}_2}{4}\right)\cosh^4(h)+\left(-\frac{1}{4}-\frac{\hat{P}_1}{2}+\frac{\hat{P}_2}{4}\right)\cosh^2(h)+\frac{\hat{P}_1}{4}+\frac{1}{4}\right)\left|\hat{P}_1+1\right|\right.\right.\right. \\
&\left.\left.\left.+\frac{1}{2}\left(\left(\hat{P}_2+1\right)\cosh^4(h)+\left(\hat{P}_1+1\right)\cosh^2(h)-\frac{\hat{P}_1}{2}-\frac{1}{2}\right)\left(\hat{P}_1+1\right)\right)\right)\right) \\
&\times\left(\left(\left(\left(\frac{1}{2}+\hat{P}_1-\frac{\hat{P}_2}{2}\right)\cosh^2(h)-\frac{\hat{P}_1}{2}-\frac{1}{2}\right)\omega\sinh(h)\cosh(h)\right)\right)^{-1}
\end{aligned} \tag{1.1}$$

$$\mathrm{KIN}_{3,3} = \frac{3}{8}\,\frac{\mathrm{i}\left(3\cosh^2(h)-1\right)\left(\cosh^2(h)\hat{P}_2+\cosh^2(h)+2\hat{P}_1+2\right)\omega}{\left(2\cosh^2(h)\hat{P}_1-\cosh^2(h)\hat{P}_2+\cosh^2(h)-\hat{P}_1-1\right)\sinh^2(h)} \tag{1.2}$$

$$
\begin{aligned}
&\mathrm{DYN}_{3,1} = \\
&-\Bigg(\Bigg(\Big(\hat{P}_1+1\Big)\cosh^4(h)+\Bigg(-\frac{3}{4}\hat{P}_1+\frac{\hat{P}_2}{4}-\frac{1}{2}\Bigg)\cosh^2(h)-\frac{\hat{P}_1}{8}-\frac{1}{8}\Bigg)\hat{P}_1\Big|\hat{P}_1+1\Big| \\
&+\frac{1}{2}\Bigg(\Bigg(\Bigg(-\hat{P}_1+\frac{\hat{P}_2}{4}-\frac{3}{4}\Bigg)\Big|\hat{P}_1\Big|^2+\Bigg(\hat{P}_1^2+\Bigg(-\frac{3}{2}\hat{P}_2-\frac{1}{2}\Bigg)\hat{P}_1-\frac{5}{4}\hat{P}_2-\frac{5}{4}\Bigg)\hat{P}_1\Bigg)\cosh^2(h)-\frac{5}{2}\Big(\hat{P}_1+1\Big)^2\hat{P}_1\Bigg)\Big(\cosh(h)+1\Big)\Big(\cosh(h)-1\Big)\Bigg) \\
&\times\Big(\sinh^2(h)\hat{P}_1\Big(2\cosh^2(h)\hat{P}_1-\cosh^2(h)\hat{P}_2+\cosh^2(h)-\hat{P}_1-1\Big)\Big)^{-1}
\end{aligned}
\tag{1.3}
$$

$$
\mathrm{DYN}_{3,3} = -\frac{1}{8}\frac{\Big(4\cosh^4(h)\hat{P}_1-7\cosh^4(h)\hat{P}_2-3\cosh^4(h)-18\cosh^2(h)\hat{P}_1+9\cosh^2(h)\hat{P}_2-9\cosh^2(h)+15\hat{P}_1+15\Big)\Big(\hat{P}_1+1\Big)}{\Big(2\cosh^2(h)\hat{P}_1-\cosh^2(h)\hat{P}_2+\cosh^2(h)-\hat{P}_1-1\Big)\sinh^2(h)}
\tag{1.4}
$$

Recall that $\mathrm{COMB}_{3,1}$ is defined in **??** as the combination of $\mathrm{KIN}_{3,1}$ and $\mathrm{DYN}_{3,1}$ formed by eliminating $\hat{\phi}_{3,1}$. This grouping appears often in the 4th-order coefficients and governs the higher-order corrections to the phase speed **??** and growth rate **??**. Therefore, we give its expression here.

$$
\begin{aligned}
&\mathrm{COMB}_{3,1} = \qquad (1.5) \\
&\frac{1}{6}\Big(\hat{P}_1+1\Big)\Bigg(\frac{1}{2}\hat{P}_1\Big(8\cosh^4(h)\hat{P}_1-2\cosh^4(h)\hat{P}_2+6\cosh^4(h)-6\cosh^2(h)\hat{P}_1+2\cosh^2(h)\hat{P}_2-4\cosh^2(h)+\hat{P}_1+1\Big)\Big|\hat{P}_1+1\Big| \\
&+\frac{1}{4}\Big(\Big(-8\hat{P}_1+2\hat{P}_2-6\Big)\Big|\hat{P}_1\Big|^2+4\hat{P}_1^3-2\hat{P}_1^2\hat{P}_2-2\hat{P}_1^2-4\hat{P}_1\Big)\cosh^4(h)-\frac{5}{2}\Bigg(\Bigg(-\frac{2}{5}-\frac{3}{5}\hat{P}_1+\frac{\hat{P}_2}{5}\Bigg)\Big|\hat{P}_1\Big|^2+\Bigg(\hat{P}_1\Bigg(\hat{P}_1-\frac{3}{5}\hat{P}_2\Bigg)+\frac{4}{5}\hat{P}_1
\end{aligned}
$$

$$- \frac{2}{5}\hat{P}_2\Big)\hat{P}_1\Bigg)\cosh^2(h) + 2\Big(\hat{P}_1^2 - \frac{1}{8}\Big|\hat{P}_1\Big|^2 + \frac{7\hat{P}_1}{8}\Big)\Big(\hat{P}_1 + 1\Big)\Bigg)\Bigg(\Big(\Big(1 - \hat{P}_2 + 2\hat{P}_1\Big)\cosh^2(h)^2 - \hat{P}_1 - 1\Big)\cosh(h)\hat{P}_1\omega\Big(\hat{P}_1 - \frac{1}{3}\Big|\hat{P}_1 + 1\Big|$$

$$+ 1\Big)\sinh(h)\Bigg)^{-1} \tag{1.6}$$

## 2. The $\mathcal{O}\left(\varepsilon^4\right)$ coefficients

Here, we give the full expressions for fourth-order coefficients, $\mathrm{KIN}_{4,m}$ and $\mathrm{DYN}_{4,m}$, defined in **????** as the coefficients of the lower-order terms' $m$-th harmonics in the kinematic and dynamic boundary conditions, respectively.

$$\mathrm{KIN}_{4,2} =$$

$$\frac{1}{2}\Big(\hat{P}_1 + 1\Big)\Bigg(-2\Big(\cosh^2(h) + \frac{1}{2}\Big)\Bigg(\frac{1}{2}\mathrm{i}\sinh(h)\mathrm{COMB}_{3,1}\Bigg(\Bigg(\frac{2}{3} + \hat{P}_1 - \frac{\hat{P}_3}{3}\Bigg)\cosh^2(h) - \frac{2}{3} - \frac{3}{4}\hat{P}_1 + \frac{\hat{P}_3}{12}\Bigg)\cosh(h)\omega + \mathrm{i}\Bigg(\Bigg(\frac{2}{3} + \hat{P}_1$$

$$- \frac{\hat{P}_3}{3}\Bigg)\Bigg(\frac{9}{8} + \hat{P}_1 + \frac{\hat{P}_2}{8}\Bigg)\cosh^4(h) - \frac{\Big(13\hat{P}_1 + 13\Big)\cosh^2(h)}{8}\Bigg(\frac{29}{39} + \hat{P}_1 - \frac{10\hat{P}_3}{39}\Bigg) + \frac{3}{4}\Big(\hat{P}_1 + 1\Big)\Bigg(\frac{31}{36} + \hat{P}_1 - \frac{5\hat{P}_3}{36}\Bigg)\Bigg)\Bigg)\hat{P}_1\Big|\hat{P}_1 + 1\Big|$$

$$+ 3\mathrm{i}\sinh(h)\Big(\hat{P}_1 + 1\Big)\Big(\cosh^2(h) + \frac{1}{2}\Big)\mathrm{COMB}_{3,1}\hat{P}_1\Bigg(\Bigg(\frac{2}{3} + \hat{P}_1 - \frac{\hat{P}_3}{3}\Bigg)\cosh^2(h) - \frac{2}{3} - \frac{3}{4}\hat{P}_1 + \frac{\hat{P}_3}{12}\Bigg)\cosh(h)\omega + \mathrm{i}\Bigg(\Bigg(\Bigg(\frac{25}{36} + \hat{P}_1^2$$

$$+ \Bigg(-\frac{5\hat{P}_3}{18} - \frac{\hat{P}_2}{12}\Bigg)\hat{P}_1 + \frac{1}{18}\hat{P}_2\hat{P}_3 + \frac{59\hat{P}_1}{36} - \frac{2}{9}\hat{P}_3 - \frac{\hat{P}_2}{36}\Bigg)\Big|\hat{P}_1\Big|^2 - \frac{1}{2}\Bigg(-\frac{5}{9} + \Bigg(\hat{P}_1^2 + \Bigg(\frac{2}{3}\hat{P}_3 - 2\hat{P}_2\Bigg)\hat{P}_1 + \frac{7}{6}\hat{P}_2\hat{P}_3\Bigg)\hat{P}_1 - \frac{1}{3}\hat{P}_1^2$$

$$+\left(\frac{55\hat{P}_3}{18}-\frac{8}{3}\hat{P}_2\right)\hat{P}_1+\frac{19\hat{P}_2\hat{P}_3}{18}-\frac{16\hat{P}_1}{9}+\frac{41\hat{P}_3}{18}-\frac{7\hat{P}_2}{9}\right)\hat{P}_1\Bigg)\cosh^6(h)+\Bigg(-\frac{17\left|\hat{P}_1\right|^2}{12}\Big(\frac{10}{17}+\hat{P}_1^2+\left(-\frac{23\hat{P}_3}{102}-\frac{4\hat{P}_2}{17}\right)\hat{P}_1+\frac{5\hat{P}_2\hat{P}_3}{102}$$

$$+\frac{157\hat{P}_1}{102}-\frac{3\hat{P}_3}{17}-\frac{19\hat{P}_2}{102}\Big)+\frac{25\hat{P}_1}{8}\Big(-\frac{8}{75}+\hat{P}_1\left(\hat{P}_1^2+\left(-\frac{18\hat{P}_2}{25}-\frac{4\hat{P}_3}{15}\right)\hat{P}_1+\frac{11\hat{P}_2\hat{P}_3}{75}\right)+\frac{39\hat{P}_1^2}{25}+\left(-\frac{64\hat{P}_3}{225}-\frac{89\hat{P}_2}{75}\right)\hat{P}_1+\frac{28\hat{P}_2\hat{P}_3}{225}$$

$$+\frac{107\hat{P}_1}{225}-\frac{\hat{P}_3}{25}-\frac{22\hat{P}_2}{45}\Big)\Bigg)\cosh^4(h)+\left(\frac{23\left|\hat{P}_1\right|^2}{48}\left(\frac{16}{23}+\hat{P}_1^2+\left(-\frac{5\hat{P}_3}{69}-\frac{6\hat{P}_2}{23}\right)\hat{P}_1+\frac{2\hat{P}_2\hat{P}_3}{69}+\frac{5}{3}\hat{P}_1-\frac{\hat{P}_3}{23}-\frac{16\hat{P}_2}{69}\right)-\frac{73\hat{P}_1}{16}\left(\frac{104}{219}\right.\right.$$

$$\left.\left.+\left(\hat{P}_1^2+\left(-\frac{18\hat{P}_2}{73}-\frac{17\hat{P}_3}{73}\right)\hat{P}_1+\frac{2\hat{P}_2\hat{P}_3}{73}\right)\hat{P}_1+\frac{529\hat{P}_1^2}{219}+\left(-\frac{283\hat{P}_3}{657}-\frac{32\hat{P}_2}{73}\right)\hat{P}_1+\frac{16\hat{P}_2\hat{P}_3}{657}+\frac{1244\hat{P}_1}{657}-\frac{44\hat{P}_3}{219}-\frac{128\hat{P}_2}{657}\right)\right)\cosh^2(h)$$

$$+\frac{3\hat{P}_1+3}{32}\left(\left(\frac{25}{27}+\hat{P}_1-\frac{2\hat{P}_3}{27}\right)\left|\hat{P}_1\right|^2+20\left(\frac{239}{270}+\hat{P}_1\left(\hat{P}_1-\frac{29\hat{P}_3}{180}\right)+\frac{17\hat{P}_1}{9}-\frac{89\hat{P}_3}{540}\right)\hat{P}_1\right)\Bigg)\Bigg)$$

$$\times\left(\omega\sinh^2(h)\left(\left(\frac{1}{2}+\hat{P}_1-\frac{\hat{P}_2}{2}\right)\cosh^2(h)-\frac{\hat{P}_1}{2}-\frac{1}{2}\right)\hat{P}_1\left(-3\hat{P}_1+\left|\hat{P}_1+1\right|-3\right)\left(\left(\frac{2}{3}+\hat{P}_1-\frac{\hat{P}_3}{3}\right)\cosh^2(h)-\frac{2}{3}-\frac{3}{4}\hat{P}_1+\frac{\hat{P}_3}{12}\right)\right)^{-1} \tag{2.1}$$

$\mathrm{KIN}_{4,4} =$

$$\frac{1}{4}\big(\mathrm{i}\cosh(h)\omega\big)\Bigg(\Bigg(\frac{8}{9}+\hat{P}_1^3+\left(-\frac{4}{3}\hat{P}_2+\frac{5}{9}\hat{P}_3\right)\hat{P}_1^2+\frac{2}{3}\left(\frac{2}{3}\hat{P}_3+\hat{P}_2\right)\hat{P}_2\hat{P}_1-\frac{4}{9}\hat{P}_2^2\hat{P}_3+\frac{20\hat{P}_1^2}{9}+\left(-\frac{8\hat{P}_2}{9}+\frac{14\hat{P}_3}{9}\right)\hat{P}_1+\frac{2}{9}\hat{P}_2\big(-2\hat{P}_3+\hat{P}_2\big)$$

$$+\frac{23\hat{P}_1}{9}-\frac{4}{9}\hat{P}_2+\frac{5}{9}\hat{P}_3\Bigg)\cosh^8(h)+\Bigg(-\frac{5}{9}-\frac{59\hat{P}_1^3}{12}+\Bigg(\frac{85\hat{P}_2}{12}+\frac{11\hat{P}_3}{12}\Bigg)\hat{P}_1^2-\frac{49\hat{P}_2\hat{P}_1}{24}\Bigg(\frac{146\hat{P}_3}{147}+\hat{P}_2\Bigg)+\frac{31\hat{P}_2^2\hat{P}_3}{72}-\frac{27\hat{P}_1^2}{4}$$

$$+\Bigg(\frac{145\hat{P}_2}{18}-\frac{7\hat{P}_3}{36}\Bigg)\hat{P}_1-\frac{29\hat{P}_2}{18}\Bigg(\hat{P}_2+\frac{21\hat{P}_3}{29}\Bigg)-\frac{203\hat{P}_1}{72}+\frac{11\hat{P}_2}{6}-\frac{49\hat{P}_3}{72}\Bigg)\cosh^6(h)+\Bigg(\frac{11}{6}+\frac{81\hat{P}_1^3}{8}+\Bigg(-\frac{33\hat{P}_2}{4}-\frac{185\hat{P}_3}{72}\Bigg)\hat{P}_1^2$$

$$+\hat{P}_2\Bigg(\hat{P}_2+\frac{59\hat{P}_3}{36}\Bigg)\hat{P}_1-1/9\hat{P}_2^2\hat{P}_3+\frac{176\hat{P}_1^2}{9}+\Bigg(-\frac{463\hat{P}_2}{36}-7/2\hat{P}_3\Bigg)\hat{P}_1+\frac{8\hat{P}_2}{9}\Bigg(\hat{P}_2+\frac{51\hat{P}_3}{32}\Bigg)+\frac{91\hat{P}_1}{8}-\frac{29\hat{P}_2}{6}-\frac{25\hat{P}_3}{24}\Bigg)\cosh^4(h)$$

$$-\frac{\left(361\hat{P}_1+361\right)\cosh^2(h)^2}{48}\Bigg(\frac{530}{1083}+\hat{P}_1^2+\Bigg(-\frac{129\hat{P}_2}{361}-\frac{73\hat{P}_3}{361}\Bigg)\hat{P}_1+\frac{53\hat{P}_2\hat{P}_3}{1083}+\frac{520\hat{P}_1}{361}-\frac{334\hat{P}_2}{1083}-\frac{166\hat{P}_3}{1083}\Bigg)+\frac{57\left(\hat{P}_1+1\right)^2}{32}\Bigg(\frac{436}{513}$$

$$+\hat{P}_1-\frac{77\hat{P}_3}{513}\Bigg)\Bigg)\Bigg(\sinh^3(h)\Bigg(\Bigg(\frac{2}{3}+\hat{P}_1-\frac{\hat{P}_3}{3}\Bigg)\cosh^2(h)-\frac{2}{3}-\frac{3}{4}\hat{P}_1+\frac{\hat{P}_3}{12}\Bigg)\Bigg(\Bigg(\frac{1}{2}+\hat{P}_1-\frac{\hat{P}_2}{2}\Bigg)\cosh^2(h)-\frac{\hat{P}_1}{2}-\frac{1}{2}\Bigg)^2\Bigg)^{-1} \tag{2.2}$$

$$\mathrm{KIN}_{4,0}=0 \tag{2.3}$$

$$\mathrm{DYN}_{4,2}=$$

$$-\frac{1}{4}\Bigg(\Bigg(-\mathrm{COMB}_{3,1}\sinh(h)\Bigg(\left(\hat{P}_2+1\right)\cosh^4(h)+\left(\hat{P}_1+1\right)\cosh^2(h)-\frac{\hat{P}_1}{2}-\frac{1}{2}\Bigg)\hat{P}_1\Bigg(\Bigg(\frac{2}{3}+\hat{P}_1-\frac{\hat{P}_3}{3}\Bigg)\cosh^2(h)-\frac{2}{3}-\frac{3}{4}\hat{P}_1$$

$$+\frac{\hat{P}_3}{12}\Bigg)\cosh(h)\omega-\frac{1}{6}\Bigg(\Bigg(\Bigg(\frac{7}{6}+\Bigg(\hat{P}_2+\frac{\hat{P}_3}{3}\Bigg)\hat{P}_1-\frac{1}{6}\hat{P}_2\hat{P}_3+\frac{4}{3}\hat{P}_1+\frac{5}{6}\hat{P}_2+\frac{\hat{P}_3}{6}\Bigg)\left|\hat{P}_1\right|^2+15\hat{P}_1\Bigg(\frac{59}{45}+\hat{P}_1\Bigg(\hat{P}_1^2+\Bigg(-\frac{\hat{P}_2}{2}+\frac{4\hat{P}_3}{15}\Bigg)\hat{P}_1+\frac{7\hat{P}_2\hat{P}_3}{15}\Bigg)$$

$$+\frac{83\hat{P}_1^2}{30}+\Bigg(-\frac{7\hat{P}_2}{15}+\frac{46\hat{P}_3}{45}\Bigg)\hat{P}_1+\frac{41\hat{P}_2\hat{P}_3}{90}+\frac{\hat{P}_2}{45}+\frac{67\hat{P}_3}{90}+\frac{139\hat{P}_1}{45}\Bigg)\Bigg)\cosh^8(h)-\frac{1}{6}\Bigg(\frac{7}{2}\Bigg(\frac{5}{21}+\hat{P}_1^2+\Bigg(-\frac{13\hat{P}_2}{28}-\frac{17\hat{P}_3}{42}\Bigg)\hat{P}_1+\frac{3\hat{P}_2\hat{P}_3}{28}$$

$$+\frac{95\hat{P}_1}{84}-\frac{5\hat{P}_2}{14}-\frac{25\hat{P}_3}{84}\Bigg)\left|\hat{P}_1\right|^2-\frac{165\hat{P}_1}{4}\Bigg(\frac{2}{9}+\hat{P}_1\Bigg(\hat{P}_1^2+\Bigg(-\frac{8\hat{P}_3}{55}-\frac{93\hat{P}_2}{110}\Bigg)\hat{P}_1+\frac{7\hat{P}_2\hat{P}_3}{30}\Bigg)+\frac{127\hat{P}_1^2}{66}+\Bigg(-\frac{78\hat{P}_2}{55}-\frac{23\hat{P}_3}{990}\Bigg)\hat{P}_1+\frac{37\hat{P}_2\hat{P}_3}{165}$$

$$+\frac{52\hat{P}_1}{45}-\frac{32\hat{P}_2}{55}+\frac{56\hat{P}_3}{495}\Bigg)\Bigg)\cosh^6(h)-\frac{1}{6}\Bigg(-\frac{11}{2}\Bigg(\frac{7}{11}+\hat{P}_1^2+\Bigg(-\frac{2}{11}\hat{P}_2-\frac{5\hat{P}_3}{22}\Bigg)\hat{P}_1+\frac{1}{22}\hat{P}_2\hat{P}_3+\frac{35\hat{P}_1}{22}-\frac{3\hat{P}_2}{22}-\frac{2}{11}\hat{P}_3\Bigg)\left|\hat{P}_1\right|^2+63\hat{P}_1\Bigg(\frac{2}{7}$$

$$+\Bigg(\hat{P}_1^2+\Bigg(-\frac{16\hat{P}_3}{63}-\frac{13\hat{P}_2}{28}\Bigg)\hat{P}_1+\frac{5\hat{P}_2\hat{P}_3}{84}\Bigg)\hat{P}_1+\frac{79\hat{P}_1^2}{36}+\Bigg(-\frac{215\hat{P}_2}{252}-\frac{3}{7}\hat{P}_3\Bigg)\hat{P}_1+\frac{1}{18}\hat{P}_2\hat{P}_3+\frac{187\hat{P}_1}{126}-\frac{11\hat{P}_2}{28}-\frac{5\hat{P}_3}{28}\Bigg)\Bigg)\cosh^4(h)$$

$$-\frac{1}{6}\Bigg(2\Bigg(\frac{73}{96}+\hat{P}_1^2+\Bigg(-\frac{3}{16}\hat{P}_2-\frac{7\hat{P}_3}{96}\Bigg)\hat{P}_1+\frac{1}{48}\hat{P}_2\hat{P}_3+\frac{167\hat{P}_1}{96}-\frac{\hat{P}_2}{6}-\frac{5\hat{P}_3}{96}\Bigg)\left|\hat{P}_1\right|^2-\frac{747\hat{P}_1}{16}\Bigg(\frac{1538}{2241}+\Bigg(\hat{P}_1^2+\Bigg(-\frac{18\hat{P}_3}{83}-\frac{6\hat{P}_2}{83}\Bigg)\hat{P}_1$$

$$+\frac{2\hat{P}_2\hat{P}_3}{249}\Bigg)\hat{P}_1+\frac{1993\hat{P}_1^2}{747}+\Bigg(-\frac{32\hat{P}_2}{249}-\frac{947\hat{P}_3}{2241}\Bigg)\hat{P}_1+\frac{16\hat{P}_2\hat{P}_3}{2241}+\frac{5278\hat{P}_1}{2241}-\frac{128\hat{P}_2}{2241}-\frac{463\hat{P}_3}{2241}\Bigg)\Bigg)\cosh^2(h)+\frac{1}{288}\Bigg(\Bigg(\hat{P}_3+1\Bigg)\left|\hat{P}_1\right|^2-486\hat{P}_1^3$$

$$+72\hat{P}_1^2\hat{P}_3 - 900\hat{P}_1^2 + 73\hat{P}_1\hat{P}_3 - 413\hat{P}_1\Bigg)\left(\hat{P}_1+1\right)\Bigg)\left|\hat{P}_1+1\right| - 3\left(\hat{P}_1+1\right)\Bigg(-\mathrm{COMB}_{3,1}\sinh(h)\Bigg(\left(\hat{P}_2+1\right)\cosh^4(h) + \left(\hat{P}_1+1\right)\cosh^2(h)$$

$$-\frac{\hat{P}_1}{2} - \frac{1}{2}\Bigg)\hat{P}_1\Bigg(\Bigg(\left(\frac{2}{3} + \hat{P}_1 - \frac{\hat{P}_3}{3}\right)\cosh^2(h) - \frac{2}{3} - \frac{3}{4}\hat{P}_1 + \frac{\hat{P}_3}{12}\Bigg)\cosh(h)\omega - \frac{1}{3}\Bigg(\left(\frac{3}{2} + \hat{P}_1 + \frac{\hat{P}_2}{2}\right)\left(\frac{7}{6} + \hat{P}_1 + \frac{\hat{P}_3}{6}\right)\left|\hat{P}_1\right|^2 + \Bigg(\frac{11}{6} + \hat{P}_1\Bigg(\hat{P}_1^2$$

$$+\left(\frac{7}{6}\hat{P}_3 - \frac{17\hat{P}_2}{4}\right)\hat{P}_1 + \frac{13\hat{P}_2\hat{P}_3}{6}\Bigg) + \frac{11\hat{P}_1^2}{12} + \left(-\frac{35\hat{P}_2}{6} + \frac{14}{3}\hat{P}_3\right)\hat{P}_1 + \frac{9}{4}\hat{P}_2\hat{P}_3 - \frac{3}{2}\hat{P}_2 + \frac{43\hat{P}_3}{12} + \frac{5}{3}\hat{P}_1\Bigg)\hat{P}_1\Bigg)\cosh^8(h) - \frac{1}{3}\Bigg(\frac{1}{2}\Bigg(-\frac{2}{3} + \hat{P}_1^2$$

$$+\left(-\frac{5}{4}\hat{P}_3 - \frac{\hat{P}_2}{2}\right)\hat{P}_1 + \frac{1}{12}\hat{P}_2\hat{P}_3 + \frac{\hat{P}_1}{4} - \frac{5\hat{P}_2}{12} - \frac{7}{6}\hat{P}_3\Bigg)\left|\hat{P}_1\right|^2 - \frac{49\hat{P}_1}{4}\Bigg(\frac{1}{21} + \Bigg(\hat{P}_1^2 + \left(-\frac{67\hat{P}_3}{294} - \frac{51\hat{P}_2}{49}\right)\hat{P}_1 + \frac{85\hat{P}_2\hat{P}_3}{294}\Bigg)\hat{P}_1 + \frac{71\hat{P}_1^2}{42}$$

$$+\left(-\frac{521\hat{P}_2}{294} - \frac{17\hat{P}_3}{147}\right)\hat{P}_1 + \frac{2}{7}\hat{P}_2\hat{P}_3 + \frac{109\hat{P}_1}{147} - \frac{36\hat{P}_2}{49} + \frac{16\hat{P}_3}{147}\Bigg)\Bigg)\cosh^6(h) - \frac{1}{3}\Bigg(-\frac{35\left|\hat{P}_1\right|^2}{16}\Bigg(\frac{4}{5} + \hat{P}_1^2 + \frac{4\hat{P}_2\hat{P}_3}{105} - \frac{5\hat{P}_1\hat{P}_3}{21} + \frac{37\hat{P}_1}{21} + \frac{4\hat{P}_2}{105}$$

$$-\frac{\hat{P}_3}{5}\Bigg) + \frac{425\hat{P}_1}{16}\Bigg(\frac{148}{425} + \hat{P}_1\Bigg(\hat{P}_1^2 + \left(-\frac{111\hat{P}_3}{425} - \frac{162\hat{P}_2}{425}\right)\hat{P}_1 + \frac{24\hat{P}_2\hat{P}_3}{425}\Bigg) + \frac{967\hat{P}_1^2}{425} + \left(-\frac{12\hat{P}_2}{17} - \frac{569\hat{P}_3}{1275}\right)\hat{P}_1 + \frac{4\hat{P}_2\hat{P}_3}{75} + \frac{122\hat{P}_1}{75} - \frac{418\hat{P}_2}{1275}$$

$$- \frac{16\hat{P}_3}{85}\Bigg)\Bigg)\Bigg) \cosh^4(h) - \frac{1}{3}\Bigg(\frac{17\left|\hat{P}_1\right|^2}{32}\Bigg(\frac{15}{17} + \hat{P}_1^2 + \left(\frac{10\hat{P}_3}{51} - \frac{6\hat{P}_2}{17}\right)\hat{P}_1 + \frac{2\hat{P}_2\hat{P}_3}{51} + \frac{94\hat{P}_1}{51} - \frac{16\hat{P}_2}{51} + \frac{4\hat{P}_3}{17}\Bigg) - \frac{597\hat{P}_1}{32}\Bigg(\frac{1274}{1791}$$

$$+ \hat{P}_1\Bigg(\hat{P}_1^2 + \left(-\frac{328\hat{P}_3}{1791} - \frac{18\hat{P}_2}{199}\right)\hat{P}_1 + \frac{2\hat{P}_2\hat{P}_3}{199}\Bigg) + \frac{4832\hat{P}_1^2}{1791} + \left(-\frac{32\hat{P}_2}{199} - \frac{72\hat{P}_3}{199}\right)\hat{P}_1 + \frac{16\hat{P}_2\hat{P}_3}{1791} + \frac{1439\hat{P}_1}{597} - \frac{128\hat{P}_2}{1791} - \frac{322\hat{P}_3}{1791}\Bigg)\Bigg) \cosh^2(h)$$

$$- \frac{1}{16}\Bigg(\left(\frac{7}{9} + \hat{P}_1 - \frac{2}{9}\hat{P}_3\right)\left|\hat{P}_1\right|^2 + 18\hat{P}_1\Bigg(\frac{293}{324} + \hat{P}_1\left(\hat{P}_1 - \frac{5\hat{P}_3}{36}\right) + \frac{23\hat{P}_1}{12} - \frac{49\hat{P}_3}{324}\Bigg)\Bigg)\left(\hat{P}_1 + 1\right)\Bigg)\Bigg)\Bigg(\Big(\left(\cosh(h) + 1\right)\sinh(h)\Big(-3\hat{P}_1$$

$$+ \left|\hat{P}_1 + 1\right| - 3\Big)\Big(\Bigg(\left(\frac{1}{2} + \hat{P}_1 - \frac{\hat{P}_2}{2}\right)\cosh^2(h) - \frac{\hat{P}_1}{2} - \frac{1}{2}\Bigg)\left(\cosh(h) - 1\right)\hat{P}_1\Bigg(\left(\frac{2}{3} + \hat{P}_1 - \frac{\hat{P}_3}{3}\right)\cosh^2(h) - \frac{2}{3} - \frac{3}{4}\hat{P}_1 + \frac{\hat{P}_3}{12}\Bigg)\cosh(h)\Bigg)^{-1} \tag{2.4}$$

$\mathrm{DYN}_{4,4} =$

$$\frac{1}{8}\left(\hat{P}_1 + 1\right)\Bigg(\Bigg(\frac{4}{9} + \hat{P}_1^3 + \left(\frac{7\hat{P}_3}{9} - \frac{19\hat{P}_2}{6}\right)\hat{P}_1^2 + \frac{\hat{P}_2\hat{P}_1}{12}\left(\frac{38\hat{P}_3}{3} + 19\hat{P}_2\right) - \frac{29\hat{P}_2^2\hat{P}_3}{36} + \frac{11\hat{P}_1^2}{18} + \left(-\frac{19\hat{P}_2}{9} + \frac{47\hat{P}_3}{18}\right)\hat{P}_1 + \frac{7\hat{P}_2}{9}\left(-\frac{5}{7}\hat{P}_3 + \hat{P}_2\right)$$

$$- \frac{5}{9}\hat{P}_2 + \frac{37\hat{P}_3}{36} + \frac{31\hat{P}_1}{36}\Bigg)\cosh^{10}(h) + \Bigg(-\frac{5}{9} - \frac{115\hat{P}_1^3}{12} + \left(\frac{46\hat{P}_2}{3} + \frac{53\hat{P}_3}{36}\right)\hat{P}_1^2 - \frac{14}{3}\hat{P}_2\left(\hat{P}_2 + \frac{20\hat{P}_3}{21}\right)\hat{P}_1 + \frac{4}{3}\hat{P}_2^2\hat{P}_3 - \frac{215\hat{P}_1^2}{18} + \Bigg(\frac{152\hat{P}_2}{9}$$

$$- \frac{3}{2}\hat{P}_3\Bigg)\hat{P}_1 - \frac{10}{3}\hat{P}_2\left(\frac{8\hat{P}_3}{15} + \hat{P}_2\right) - \frac{17\hat{P}_1}{4} + \frac{38\hat{P}_2}{9} - \frac{59\hat{P}_3}{36}\Bigg)\cosh^8(h) + \Bigg(\frac{145}{36} + \frac{581\hat{P}_1^3}{24} + \left(-\frac{1067\hat{P}_2}{48} - \frac{439\hat{P}_3}{72}\right)\hat{P}_1^2 + \frac{347\hat{P}_2\hat{P}_1}{96}\Bigg(\hat{P}_2$$

$$+ \frac{482\hat{P}_3}{347}\Bigg) - \frac{47\hat{P}_2^2\hat{P}_3}{96} + \frac{6379\hat{P}_1^2}{144} + \left(-\frac{773\hat{P}_2}{24} - \frac{1033\hat{P}_3}{144}\right)\hat{P}_1 + \frac{25\hat{P}_2}{8}\left(\frac{97\hat{P}_3}{75} + \hat{P}_2\right) + \frac{7087\hat{P}_1}{288} - \frac{263\hat{P}_2}{24} - \frac{451\hat{P}_3}{288}\Bigg)\cosh^6(h)$$

$$+ \Bigg(-\frac{671}{72} - \frac{2257\hat{P}_1^3}{96} + \left(\frac{363\hat{P}_2}{32} + \frac{1447\hat{P}_3}{288}\right)\hat{P}_1^2 - \frac{1}{2}\hat{P}_2\left(\hat{P}_2 + \frac{499\hat{P}_3}{144}\right)\hat{P}_1 + \frac{1}{18}\hat{P}_2^2\hat{P}_3 - \frac{15599\hat{P}_1^2}{288} + \left(\frac{5747\hat{P}_2}{288} + \frac{2395\hat{P}_3}{288}\right)\hat{P}_1$$

$$- \frac{4}{9}\hat{P}_2\left(\frac{467\hat{P}_3}{128} + \hat{P}_2\right) - \frac{1441\hat{P}_1}{36} + \frac{157\hat{P}_2}{18} + \frac{241\hat{P}_3}{72}\Bigg)\cosh^4(h) + \frac{\left(26\hat{P}_1 + 26\right)\cosh^2(h)}{3}\Bigg(\frac{1753}{2496} + \hat{P}_1^2 + \left(-\frac{129\hat{P}_2}{832} - \frac{103\hat{P}_3}{624}\right)\hat{P}_1$$

$$+ \frac{7\hat{P}_2\hat{P}_3}{312} + \frac{4193\hat{P}_1}{2496} - \frac{331\hat{P}_2}{2496} - \frac{89\hat{P}_3}{624}\Bigg) - \frac{105\left(\hat{P}_1 + 1\right)^2}{128}\left(\frac{788}{945} + \hat{P}_1 - \frac{157\hat{P}_3}{945}\right)\Bigg)$$

$$\times \left(\cosh(h)\left(\left(\frac{2}{3} + \hat{P}_1 - \frac{\hat{P}_3}{3}\right)\cosh^2(h) - \frac{2}{3} - \frac{3}{4}\hat{P}_1 + \frac{\hat{P}_3}{12}\right)\left(\left(\frac{1}{2} + \hat{P}_1 - \frac{\hat{P}_2}{2}\right)\cosh^2(h) - \frac{\hat{P}_1}{2} - \frac{1}{2}\right)^2 \sinh^3(h)\right)^{-1} \tag{2.5}$$

$\mathrm{DYN}_{4,0} =$

$$\frac{\mathrm{i}}{\mathrm{COMB}_{3,1}}\Bigg(\Bigg(-\frac{1}{6}\Bigg(13\mathrm{i}\hat{P}_1\cosh(h)\left(\left(\hat{P}_1 - \frac{\hat{P}_2}{2} + \frac{1}{2}\right)\cosh^2(h) - \frac{\hat{P}_1}{2} - \frac{1}{2}\right)\sinh(h)\left(\left(2\left|\hat{P}_1\right|^2\hat{P}_2 + \hat{P}_1\left(-\left|\hat{P}_2\right|^2 + \hat{P}_2\right)\right)\cosh^2(h)\right.$$

$$\times -\hat{P}_2\left(\left|\hat{P}_1\right|^2+\hat{P}_1\right)\right)\omega\Bigg)\Bigg(\Bigg(\Bigg(-\frac{3}{13}\mathrm{COMB}_{3,1}{}^2\left|\hat{P}_1\right|^4+\hat{P}_1\left(\left(\hat{P}_1+1\right)\left|\mathrm{COMB}_{3,1}\right|^2-\mathrm{COMB}_{3,1}{}^2\left(\hat{P}_1+\frac{19}{13}\right)\right)\left|\hat{P}_1\right|^2+\hat{P}_1^2\Bigg(\Bigg(\frac{3}{13}\hat{P}_1^2+\frac{19\hat{P}_1}{13}$$

$$+\frac{16}{13}\right)\left|\mathrm{COMB}_{3,1}\right|^2-\mathrm{COMB}_{3,1}{}^2\left(\hat{P}_1+\frac{16}{13}\right)\Bigg)\Bigg)\cosh^2(h)-\frac{3}{13}\left(\hat{P}_1^2+\left|\hat{P}_1\right|^2+2\hat{P}_1\right)\Bigg(-\mathrm{COMB}_{3,1}{}^2\left|\hat{P}_1\right|^2+\Bigg(\left(\hat{P}_1+1\right)\left|\mathrm{COMB}_{3,1}\right|^2$$

$$-\mathrm{COMB}_{3,1}{}^2\Bigg)\hat{P}_1\Bigg)\Bigg)+\frac{19\mathrm{i}\left(\hat{P}_1+1\right)\hat{P}_1\mathrm{COMB}_{3,1}}{24}\Bigg(\Bigg(\frac{100\hat{P}_2\left|\hat{P}_1\right|^6}{19}\left(\frac{67}{100}+\hat{P}_1-\frac{33\hat{P}_2}{100}\right)+\frac{146\hat{P}_1\left|\hat{P}_1\right|^4}{19}\Bigg(-\frac{65\left|\hat{P}_2\right|^2}{146}\left(\frac{101}{130}+\hat{P}_1-\frac{29\hat{P}_2}{130}\right)$$

$$+\left(\frac{411}{292}+\hat{P}_1\left(\hat{P}_1-\frac{91\hat{P}_2}{146}\right)+\frac{218\hat{P}_1}{73}-\frac{351\hat{P}_2}{292}\right)\hat{P}_2\Bigg)+\frac{26\hat{P}_1^2\left|\hat{P}_1\right|^2}{19}\Bigg(-\frac{34\left|\hat{P}_2\right|^2}{13}\left(\frac{147}{68}+\hat{P}_1\left(\hat{P}_1-\frac{11\hat{P}_2}{34}\right)+\frac{61\hat{P}_1}{17}-\frac{3}{4}\hat{P}_2\right)$$

$$+\hat{P}_2\left(\frac{87}{13}+\hat{P}_1^2\left(\hat{P}_1-\frac{8\hat{P}_2}{13}\right)+11\hat{P}_1\left(\hat{P}_1-\frac{96\hat{P}_2}{143}\right)+\frac{252\hat{P}_1}{13}-\frac{123\hat{P}_2}{13}\right)\Bigg)+\hat{P}_1^3\Bigg(-\frac{7\left|\hat{P}_2\right|^2}{19}\Bigg(\frac{100}{7}+\hat{P}_1^2\left(\hat{P}_1-\frac{\hat{P}_2}{2}\right)+\frac{171\hat{P}_1}{14}\left(\hat{P}_1-\frac{58\hat{P}_2}{171}\right)$$

$$+\frac{193\hat{P}_1}{7}-\frac{40\hat{P}_2}{7}\Bigg)\left(\frac{32}{19}+\hat{P}_1^2\left(\hat{P}_1-\frac{25\hat{P}_2}{38}\right)+\frac{245\hat{P}_1}{38}\left(\hat{P}_1-\frac{188\hat{P}_2}{245}\right)+\frac{154\hat{P}_1}{19}-\frac{100\hat{P}_2}{19}\right)\hat{P}_2\Bigg)\Bigg)\cosh^8(h)+\Bigg(-\frac{260\hat{P}_2\left|\hat{P}_1\right|^6}{19}\Bigg(\frac{97}{130}+\hat{P}_1$$

$$-\frac{33\hat{P}_2}{130}\Bigg)-\frac{425\hat{P}_1\left|\hat{P}_1\right|^4}{19}\Bigg(\left(-\frac{167\hat{P}_1}{850}+\frac{79\hat{P}_2}{1700}-\frac{3}{20}\right)\left|\hat{P}_2\right|^2+\left(\frac{3267}{1700}+\hat{P}_1\left(\hat{P}_1-\frac{253\hat{P}_2}{850}\right)+\frac{284\hat{P}_1}{85}-\frac{1219\hat{P}_2}{1700}\right)\hat{P}_2\Bigg)$$

$$
-\frac{59\hat{P}_1^2\left|\hat{P}_1\right|^2}{19}\left(-\frac{229\left|\hat{P}_2\right|^2}{118}\left(\frac{420}{229}+\hat{P}_1\left(\hat{P}_1-\frac{64\hat{P}_2}{229}\right)+\frac{728\hat{P}_1}{229}-\frac{143\hat{P}_2}{229}\right)+\left(\frac{2017}{118}+\hat{P}_1^2\left(\hat{P}_1-\frac{39\hat{P}_2}{118}\right)+\frac{893\hat{P}_1}{59}\left(\hat{P}_1-\frac{260\hat{P}_2}{893}\right)+\frac{2001\hat{P}_1}{59}\right.\right.
$$

$$
\left.\left.-\frac{399\hat{P}_2}{59}\right)\hat{P}_2\right)-\frac{45\hat{P}_1^3}{19}\left(-\frac{14\left|\hat{P}_2\right|^2}{45}\left(\frac{78}{7}+\hat{P}_1^2\left(\hat{P}_1-\frac{17\hat{P}_2}{56}\right)+\frac{87\hat{P}_1}{8}\left(\hat{P}_1-\frac{54\hat{P}_2}{203}\right)+\frac{157\hat{P}_1}{7}-4\hat{P}_2\right)+\left(\frac{376}{45}+\hat{P}_1^2\left(\hat{P}_1-\frac{61\hat{P}_2}{180}\right)\right.\right.
$$

$$
\left.\left.\left.+\frac{1721\hat{P}_1}{180}\left(\hat{P}_1-\frac{500\hat{P}_2}{1721}\right)+\frac{323\hat{P}_1}{18}-\frac{52\hat{P}_2}{15}\right)\hat{P}_2\right)\right)\cosh^6(h)+\left(\frac{315\hat{P}_2\left|\hat{P}_1\right|^6}{38}\left(\frac{517}{630}+\hat{P}_1-\frac{113\hat{P}_2}{630}\right)+\frac{261\hat{P}_1\left|\hat{P}_1\right|^4}{19}\left(-\frac{25\left|\hat{P}_2\right|^2}{522}\left(\frac{43}{50}+\hat{P}_1\right.\right.\right.
$$

$$
\left.\left.-\frac{7\hat{P}_2}{50}\right)+\left(\frac{206}{87}+\hat{P}_1\left(\hat{P}_1-\frac{20\hat{P}_2}{261}\right)+\frac{962\hat{P}_1}{261}-\frac{103\hat{P}_2}{261}\right)\hat{P}_2\right)+\frac{67\hat{P}_1^2\left|\hat{P}_1\right|^2}{38}\left(-\frac{127\left|\hat{P}_2\right|^2}{134}\left(\frac{223}{127}+\hat{P}_1\left(\hat{P}_1+\frac{10\hat{P}_2}{127}\right)+\frac{364\hat{P}_1}{127}-\frac{4\hat{P}_2}{127}\right)\right.
$$

$$
\left.+\hat{P}_2\left(\frac{3379}{134}+\hat{P}_1^2\left(\hat{P}_1-\frac{11\hat{P}_2}{134}\right)+\frac{1176\hat{P}_1}{67}\left(\hat{P}_1-\frac{4\hat{P}_2}{49}\right)+\frac{2961\hat{P}_1}{67}-\frac{253\hat{P}_2}{67}\right)\right)+\frac{107\hat{P}_1^3}{76}\left(-\frac{27\left|\hat{P}_2\right|^2}{107}\left(\frac{68}{9}+\hat{P}_1^2\left(\hat{P}_1-\frac{\hat{P}_2}{9}\right)\right.\right.
$$

$$
\left.\left.\left.+\frac{205\hat{P}_1}{27}\left(\hat{P}_1+\frac{4\hat{P}_2}{205}\right)+\frac{389\hat{P}_1}{27}\right)+\hat{P}_2\left(\frac{1400}{107}+\hat{P}_1^2\left(\hat{P}_1-\frac{8\hat{P}_2}{107}\right)+\frac{1230\hat{P}_1}{107}\left(\hat{P}_1-\frac{53\hat{P}_2}{615}\right)+\frac{2629\hat{P}_1}{107}-\frac{204\hat{P}_2}{107}\right)\right)\right)\cosh^4(h)
$$

$$
+\left(\frac{12\hat{P}_2\left|\hat{P}_1\right|^6}{19}\left(\frac{31}{48}+\hat{P}_1-\frac{17\hat{P}_2}{48}\right)+\frac{179\hat{P}_1\left|\hat{P}_1\right|^4}{76}\left(\left(-\frac{19\hat{P}_1}{179}-\frac{19}{179}\right)\left|\hat{P}_2\right|^2+\left(\frac{193}{179}+\hat{P}_1\left(\hat{P}_1-\frac{60\hat{P}_2}{179}\right)+\frac{423\hat{P}_1}{179}-\frac{111\hat{P}_2}{179}\right)\hat{P}_2\right)\right.
$$

$$
+\frac{5\hat{P}_1^2\left|\hat{P}_1\right|^2}{76}\left(\left(-12\hat{P}_1^2-\frac{158\hat{P}_1}{5}-\frac{98}{5}\right)\left|\hat{P}_2\right|^2+\left(47+\hat{P}_1^2\left(\hat{P}_1-\frac{3}{5}\hat{P}_2\right)+62\hat{P}_1\left(\hat{P}_1-\frac{63\hat{P}_2}{155}\right)+\frac{591\hat{P}_1}{5}-\frac{174\hat{P}_2}{5}\right)\hat{P}_2\right)
$$

$$
+\frac{1}{19}\hat{P}_1^3\left(\left(-\frac{1}{4}\hat{P}_1^3-\frac{63\hat{P}_1^2}{4}-\frac{71\hat{P}_1}{2}-20\right)\left|\hat{P}_2\right|^2+\left(18+\hat{P}_1^2\left(\hat{P}_1-\frac{3}{4}\hat{P}_2\right)+32\hat{P}_1\left(\hat{P}_1-\frac{33\hat{P}_2}{64}\right)+\frac{213\hat{P}_1}{4}-20\hat{P}_2\right)\hat{P}_2\right)\Bigg)\cosh^2(h)
$$

$$
-\frac{\left(7\hat{P}_1+7\right)\left(\left|\hat{P}_1\right|^2+\hat{P}_1\right)\hat{P}_2}{76}\left(\frac{89\left|\hat{P}_1\right|^4}{14}+\left(\frac{233\hat{P}_1^2}{14}+\frac{411\hat{P}_1}{14}\right)\left|\hat{P}_1\right|^2+\hat{P}_1^2\left(\hat{P}_1^2+\frac{261\hat{P}_1}{14}+24\right)\right)\Bigg)\Bigg)\left|\hat{P}_1+1\right|
$$

$$
+\frac{1}{12}\mathrm{i}\hat{P}_1\sinh(h)\cosh(h)\left(\hat{P}_1+1\right)\left(2\cosh^2(h)\hat{P}_1-\cosh^2(h)\hat{P}_2+\cosh^2(h)-\hat{P}_1-1\right)\left(\left|\hat{P}_1\right|^2+\hat{P}_1\right)
$$

$$
\times\left(2\left|\hat{P}_1\right|^2\cosh^2(h)\hat{P}_2-\left|\hat{P}_2\right|^2\cosh^2(h)\hat{P}_1-\left|\hat{P}_1\right|^2\hat{P}_2+\hat{P}_2\hat{P}_1\cosh^2(h)-\hat{P}_1\hat{P}_2\right)\Big(-13\mathrm{COMB}_{3,1}{}^2\left|\hat{P}_1\right|^2\cosh^2(h)
$$

$$
-3\mathrm{COMB}_{3,1}{}^2\cosh^2(h)\hat{P}_1^2+3\left|\hat{P}_1\right|^2\left|\mathrm{COMB}_{3,1}\right|^2\cosh^2(h)+13\left|\mathrm{COMB}_{3,1}\right|^2\cosh^2(h)\hat{P}_1^2+10\mathrm{COMB}_{3,1}{}^2\left|\hat{P}_1\right|^2
$$

$$
-16\mathrm{COMB}_{3,1}{}^2\cosh^2(h)\hat{P}_1+16\left|\mathrm{COMB}_{3,1}\right|^2\cosh^2(h)\hat{P}_1-10\left|\mathrm{COMB}_{3,1}\right|^2\hat{P}_1^2+10\mathrm{COMB}_{3,1}{}^2\hat{P}_1-10\left|\mathrm{COMB}_{3,1}\right|^2\hat{P}_1\Big)\omega
$$

$$
-\frac{67\mathrm{i}\left(\hat{P}_1+1\right)\left(\left|\hat{P}_1\right|^2+\hat{P}_1\right)\mathrm{COMB}_{3,1}}{24}\Bigg(\Bigg(\frac{26\hat{P}_2\left|\hat{P}_1\right|^6}{67}\left(\frac{19}{26}+\hat{P}_1-\frac{7\hat{P}_2}{26}\right)+\frac{146\hat{P}_1\left|\hat{P}_1\right|^4}{67}\left(-\frac{8\left|\hat{P}_2\right|^2}{73}\left(\frac{25}{32}+\hat{P}_1-\frac{7\hat{P}_2}{32}\right)\right.
$$

$$+\left(\frac{245}{292}+\hat{P}_1\left(\hat{P}_1-\frac{34\hat{P}_2}{73}\right)+\frac{143\hat{P}_1}{73}-\frac{171\hat{P}_2}{292}\right)\hat{P}_2\Bigg)+\frac{100\hat{P}_1^2\left|\hat{P}_1\right|^2}{67}\Bigg(-\frac{91\left|\hat{P}_2\right|^2}{100}\left(\frac{94}{91}+\hat{P}_1\left(\hat{P}_1-\frac{22\hat{P}_2}{91}\right)+\frac{192\hat{P}_1}{91}-\frac{29\hat{P}_2}{91}\right)$$

$$+\left(\frac{77}{50}+\hat{P}_1^2\left(\hat{P}_1-\frac{13\hat{P}_2}{20}\right)+\frac{109\hat{P}_1}{25}\left(\hat{P}_1-\frac{61\hat{P}_2}{109}\right)+\frac{126\hat{P}_1}{25}-\frac{193\hat{P}_2}{100}\right)\hat{P}_2\Bigg)+\hat{P}_1^3\Bigg(-\frac{33\left|\hat{P}_2\right|^2}{67}\Bigg(\frac{100}{33}+\hat{P}_1^2\left(\hat{P}_1-\frac{29\hat{P}_2}{66}\right)$$

$$+\frac{117\hat{P}_1}{22}\left(\hat{P}_1-\frac{34\hat{P}_2}{117}\right)+\frac{82\hat{P}_1}{11}-\frac{40\hat{P}_2}{33}\Bigg)+\hat{P}_2\left(\frac{32}{67}+\hat{P}_1^2\left(\hat{P}_1-\frac{101\hat{P}_2}{134}\right)+\frac{411\hat{P}_1}{134}\left(\hat{P}_1-\frac{98\hat{P}_2}{137}\right)+\frac{174\hat{P}_1}{67}-\frac{100\hat{P}_2}{67}\right)\Bigg)\Bigg)\Bigg)\cosh^8(h)$$

$$+\Bigg(-\frac{59\hat{P}_2\left|\hat{P}_1\right|^6}{67}\left(\frac{45}{59}+\hat{P}_1-\frac{14\hat{P}_2}{59}\right)-\frac{425\hat{P}_1\left|\hat{P}_1\right|^4}{67}\Bigg(-\frac{39\left|\hat{P}_2\right|^2}{850}\left(\frac{61}{78}+\hat{P}_1-\frac{17\hat{P}_2}{78}\right)+\Bigg(\frac{1721}{1700}+\hat{P}_1\left(\hat{P}_1-\frac{229\hat{P}_2}{850}\right)+\frac{893\hat{P}_1}{425}$$

$$-\frac{609\hat{P}_2}{1700}\Bigg)\hat{P}_2\Bigg)-\frac{260\hat{P}_1^2\left|\hat{P}_1\right|^2}{67}\Bigg(\Bigg(\left(-\frac{253\hat{P}_1}{520}\left(\hat{P}_1-\frac{64\hat{P}_2}{253}\right)-\hat{P}_1+\frac{81\hat{P}_2}{520}-\frac{25}{52}\right)\left|\hat{P}_2\right|^2+\Bigg(\frac{323}{104}+\hat{P}_1^2\left(\hat{P}_1-\frac{167\hat{P}_2}{520}\right)+\frac{71\hat{P}_1}{13}\left(\hat{P}_1-\frac{91\hat{P}_2}{355}\right)$$

$$+\frac{2001\hat{P}_1}{260}-\frac{157\hat{P}_2}{130}\Bigg)\hat{P}_2\Bigg)-\frac{194\hat{P}_1^3}{67}\Bigg(-\frac{33\left|\hat{P}_2\right|^2}{97}\left(\frac{26}{11}+\hat{P}_1^2\left(\hat{P}_1-\frac{79\hat{P}_2}{264}\right)+\frac{1219\hat{P}_1}{264}\left(\hat{P}_1-\frac{286\hat{P}_2}{1219}\right)+\frac{133\hat{P}_1}{22}-\frac{28\hat{P}_2}{33}\right)+\Bigg(\frac{188}{97}+\hat{P}_1^2\Bigg(\hat{P}_1$$

$$-\frac{255\hat{P}_2}{776}\Bigg)+\frac{3267\hat{P}_1}{776}\left(\hat{P}_1-\frac{280\hat{P}_2}{1089}\right)+\frac{2017\hat{P}_1}{388}-\frac{78\hat{P}_2}{97}\Bigg)\hat{P}_2\Bigg)\Bigg)\cosh^6(h)+\Bigg(\frac{1}{2}\left(\frac{107}{134}+\hat{P}_1-\frac{27\hat{P}_2}{134}\right)\hat{P}_2\left|\hat{P}_1\right|^6+\frac{261\hat{P}_1\left|\hat{P}_1\right|^4}{67}\Bigg(-\frac{11\left|\hat{P}_2\right|^2}{1044}$$

$$\times\left(\frac{8}{11}+\hat{P}_1-\frac{3}{11}\hat{P}_2\right)+\left(\frac{205}{174}+\hat{P}_1\left(\hat{P}_1-\frac{127\hat{P}_2}{1044}\right)+\frac{196\hat{P}_1}{87}-\frac{205\hat{P}_2}{1044}\right)\hat{P}_2\Bigg)+\frac{315\hat{P}_1^2\left|\hat{P}_1\right|^2}{134}\left(-\frac{8\left|\hat{P}_2\right|^2}{63}\left(\frac{53}{40}+\hat{P}_1\left(\hat{P}_1+\frac{\hat{P}_2}{8}\right)+\frac{12\hat{P}_1}{5}\right.\right.$$

$$\left.+\frac{\hat{P}_2}{20}\right)+\hat{P}_2\left(\frac{2629}{630}+\hat{P}_1^2\left(\hat{P}_1-\frac{5\hat{P}_2}{63}\right)+\frac{1924\hat{P}_1}{315}\left(-\frac{7\hat{P}_2}{74}+\hat{P}_1\right)+\frac{47\hat{P}_1}{5}-\frac{389\hat{P}_2}{630}\right)\Bigg)+\frac{517\hat{P}_1^3}{268}\left(-\frac{113\left|\hat{P}_2\right|^2}{517}\left(\frac{204}{113}+\hat{P}_1^2\left(\hat{P}_1-\frac{7\hat{P}_2}{113}\right)\right.\right.$$

$$\left.\left.+\frac{412\hat{P}_1}{113}\left(\hat{P}_1-\frac{\hat{P}_2}{103}\right)+\frac{506\hat{P}_1}{113}\right)+\left(\frac{1400}{517}+\hat{P}_1^2\left(\hat{P}_1-\frac{43\hat{P}_2}{517}\right)+\frac{2472\hat{P}_1}{517}\left(\hat{P}_1-\frac{223\hat{P}_2}{2472}\right)+\frac{3379\hat{P}_1}{517}-\frac{204\hat{P}_2}{517}\right)\hat{P}_2\right)\Bigg)\cosh^4(h)$$

$$+\left(\frac{\left(4+5\hat{P}_1-\hat{P}_2\right)\hat{P}_2\left|\hat{P}_1\right|^6}{268}+\frac{179\hat{P}_1\left|\hat{P}_1\right|^4}{268}\left(\left(-\frac{3\hat{P}_1}{179}-\frac{3}{179}\right)\left|\hat{P}_2\right|^2+\left(\frac{128}{179}+\hat{P}_1\left(\hat{P}_1-\frac{60\hat{P}_2}{179}\right)+\frac{310\hat{P}_1}{179}-\frac{63\hat{P}_2}{179}\right)\hat{P}_2\right)\right.$$

$$+\frac{12\hat{P}_1^2\left|\hat{P}_1\right|^2}{67}\left(\left(-\frac{5}{4}\hat{P}_1^2-\frac{21\hat{P}_1}{8}-\frac{11}{8}\right)\left|\hat{P}_2\right|^2+\hat{P}_2\left(\frac{71}{16}+\hat{P}_1^2\left(\hat{P}_1-\frac{19\hat{P}_2}{48}\right)+\frac{141\hat{P}_1}{16}\left(\hat{P}_1-\frac{158\hat{P}_2}{423}\right)+\frac{197\hat{P}_1}{16}-\frac{71\hat{P}_2}{24}\right)\right)$$

$$\left.+\frac{31\hat{P}_1^3}{268}\left(\left(-\frac{17\hat{P}_1^3}{31}-\frac{111\hat{P}_1^2}{31}-\frac{174\hat{P}_1}{31}-\frac{80}{31}\right)\left|\hat{P}_2\right|^2+\hat{P}_2\left(\frac{72}{31}+\hat{P}_1^2\left(\hat{P}_1-\frac{19\hat{P}_2}{31}\right)+\frac{193\hat{P}_1}{31}\left(\hat{P}_1-\frac{98\hat{P}_2}{193}\right)+\frac{235\hat{P}_1}{31}-\frac{80\hat{P}_2}{31}\right)\right)\right)\cosh^2(h)$$

$$- \frac{\left(89\hat{P}_1 + 89\right)\left(\left|\hat{P}_1\right|^2 + \hat{P}_1\right)\hat{P}_2}{536}\left(\frac{14\left|\hat{P}_1\right|^4}{89} + \left(\frac{233\hat{P}_1^2}{89} + \frac{261\hat{P}_1}{89}\right)\left|\hat{P}_1\right|^2 + \hat{P}_1^2\left(\hat{P}_1^2 + \frac{411\hat{P}_1}{89} + \frac{336}{89}\right)\right)\Bigg)\Bigg)$$

$$\times\Bigg(\left(\hat{P}_1 + 1\right)\cosh(h)\left(\left(2\left|\hat{P}_1\right|^2\hat{P}_2 + \hat{P}_1\left(-\left|\hat{P}_2\right|^2 + \hat{P}_2\right)\right)\cosh^2(h) - \hat{P}_2\left(\left|\hat{P}_1\right|^2 + \hat{P}_1\right)\right)\left(-\frac{1}{3}\hat{P}_1\left(13\left|\hat{P}_1\right|^2 + 3\hat{P}_1^2 + 16\hat{P}_1\right)\left|\hat{P}_1 + 1\right|\right.$$

$$\left.+ \frac{1}{3}\left(\left|\hat{P}_1\right|^2 + \hat{P}_1\right)\left(3\left|\hat{P}_1\right|^2 + 13\hat{P}_1^2 + 16\hat{P}_1\right)\right)\hat{P}_1\left(\cosh(h) + 1\right)\left(\left(1 + 2\hat{P}_1 - \hat{P}_2\right)\cosh^2(h) - \hat{P}_1 - 1\right)\sinh(h)\left(\cosh(h) - 1\right)\Bigg)^{-1} \tag{2.6}$$

## 3. The full $C_{4,2}$ expression

Here, we give the full expression for $C_{4,2}$, defined in **??** as the coefficient of the $\mathcal{O}\left(\varepsilon^4\right)$ correction to the $m = 2$ harmonic of $\eta$.

$$C_{4,2} =$$

$$- 8\cosh(h)\Bigg(\Bigg(-\frac{1}{72}\left(\hat{P}_1\mathrm{COMB}_{3,1}\sinh(h)\cosh(h)\left(12\cosh^2(h)\hat{P}_1 - 4\cosh^2(h)\hat{P}_3 + 8\cosh^2(h) - 9\hat{P}_1 + \hat{P}_3 - 8\right)\left(6\cosh^4(h)\hat{P}_1\right.\right.$$

$$+\cosh^4(h)\hat{P}_2+7\cosh^4(h)+\cosh^2(h)\hat{P}_1+\cosh^2(h)-2\hat{P}_1-2\Big)\omega\Big)-\frac{4}{3}\Bigg(\frac{1}{8}\Bigg(\frac{8}{9}+\hat{P}_1^2+\left(\frac{\hat{P}_2}{12}-\frac{2}{9}\hat{P}_3\right)\hat{P}_1+\frac{1}{36}\hat{P}_2\hat{P}_3+\frac{67\hat{P}_1}{36}+\frac{\hat{P}_2}{9}$$

$$-\frac{7\hat{P}_3}{36}\Bigg)\left|\hat{P}_1\right|+\Bigg(\frac{145}{144}+\Bigg(\hat{P}_1^2+\left(-\frac{5\hat{P}_3}{24}+\frac{\hat{P}_2}{16}\right)\hat{P}_1+\frac{1}{24}\hat{P}_2\hat{P}_3\Bigg)\hat{P}_1+\frac{143\hat{P}_1^2}{48}+\left(\frac{17\hat{P}_2}{96}-\frac{29\hat{P}_3}{72}\right)\hat{P}_1+\frac{13\hat{P}_2\hat{P}_3}{288}+\frac{17\hat{P}_2}{144}-\frac{55\hat{P}_3}{288}$$

$$+\frac{859\hat{P}_1}{288}\Bigg)\hat{P}_1\Bigg)\cosh^8(h)-\frac{4}{3}\Bigg(-\frac{1}{6}\Bigg(\frac{25}{32}+\hat{P}_1^2+\left(-\frac{5\hat{P}_2}{64}-\frac{\hat{P}_3}{6}\right)\hat{P}_1+\frac{5\hat{P}_2\hat{P}_3}{192}+\frac{337\hat{P}_1}{192}-\frac{5\hat{P}_2}{96}-\frac{9\hat{P}_3}{64}\Bigg)\left|\hat{P}_1\right|^2-\frac{53\hat{P}_1}{32}\Bigg(\frac{37}{53}+\Bigg(\hat{P}_1^2+$$

$$\left(-\frac{34\hat{P}_3}{159}-\frac{49\hat{P}_2}{212}\right)\hat{P}_1+\frac{41\hat{P}_2\hat{P}_3}{636}\Bigg)\hat{P}_1+\frac{563\hat{P}_1^2}{212}+\left(-\frac{43\hat{P}_2}{106}-\frac{725\hat{P}_3}{1908}\right)\hat{P}_1+\frac{32\hat{P}_2\hat{P}_3}{477}+\frac{2243\hat{P}_1}{954}-\frac{82\hat{P}_2}{477}-\frac{26\hat{P}_3}{159}\Bigg)\Bigg)\cosh^6(h)$$

$$-\frac{4}{3}\Bigg(\frac{13\left|\hat{P}_1\right|^2}{384}\Bigg(\frac{8}{13}+\hat{P}_1^2+\left(-\frac{6\hat{P}_2}{13}+\frac{2\hat{P}_3}{39}\right)\hat{P}_1+\frac{1}{39}\hat{P}_2\hat{P}_3+\frac{62\hat{P}_1}{39}-\frac{17\hat{P}_2}{39}+\frac{\hat{P}_3}{13}\Bigg)+\frac{59\hat{P}_1}{64}\Bigg(\frac{30}{59}+\hat{P}_1\Bigg(\hat{P}_1^2+\left(-\frac{21\hat{P}_3}{118}-\frac{45\hat{P}_2}{118}\right)\hat{P}_1$$

$$+\frac{8\hat{P}_2\hat{P}_3}{177}\Bigg)+\frac{877\hat{P}_1^2}{354}+\left(-\frac{130\hat{P}_2}{177}-\frac{164\hat{P}_3}{531}\right)\hat{P}_1+\frac{49\hat{P}_2\hat{P}_3}{1062}+\frac{1054\hat{P}_1}{531}-\frac{187\hat{P}_2}{531}-\frac{23\hat{P}_3}{177}\Bigg)\Bigg)\cosh^4(h)+\frac{\left(19\hat{P}_1+19\right)\cosh^2(h)}{96}\Bigg(-\frac{3\left|\hat{P}_1\right|^2}{19}$$

$$\times\left(\frac{25}{27}+\hat{P}_1-\frac{2\hat{P}_3}{27}\right)+\left(\frac{113}{171}+\left(\hat{P}_1-\frac{11\hat{P}_3}{57}\right)\hat{P}_1+\frac{94\hat{P}_1}{57}-\frac{31\hat{P}_3}{171}\right)\hat{P}_1\Bigg)+\frac{1}{16}\left(\hat{P}_1+1\right)^2\left(\frac{1}{8}\left|\hat{P}_1\right|^2+\left(\frac{35}{36}+\hat{P}_1-\frac{11\hat{P}_3}{72}\right)\hat{P}_1\right)\Bigg)\left|\hat{P}_1+1\right|$$

$$+\frac{1}{288}\mathrm{COMB}_{3,1}\sinh(h)\cosh(h)\left(\hat{P}_1+1\right)\left(12\cosh^2(h)\hat{P}_1-4\cosh^2(h)\hat{P}_3+8\cosh^2(h)-9\hat{P}_1+\hat{P}_3-8\right)\Big(4\cosh^4(h)\left|\hat{P}_1\right|^2+36\cosh^4(h)\hat{P}_1^2$$

$$+12\cosh^4(h)\hat{P}_1\hat{P}_2+52\cosh^4(h)\hat{P}_1+12\cosh^2(h)\hat{P}_1^2+12\cosh^2(h)\hat{P}_1-\left|\hat{P}_1\right|^2-15\hat{P}_1^2-16\hat{P}_1\Big)\omega-\frac{1}{12}\left(\hat{P}_1+1\right)\Bigg(\bigg(-12\bigg(\frac{8}{9}+\hat{P}_1^2+\bigg(\frac{\hat{P}_2}{12}$$

$$-\frac{2}{9}\hat{P}_3\bigg)\hat{P}_1+\frac{1}{36}\hat{P}_2\hat{P}_3+\frac{67\hat{P}_1}{36}+\frac{\hat{P}_2}{9}-\frac{7\hat{P}_3}{36}\bigg)\left|\hat{P}_1\right|+\hat{P}_1\bigg(-\frac{25}{3}+\left(\hat{P}_1-\frac{\hat{P}_3}{3}\right)\left(\hat{P}_1+\frac{5}{2}\hat{P}_2\right)\hat{P}_1-\frac{41\hat{P}_1^2}{6}+\left(\frac{19\hat{P}_2}{6}+\frac{7}{6}\hat{P}_3\right)\hat{P}_1-\frac{7}{6}\hat{P}_2\hat{P}_3+\frac{\hat{P}_2}{3}$$

$$+\frac{7}{6}\hat{P}_3-\frac{95\hat{P}_1}{6}\bigg)\bigg)\cosh^8(h)+\Bigg(17\left(\frac{151}{204}+\hat{P}_1^2+\left(-\frac{7\hat{P}_2}{68}-\frac{19\hat{P}_3}{102}\right)\hat{P}_1+\frac{1}{34}\hat{P}_2\hat{P}_3+\frac{349\hat{P}_1}{204}-\frac{5\hat{P}_2}{68}-\frac{8\hat{P}_3}{51}\right)\left|\hat{P}_1\right|^2+\frac{17\hat{P}_1}{4}\bigg(\frac{106}{51}+\bigg(\hat{P}_1^2+$$

$$\left(-\frac{19\hat{P}_3}{51}-\frac{36\hat{P}_2}{17}\right)\hat{P}_1+\frac{31\hat{P}_2\hat{P}_3}{51}\bigg)\hat{P}_1+\frac{230\hat{P}_1^2}{51}+\left(-\frac{206\hat{P}_2}{51}-\frac{15\hat{P}_3}{17}\right)\hat{P}_1+\frac{37\hat{P}_2\hat{P}_3}{51}+\frac{93\hat{P}_1}{17}-\frac{92\hat{P}_2}{51}-\frac{20\hat{P}_3}{51}\bigg)\Bigg)\cosh^6(h)+\Bigg(-\frac{19\left|\hat{P}_1\right|^2}{4}$$

$$
\times\left(\frac{34}{57}+\hat{P}_1^2+\left(-\frac{15\hat{P}_2}{38}-\frac{5\hat{P}_3}{114}\right)\hat{P}_1+\frac{2\hat{P}_2\hat{P}_3}{57}+\frac{89\hat{P}_1}{57}-\frac{41\hat{P}_2}{114}-\frac{\hat{P}_3}{114}\right)-\left(\frac{131\hat{P}_1}{8}\left(\frac{200}{393}+\left(\hat{P}_1^2+\left(-\frac{40\hat{P}_3}{131}-\frac{54\hat{P}_2}{131}\right)\hat{P}_1+\frac{7\hat{P}_2\hat{P}_3}{131}\right)\hat{P}_1\right.\right.
$$

$$
\left.\left.+\frac{337\hat{P}_1^2}{131}+\left(-\frac{116\hat{P}_2}{131}-\frac{224\hat{P}_3}{393}\right)\hat{P}_1+\frac{25\hat{P}_2\hat{P}_3}{393}+\frac{814\hat{P}_1}{393}-\frac{182\hat{P}_2}{393}-\frac{100\hat{P}_3}{393}\right)\right)\cosh^4(h)+\frac{\left(99\hat{P}_1+99\right)\cosh^2(h)}{8}\left(-\frac{29\left|\hat{P}_1\right|^2}{198}\left(\frac{80}{87}+\hat{P}_1\right.\right.
$$

$$
\left.-\frac{7\hat{P}_3}{87}\right)+\hat{P}_1\left(\frac{71}{99}+\hat{P}_1\left(\hat{P}_1-\frac{4\hat{P}_3}{27}\right)+\frac{1013\hat{P}_1}{594}-\frac{3\hat{P}_3}{22}\right)\Bigg)-\frac{9\hat{P}_1+9}{8}\left(-\frac{7\left|\hat{P}_1\right|^2}{12}\left(\hat{P}_1-\frac{4\hat{P}_3}{63}+\frac{59}{63}\right)+\left(\hat{P}_1\left(\hat{P}_1-\frac{\hat{P}_3}{36}\right)+\frac{25\hat{P}_1}{18}+\frac{\hat{P}_3}{108}\right.\right.
$$

$$
\left.\left.\left.\left.+\frac{23}{54}\right)\hat{P}_1\right)\right)\right)\left(\left(\frac{1}{18}\left(-54\hat{P}_1^2+4\hat{P}_1\hat{P}_2-2\left|\hat{P}_1\right|^2-52\hat{P}_1\right)\cosh^2(h)+\frac{3}{2}\hat{P}_1^2+\frac{1}{18}\left|\hat{P}_1\right|^2+\frac{14\hat{P}_1}{9}\right)\left|\hat{P}_1+1\right|+\frac{1}{6}\left(\hat{P}_1+1\right)\left(6\cosh^2(h)\left|\hat{P}_1\right|^2\right.\right.
$$

$$
\left.\left.+18\cosh^2(h)\hat{P}_1^2-4\hat{P}_2\hat{P}_1\cosh^2(h)+20\cosh^2(h)\hat{P}_1-3\left|\hat{P}_1\right|^2-9\hat{P}_1^2-12\hat{P}_1\right)\right)^{-1}\left(\left(\left(1-\hat{P}_2+2\hat{P}_1\right)\cosh^2(h)-\hat{P}_1-1\right)\sinh(h)\right.
$$

$$
\left.\times\left(\cosh(h)+1\right)\left(\cosh(h)-1\right)\left(\left(8+12\hat{P}_1-4\hat{P}_3\right)\cosh^2(h)-8-9\hat{P}_1+\hat{P}_3\right)\right)^{-1} \tag{3.1}
$$